\documentclass[twocolumn]{aastex63}

\pdfoutput=1 %for arXiv submission
\usepackage{appendix}
\usepackage{amsmath,amstext}
\usepackage{amssymb,latexsym}
\usepackage[T1]{fontenc}
\usepackage[figure,figure*]{hypcap}
\usepackage{soul}
\newcommand{\fave}{\langle F \rangle}
\newcommand{\fluxcgs}{10$^9$ erg s$^{-1}$ cm$^{-2}$}

\providecommand{\logg}{cm\,s$^{-2}$}
\newcommand{\mos}{\,m\,s$^{-1}$}
\newcommand{\kms}{\,km\,s$^{-1}$}
\newcommand{\degrees}{$^{\rm{o}}$}
\newcommand{\Msun}{\mbox{M$_{\odot}$}}
\newcommand{\Rsun}{\mbox{R$_{\odot}$}}
\newcommand{\Mjup}{\mbox{M$_{\rm J}$}}
\newcommand{\Rjup}{\mbox{R$_{\rm J}$}}
\newcommand{\Lsun}{\mbox{L$_{\odot}$}}
\newcommand{\Mearth}{\mbox{M$_{\oplus}$}}
\newcommand{\Rearth}{\mbox{R$_{\oplus}$}}

\providecommand{\densitycgs}{g\,cm$^{-3}$}

\accepted{27 August 2021 in The Astronomical Journal}

\shorttitle{TOI-1431\,\MakeLowercase{b}/MASCARA-5\,\MakeLowercase{b}}
\shortauthors{Addison et al.}

\graphicspath{{./}{figures/}}
\begin{document}

\title{TOI-1431\,b/MASCARA-5\,b: A Highly Irradiated Ultra-Hot Jupiter Orbiting One of the Hottest \& Brightest Known Exoplanet Host Stars}

\correspondingauthor{Brett C. Addison}
\email{Brett.Addison@usq.edu.au}

\author[0000-0003-3216-0626]{Brett C. Addison}
\affiliation{University of Southern Queensland, Centre for Astrophysics, USQ Toowoomba, West Street, QLD 4350 Australia}

%main contributing authors. Working on author order.
\author[0000-0001-7880-594X]{Emil Knudstrup}
\affiliation{Stellar Astrophysics Centre, Department of Physics and Astronomy, Aarhus University, Ny Munkegade 120, DK-8000 Aarhus C, Denmark}
%emil@phys.au.dk 

\author[0000-0001-9665-8429]{Ian Wong}
%iwong@mit.edu 
\affiliation{Department of Earth, Atmospheric, and Planetary Sciences, Massachusetts Institute of Technology, Cambridge, MA 02139, USA}
\affiliation{51 Pegasi b Fellow}

\author{Guillaume H\'{e}brard}
\affiliation{Institut dAstrophysique de Paris, UMR7095 CNRS, Universite Pierre \& Marie Curie, 98bis boulevard Arago, 75014 Paris, France}
%hebrard@iap.fr 

\author{Patrick Dorval}
\affiliation{Leiden Observatory, Leiden University, Postbus 9513, 2300 RA Leiden, NL}
%dorval@strw.leidenuniv.nl

\author{Ignas Snellen}
\affiliation{Leiden Observatory, Leiden University, Postbus 9513, 2300 RA Leiden, NL}
%snellen@strw.leidenuniv.nl 

\author[0000-0003-1762-8235]{Simon Albrecht}
\affiliation{Stellar Astrophysics Centre, Department of Physics and Astronomy, Aarhus University, Ny Munkegade 120, DK-8000 Aarhus C, Denmark}
%albrecht@phys.au.dk 

\author[0000-0003-3355-1223]{Aaron Bello-Arufe}
\affiliation{DTU Space, National Space Institute, Technical University of Denmark, Elektrovej 328, DK-2800 Kgs. Lyngby, Denmark}
%aarb@space.dtu.dk

%Sophie data contributors:
\author[0000-0003-3208-9815]{Jose-Manuel Almenara}
%jose-manuel.almenara-villa@univ-grenoble-alpes.fr
\affiliation{Universit\'e Grenoble Alpes, CNRS, IPAG, 38000 Grenoble, France}

\author{Isabelle Boisse}
%isabelle.boisse@lam.fr
\affiliation{Aix Marseille Univ, CNRS, CNES, LAM, Marseille, France}

\author{Xavier Bonfils}
%xavier.bonfils@univ-grenoble-alpes.fr
\affiliation{Universit\'e Grenoble Alpes, CNRS, IPAG, 38000 Grenoble, France}

\author{Shweta Dalal}
%shweta.dalal@iap.fr
\affiliation{Institut dAstrophysique de Paris, UMR7095 CNRS, Universite Pierre \& Marie Curie, 98bis boulevard Arago, 75014 Paris, France}

\author[0000-0001-7918-0355]{Olivier D. S. Demangeon}
%olivier.demangeon@astro.up.pt
\affiliation{Instituto de Astrof{\'\i}sica e Ci\^encias do Espa\c{c}o, Universidade do Porto, CAUP, Rua das Estrelas, 4150-762 Porto, Portugal}
\affiliation{Centro de Astrof\'{\i}sica da Universidade do Porto, Rua das Estrelas, 4150-762 Porto, Portugal}
\affiliation{Departamento de F\'{\i}sica e Astronomia, Faculdade de Ci\^encias, Universidade do Porto, Rua Campo Alegre, 4169-007, Porto, Portugal}

\author[0000-0003-3477-2466]{Sergio Hoyer}
%sergio.hoyer@lam.fr
\affiliation{Aix Marseille Univ, CNRS, CNES, LAM, Marseille, France}

\author{Flavien Kiefer}
%kiefer@iap.fr
\affiliation{LESIA, Observatoire de Paris, Université PSL, CNRS, Sorbonne Université, Université de Paris, 5 place Jules Janssen, 92195 Meudon, France}

\author[0000-0003-4422-2919]{N. C. Santos}
%Nuno.Santos@astro.up.pt
\affiliation{Instituto de Astrof{\'\i}sica e Ci\^encias do Espa\c{c}o, Universidade do Porto, CAUP, Rua das Estrelas, 4150-762 Porto, Portugal}
\affiliation{Departamento de F\'{\i}sica e Astronomia, Faculdade de Ci\^encias, Universidade do Porto, Rua Campo Alegre, 4169-007, Porto, Portugal}

%FIES data contributors:
\author[0000-0002-7031-7754]{Grzegorz Nowak}
\affiliation{Instituto de Astrof\'isica de Canarias (IAC), E-38200 La Laguna, Tenerife, Spain}
\affiliation{Dept. Astrof\'isica, Universidad de La Laguna (ULL), E-38206 La Laguna, Tenerife, Spain}
%gnowak@iac.es

\author{Rafael Luque}
\affiliation{Instituto de Astrof\'isica de Canarias (IAC), E-38200 La Laguna, Tenerife, Spain}
\affiliation{Dept. Astrof\'isica, Universidad de La Laguna (ULL), E-38206 La Laguna, Tenerife, Spain}
%rluque@iac.es 

\author{Monika Stangret}
\affiliation{Instituto de Astrof\'isica de Canarias (IAC), E-38200 La Laguna, Tenerife, Spain}
\affiliation{Dept. Astrof\'isica, Universidad de La Laguna (ULL), E-38206 La Laguna, Tenerife, Spain}
%mstangret@iac.es

\author{Enric Palle}
\affiliation{Instituto de Astrof\'isica de Canarias (IAC), E-38200 La Laguna, Tenerife, Spain}
\affiliation{Dept. Astrof\'isica, Universidad de La Laguna (ULL), E-38206 La Laguna, Tenerife, Spain}
%epalle@iac.es

\author[0000-0003-1001-0707]{Ren\'{e} Tronsgaard}
\affiliation{DTU Space, National Space Institute, Technical University of Denmark, Elektrovej 328, DK-2800 Kgs. Lyngby, Denmark}
%rtr@space.dtu.dk

\author[0000-0002-0865-3650]{Victoria Antoci}
\affiliation{DTU Space, National Space Institute, Technical University of Denmark, Elektrovej 328, DK-2800 Kgs. Lyngby, Denmark}
%antoci@space.dtu.dk

\author[0000-0003-1605-5666]{Lars A. Buchhave}
\affiliation{DTU Space, National Space Institute, Technical University of Denmark, Elektrovej 328, DK-2800 Kgs. Lyngby, Denmark}
%buchhave@space.dtu.dk

%Assistance with Allesfitter and phase curve analysis
\author[0000-0002-3164-9086]{Maximilian N.\ G{\"u}nther}
%maxgue@mit.edu
\affiliation{Department of Physics and Kavli Institute for Astrophysics and Space Research, Massachusetts Institute of Technology, Cambridge, MA 02139, USA}
\affiliation{Juan Carlos Torres Fellow}

\author[0000-0002-6939-9211]{Tansu Daylan}
%tdaylan@mit.edu
\affiliation{Department of Physics and Kavli Institute for Astrophysics and Space Research, Massachusetts Institute of Technology, Cambridge, MA 02139, USA}
\affiliation{Kavli Fellow}

%MUSCAT2 contributors:
\author[0000-0001-9087-1245]{Felipe Murgas}
\affiliation{Instituto de Astrof\'isica de Canarias (IAC), E-38200 La Laguna, Tenerife, Spain}
\affiliation{Dept. Astrof\'isica, Universidad de La Laguna (ULL), E-38206 La Laguna, Tenerife, Spain}
%fmurgas@iac.es

\author[0000-0001-5519-1391]{Hannu Parviainen}
\affiliation{Instituto de Astrof\'isica de Canarias (IAC), E-38200 La Laguna, Tenerife, Spain}
\affiliation{Dept. Astrof\'isica, Universidad de La Laguna (ULL), E-38206 La Laguna, Tenerife, Spain}
%hpparvi@gmail.com 

\author[0000-0002-2341-3233]{Emma Esparza-Borges}
%alu0100966121@ull.edu.es
\affiliation{Instituto de Astrof\'isica de Canarias (IAC), E-38200 La Laguna, Tenerife, Spain}
\affiliation{Dept. Astrof\'isica, Universidad de La Laguna (ULL), E-38206 La Laguna, Tenerife, Spain}

\author[0000-0001-7866-8738]{Nicolas Crouzet}
%nicolas.crouzet@esa.int
\affiliation{European Space Agency (ESA), European Space Research and Technology Centre (ESTEC), Keplerlaan 1, 2201 AZ Noordwijk, The Netherlands}

\author[0000-0001-8511-2981]{Norio Narita}
\affiliation{Komaba Institute for Science, The University of Tokyo, 3-8-1 Komaba, Meguro, Tokyo 153-8902, Japan}
\affiliation{JST, PRESTO, 3-8-1 Komaba, Meguro, Tokyo 153-8902, Japan}
\affiliation{Astrobiology Center, 2-21-1 Osawa, Mitaka, Tokyo 181-8588, Japan}
\affiliation{Instituto de Astrof\'{i}sica de Canarias (IAC), 38205 La Laguna, Tenerife, Spain}
%narita@astron.s.u-tokyo.ac.jp 

\author{Akihiko Fukui}
%afukui@eps.s.u-tokyo.ac.jp
\affiliation{Department of Earth and Planetary Science, The University of Tokyo, Tokyo, Japan}
\affiliation{Instituto de Astrof\'isica de Canarias (IAC), E-38200 La Laguna, Tenerife, Spain}

\author[0000-0003-1205-5108]{Kiyoe Kawauchi}
%kawauchi@eps.s.u-tokyo.ac.jp
\affiliation{Department of Earth and Planetary Science, The University of Tokyo, Tokyo, Japan}

\author[0000-0002-7522-8195]{Noriharu Watanabe}
%noriharu.watanabe@grad.nao.ac.jp
\affiliation{Department of Astronomical Science, The Graduate University for Advanced Studies, SOKENDAI, 2-21-1 Osawa, Mitaka, Tokyo 181-8588, Japan}
\affiliation{National Astronomical Observatory of Japan, NINS, 2-21-1 Osawa, Mitaka, Tokyo 181-8588, Japan}
\affiliation{Astrobiology Center, 2-21-1 Osawa, Mitaka, Tokyo 181-8588, Japan}

%Other contributors. Will work out order

\author[0000-0003-2935-7196]{Markus Rabus}
\affiliation{Departamento de Matem\'atica y F\i'sica Aplicadas, Universidad Cat\'olica de la Sant\'isima Concepci\'on, Alonso de Rivera 2850, Concepci\'on, Chile}
\affiliation{Las Cumbres Observatory, 6740 Cortona Dr., Ste. 102, Goleta, CA 93117, USA}
\affiliation{Department of Physics, University of California, Santa Barbara, CA 93106-9530, USA}
%mrabus@lco.global

\author[0000-0002-5099-8185]{Marshall C. Johnson}
\affiliation{Las Cumbres Observatory, 6740 Cortona Dr., Ste. 102, Goleta, CA 93117, USA}
%mjohnson@lco.global

%MASCARA contributors
\author[0000-0002-6717-1977]{Gilles P. P. L. Otten}
\affiliation{Academia Sinica, Institute of Astronomy and Astrophysics, 11F Astronomy-Mathematics Building, NTU/AS campus, No. 1, Section 4, Roosevelt Rd., Taipei 10617, Taiwan}
\affiliation{Aix Marseille Univ, CNRS, CNES, LAM, Marseille, France}
%gilles.otten@gmail.com

\author[0000-0003-4787-2335]{Geert Jan Talens}
\affiliation{Institut de Recherche sur les Exoplan\`{e}tes, D\'{e}partement de Physique, Universit\'{e} de Montr\'{e}al, Montr\'{e}al, QC H3C 3J7, Canada}
%geert.jan.talens@umontreal.ca

%EXPRES data
\author[0000-0001-9749-6150]{Samuel H. C. Cabot}
\affiliation{Department of Astronomy, Yale University, 52 Hillhouse Ave, New Haven, CT 06511, USA}
%sam.cabot@yale.edu

\author[0000-0003-2221-0861]{Debra A. Fischer}
\affiliation{Department of Astronomy, Yale University, 52 Hillhouse Ave, New Haven, CT 06511, USA}

%SONG contributors
\author{Frank Grundahl}
\affiliation{Stellar Astrophysics Centre, Department of Physics and Astronomy, Aarhus University, Ny Munkegade 120, DK-8000 Aarhus C, Denmark}
%fgj@phys.au.dk 

\author{Mads Fredslund Andersen}
\affiliation{Stellar Astrophysics Centre, Department of Physics and Astronomy, Aarhus University, Ny Munkegade 120, DK-8000 Aarhus C, Denmark}
%madsfa@phys.au.dk

\author{Jens Jessen-Hansen}
\affiliation{Stellar Astrophysics Centre, Department of Physics and Astronomy, Aarhus University, Ny Munkegade 120, DK-8000 Aarhus C, Denmark}
%jjh@phys.au.dk

\author[0000-0003-3803-4823]{Pere Pall\'{e}}
%pere.l.palle@iac.es
\affiliation{Instituto de Astrof\'isica de Canarias (IAC), E-38200 La Laguna, Tenerife, Spain}
\affiliation{Dept. Astrof\'isica, Universidad de La Laguna (ULL), E-38206 La Laguna, Tenerife, Spain}

\author[0000-0002-1836-3120]{Avi Shporer}
\affiliation{Department of Physics and Kavli Institute for Astrophysics and Space Research, Massachusetts Institute of Technology, Cambridge, MA 02139, USA}
%shporeravi@gmail.com

%Please check your affiliations and provide orcid
%\author{Erica J. Gonzales}
%\affiliation{Department of Astronomy and Astrophysics, University of California, 1156 High St. Santa Cruz, CA 95064}
%Did not hear back from Erica regarding co-authorship

\author[0000-0002-5741-3047]{David R. Ciardi}
\affiliation{NASA Exoplanet Science Institute, Caltech/IPAC, 1200 East California Avenue, Pasadena, CA 91125}
%ciardi@ipac.caltech.edu

\author[0000-0003-3964-4658]{Jake T. Clark}
\affiliation{University of Southern Queensland, Centre for Astrophysics, USQ Toowoomba, West Street, QLD 4350 Australia}
%Jake.Clark@usq.edu.au

\author[0000-0001-9957-9304]{Robert A. Wittenmyer}
\affiliation{University of Southern Queensland, Centre for Astrophysics, USQ Toowoomba, West Street, QLD 4350 Australia}
%Rob.Wittenmyer@usq.edu.au

\author{Duncan J. Wright}
\affiliation{University of Southern Queensland, Centre for Astrophysics, USQ Toowoomba, West Street, QLD 4350 Australia}
%Duncan.Wright@usq.edu.au

\author[0000-0002-1160-7970]{Jonathan Horner}
\affiliation{University of Southern Queensland, Centre for Astrophysics, USQ Toowoomba, West Street, QLD 4350 Australia}
%Jonti.Horner@usq.edu.au

%TFOP SG1

%karen.collins@cfa.harvard.edu
\author[0000-0001-6588-9574]{Karen A.\ Collins}
\affiliation{Center for Astrophysics \textbar \ Harvard \& Smithsonian, 60 Garden Street, Cambridge, MA 02138, USA}

%ejensen1@swarthmore.edu
\author[0000-0002-4625-7333]{Eric L.\ N.\ Jensen}
\affiliation{Department of Physics \& Astronomy, Swarthmore College, Swarthmore PA 19081, USA}

%kielkopf@louisville.edu	
\author[0000-0003-0497-2651]{John F.\ Kielkopf} 
\affiliation{Department of Physics and Astronomy, University of Louisville, Louisville, KY 40292, USA}

%rpschwarz@comcast.net
\author[0000-0001-8227-1020]{Richard P. Schwarz}
\affiliation{Patashnick Voorheesville Observatory, Voorheesville, NY 12186, USA}

%gregorsrdoc@gmail.com	
\author{Gregor Srdoc}
\affil{Kotizarovci Observatory, Sarsoni 90, 51216 Viskovo, Croatia}

%mesutyilmaz@ankara.edu.tr
\author[0000-0002-3276-0704]{Mesut Yilmaz}
\affiliation{Ankara University, Department of Astronomy and Space Sciences, TR-06100, Ankara, Turkey}

%hvsenavci@ankara.edu.tr
\author[0000-0002-8961-277X]{Hakan Volkan Senavci}
\affiliation{Ankara University, Department of Astronomy and Space Sciences, TR-06100, Ankara, Turkey}

%bdiamond@howardcc.edu	
\author{Brendan Diamond}
\affiliation{Howard Community College, 10901 Little Patuxent Pkwy, Columbia, MD 21044, USA}

%NRES contributors:
%tbrown@lco.global
%\author{Tim Brown}
%\affiliation{Las Cumbres Observatory, 6740 Cortona Dr., Ste. 102, Goleta, CA 93117, USA}

%dharbeck@lco.global
\author[0000-0002-8590-007X]{Daniel Harbeck}
\affiliation{Las Cumbres Observatory, 6740 Cortona Dr., Ste. 102, Goleta, CA 93117, USA}

%Insights into the cooling timescales and inflation for hot/ultra-hot Jupiters
\author[0000-0002-9258-5311]{Thaddeus D. Komacek}
\affiliation{Department of Astronomy, University of Maryland, College Park, MD 20742, USA}
\affiliation{Department of the Geophysical Sciences, The University of Chicago, Chicago, IL, 60637, USA}
%tkomacek@uchicago.edu

%PDC light curve work determining origin of dip features
%jeffrey.c.smith-1@nasa.gov
\author[0000-0002-6148-7903]{Jeffrey C. Smith}
\affiliation{NASA Ames Research Center, Moffett Field, CA 94035, USA}
\affiliation{SETI Institute, 189 Bernardo Ave., Suite 200, Mountain View, CA  94043, USA}

%others

\author[0000-0002-7846-6981]{Songhu Wang}
\affiliation{Astronomy Department, Indiana University Bloomington, 727 East 3rd Street, Swain West 318, IN 4740, USA}
%sw121@iu.edu

\author[0000-0003-3773-5142]{Jason D.\ Eastman}
%jason.eastman@cfa.harvard.edu
\affiliation{Center for Astrophysics ${\rm \mid}$ Harvard {\rm \&} Smithsonian, 60 Garden Street, Cambridge, MA 02138, USA}
%jason.eastman@cfa.harvard.edu

\author[0000-0002-3481-9052]{Keivan G. Stassun}
%keivan.stassun@vanderbilt.edu
\affiliation{Department of Physics and Astronomy, Vanderbilt University, Nashville, TN 37235, USA}
\affiliation{Department of Physics, Fisk University, Nashville, TN 37208, USA}

%\author[0000-0002-8964-8377]{Samuel N. Quinn}
%squinn@cfa.harvard.edu
%\affiliation{Center for Astrophysics ${\rm \mid}$ Harvard {\rm \&} Smithsonian, 60 Garden Street, Cambridge, MA 02138, USA}
%Did not hear back from Sam regarding co-authorship

%TESS Contributing Authors 
\author[0000-0001-9911-7388]{David W. Latham}
%dlatham@cfa.harvard.edu
\affiliation{Center for Astrophysics ${\rm \mid}$ Harvard {\rm \&} Smithsonian, 60 Garden Street, Cambridge, MA 02138, USA}

% \author{George Ricker}
% %grr@space.mit.edu 
% \affiliation{MIT Kavli Institute for Astrophysics and Space
% Research, Massachusetts Institute of Technology, Cambridge, MA 02139, USA}
% \affiliation{MIT Department of Physics, Massachusetts Institute of Technology, Cambridge, MA 02139, USA}
%Did not hear back from George regarding co-authorship

\author{Roland Vanderspek}
%roland@space.mit.edu
\affiliation{MIT Kavli Institute for Astrophysics and Space
Research, Massachusetts Institute of Technology, Cambridge, MA 02139, USA}

\author[0000-0002-6892-6948]{Sara Seager}
%seager@mit.edu 
\affiliation{MIT Kavli Institute for Astrophysics and Space
Research, Massachusetts Institute of Technology, Cambridge, MA 02139, USA}
\affiliation{Earth and Planetary Science, Massachusetts Institute of Technology, 77 Massachusetts Avenue, Cambridge, MA 02139, USA}
\affiliation{Department of Aeronautics and Astronautics, Massachusetts Institute of Technology, 77 Massachusetts Avenue, Cambridge, MA 02139, USA}

\author[0000-0002-4265-047X]{Joshua N. Winn}
%jnwinn@princeton.edu 
\affiliation{Department of Astrophysical Sciences, Princeton University, Peyton Hall, 4 Ivy Lane, Princeton, NJ 08544, USA}

\author[0000-0002-4715-9460]{Jon M. Jenkins}
%jon.jenkins@nasa.gov  
\affiliation{NASA Ames Research Center, Moffett Field, CA 94035, USA}

%TSO contributors:
% \author{William Fong}
% %willfong@mit.edu
% \affiliation{MIT Kavli Institute for Astrophysics and Space
% Research, Massachusetts Institute of Technology, Cambridge, MA 02139, USA}
%Did not hear back from William regarding co-authorship

\author[0000-0002-2457-272X]{Dana R. Louie}
%danalouie@astro.umd.edu
\affiliation{University of Maryland, Department of Astronomy, College Park, MD 20742-2421, USA}

\author[0000-0002-0514-5538]{Luke G. Bouma}
%luke@astro.princeton.edu
\affiliation{Department of Astrophysical Sciences, Princeton University, 4 Ivy Lane, Princeton, NJ 08544, USA}

%SPOC contributors:
\author[0000-0002-6778-7552]{Joseph D. Twicken}
%joseph.twicken@nasa.gov
\affiliation{NASA Ames Research Center, Moffett Field, CA 94035, USA}
\affiliation{SETI Institute, 189 Bernardo Ave., Suite 200, Mountain View, CA  94043, USA}

%POC, MAST, and GI contributors
% \author{John P. Doty}
% \affiliation{Noqsi Aerospace Ltd., 15 Blanchard Avenue, Billerica, MA 01821, USA}
%jpd@noqsi.com
%Did not hear back from John regarding co-authorship

\author[0000-0001-8172-0453]{Alan M. Levine}
\affiliation{Department of Physics and Kavli Institute for Astrophysics and Space Research, Massachusetts Institute of Technology, Cambridge, MA 02139, USA}
%amlevine@mit.edu

\author{Brian McLean}
\affiliation{Space Telescope Science Institute, 3700 San Martin Drive, Baltimore, MD, 21218, USA}
%mclean@stsci.edu

\begin{abstract}

We present the discovery of a highly irradiated and moderately inflated ultra-hot Jupiter, TOI-1431\,b/MASCARA-5\,b (HD 201033\,b), first detected by NASA's Transiting Exoplanet Survey Satellite mission ({\textit {TESS}}) and the Multi-site All-Sky CAmeRA (MASCARA). The signal was established to be of planetary origin through radial velocity measurements obtained using SONG, SOPHIE, FIES, NRES, and EXPRES, which show a reflex motion of $K=294.1\pm1.1$\,\mos. A joint analysis of the {\textit {TESS}} and ground-based photometry and radial velocity measurements reveals that TOI-1431\,b has a mass of $M_{p}=3.12\pm0.18$\,$\rm{M_J}$ ($990\pm60$\,\Mearth), an inflated radius of $R_{p}=1.49\pm0.05$\,$\rm{R_J}$ ($16.7\pm0.6$\,\Rearth), and an orbital period of $P=2.650237\pm0.000003$\,d. Analysis of the spectral energy distribution of the host star reveals that the planet orbits a bright ($\mathrm{V}=8.049$\,mag) and young ($0.29^{+0.32}_{-0.19}$\,Gyr) Am type star with $T_{\rm eff}=7690^{+400}_{-250}$\,$\rm{K}$, resulting in a highly irradiated planet with an incident flux of $\fave=7.24^{+0.68}_{-0.64}\times$\,\fluxcgs\ ($5300^{+500}_{-470}\mathrm{S_{\oplus}}$) and an equilibrium temperature of $T_{eq}=2370\pm70$\,K. {\textit {TESS}} photometry also reveals a secondary eclipse with a depth of $127^{+4}_{-5}$\,ppm as well as the full phase curve of the planet's thermal emission in the red-optical. This has allowed us to measure the dayside and nightside temperature of its atmosphere as $T_\mathrm{day}=3004\pm64$\,K and $T_\mathrm{night}=2583\pm63$\,K, the second hottest measured nightside temperature. The planet's low day/night temperature contrast ($\sim$420~K) suggests very efficient heat transport between the dayside and nightside hemispheres. Given the host star brightness and estimated secondary eclipse depth of $\sim1000$\,ppm in the K-band, the secondary eclipse is potentially detectable at near-IR wavelengths with ground-based facilities, and the planet is ideal for intensive atmospheric characterization through transmission and emission spectroscopy from space missions such as with the James Webb Space Telescope and the Atmospheric Remote-sensing Infrared Exoplanet Large-survey.

\end{abstract}

\keywords{stars: individual (HD\,201033) --- techniques: radial velocities -- techniques: transits}

\section{Introduction} \label{sec:intro}

Exoplanet discoveries over the past 25 years have revealed that the diversity of alien worlds to be far greater than we had ever imagined. Planets more massive than Jupiter were found on orbits measured in mere days \citep[and became known as ``hot Jupiters'', e.g.][]{HJ1,HJ2,HJ3}, while others were found moving on highly eccentric orbits that had more in common with the Solar System's cometary bodies than its planets \citep[e.g.][]{ecc1,ecc2,ecc3,Wittenmyer2019,Bergmann2021}. And more recently, a new class of close-in gas giants has emerged that are hotter than even the coolest main-sequence stars. These super hot gas giants are known as ultra-hot Jupiters and have dayside temperatures $\gtrsim2200$\,K \citep[e.g.,][]{2018A&A...617A.110P}.

In the early years of the Exoplanet era, the majority of new exoplanet discoveries were made by radial velocity surveys \citep[e.g.][]{51Peg,RV1,RV2,2004A&A...423..385P,RV3,RV4} that mainly targeted stars considered to be `Sun-like': of spectral classes late-F, G, and K. The focus on such stars was, in part, driven by technical necessity --- such stars possess ample spectral lines for radial velocity analysis and typically rotate slowly enough that such lines are sufficiently narrow enough to yield precise radial velocity measurements. As a result, during the first two decades of the Exoplanet era, our knowledge of the diversity of planets was heavily biased towards systems with host stars similar to the Sun.

The situation changed with the launch of the \textit{Kepler} space observatory, designed to search for exoplanets using the transit technique \citep{2010Sci...327..977B}. By surveying a vast number of stars at once, \textit{Kepler} was able to get a true handle on the frequency of short-period planets around a wide variety of stars. It showed that planets are ubiquitous --- a natural byproduct of the star formation process --- and that planetary systems have a wide variety of architectures and compositions \citep[e.g.][]{Kep1,Kep2,Kep3,Kep4,Canas2019,Millholland2016,Millholland2017,2021AJ....161...36B}.

Although the transit technique is better suited to find planets around hot, massive stars than the radial velocity method, such discoveries remain challenging due to the inherent difficulty in confirming planet candidates with radial velocity measurements. This is because more massive stars tend to have rapid rotation, resulting in a significant decrease in the obtainable radial velocity precision \citep[e.g., WASP-167b/KELT-13b, KELT-17b, KELT-20b/MASCARA-2b, KELT-19Ab, HAT-P-69b, WASP-189b, HAT-P-70b, KELT-25b, KELT-26b,][transiting planets orbiting hot stars exclusively or primarily confirmed through Rossiter-McLaughlin or Doppler tomography observations]{2017MNRAS.471.2743T,kelt17,kelt20,2018A&A...612A..57T,siverd18,2018arXiv180904897A,2019AJ....158..141Z,kelt25}.

NASA's \textit{Transiting Exoplanet Survey Satellite} \citep[\textit{TESS};][]{ricker2015} was launched in 2018 and is the successor to the hugely successful \textit{Kepler} and K2 missions. While the \textit{Kepler} spacecraft could only examine a $\sim115^{\circ}$ patch of sky ($\sim0.3\%$ total area of the full sky) at any given time (performing a narrow but deep survey of the sky), \textit{TESS} is designed to survey large bands of the sky at a time (96\degrees\ by 24\degrees). As a result, over the two years of \textit{TESS}' primary mission, it surveyed the majority of the sky, obtaining high cadence observations (one target pixel image recorded every two minutes) of more than 200,000 pre-selected main-sequence dwarf stars, and low cadence observations (one full frame image every thirty minutes) of 20\,million bright stars \citep{ricker2015}. In doing so, it has enabled the search for planets around a wide variety of bright stars (including hot and more massive stars historically avoided in radial velocity surveys) --- perfect targets for ground-based follow-up observations. While we are still in the early stages of reaping the full harvest from \textit{TESS}' remarkable wealth of observations, the spacecraft has already discovered a variety of new exoplanet systems \citep[e.g.][]{TESS1,2019AJ....157...51W,TESS2,TESSBRETT,TESS5,2020AJ....160..117R,TESS3,TESS4,Jones2019,Canas2019,Gunther2019,Jordan2020,Davis2020,Brahm2020} with many planet candidates, or \textit{TESS} Objects of Interests (TOIs), awaiting confirmation \citep{2021ApJS..254...39G}. 

A great benefit of \textit{TESS}' discoveries orbiting bright stars is that those stars make ideal targets for follow-up observations, allowing us to better characterize the planets that orbit them. The most obvious route by which such characterization is achieved is through radial velocity observations, which provide measurements of the masses of the planets. Additionally, radial velocity observations also provide vital information on the planet's orbit, including its eccentricity and orbital period (critical if only one or two transits are observed). These observations can also be used to reveal the presence of otherwise undetectable, non-transiting planetary companions \citep[e.g.][]{mills19,weiss20,sozzetti21}. When radial velocity observations are made during a transit of a planet across the disk of its host star, they can be used to constrain the inclination of the planet's orbit relative to the plane of the host star's equator through the use of the `Rossiter-McLaughlin effect' \citep[e.g.,][]{RM1,RM2}. 

Such observations have yielded startling insights in recent years. Where predictions made based on our knowledge of the Solar System would suggest that exoplanets should, typically, move in roughly the same plane as their host star's equator, Rossiter-McLaughlin observations have revealed that a significant fraction of hot Jupiters move on misaligned orbits. Surprisingly, such observations have even revealed a number of planets moving on polar or retrograde orbits \citep[e.g.][]{2009PASJ...61L..35N,winn09,hirano11,albrecht12,addison13} --- a scenario markedly different from the near coplanarity of the Solar System \citep[as discussed in][and references therein]{SSRev}.

The highly excited orbits of such inclined hot Jupiters are likely linked to their origins. A number of different mechanisms have been proposed to produce hot Jupiters, ranging from the smooth and relatively sedate inward migration of the giant planets through interactions with their host star's protoplanetary disk \citep[e.g.][]{Disk1,Disk2,Disk3,Disk4}, to planet-planet scattering \citep[where close encounters between two planets throw them both onto highly eccentric orbits, dropping one inwards, towards its host, and flinging the other outwards; see e.g.,][]{Scatter1,Scatter2,Scatter3}, and even distant perturbations from binary companions \citep[through the Kozai-Lidov mechanism; see e.g.][]{KozSS,Lidov,Kozai3}. Those mechanisms that invoke a period of high orbital eccentricity then require the tidal circularization of the planet's orbit, which, for small pericenter distances, can happen on timescales far shorter than the lifetimes of the stars involved \citep{Kozai3}. 
%\citep[e.g.][]{circularise1,circularise2}. 

These three methods would naturally produce very different populations of hot Jupiters. Those formed through disk migration would be expected to orbit in essentially the same plane as their host star's equator. Those formed through planet-planet scattering would likely also congregate around such low inclinations, but would be more dispersed/excited, with scattering events able to produce at least moderate orbital inclinations for the planets involved. Planets that migrate as a result of the Kozai-Lidov mechanism, by contrast, would be expected to be found on highly misaligned orbits, as the Kozai-Lidov mechanism causes significant inclination excitation as a byproduct of the eccentricity excitation driving the planetary migration.

Given that more massive stars are statistically more likely to host binary companions \citep[e.g.][]{bin1,bin2,bin3}, especially Am stars such as TOI-1431 \citep[see,][]{2006PASP..118..419B}, and also tend to form more massive planets \citep[e.g.][]{massive1,massive2,massive3}, it would seem reasonable to expect that such stars might be more likely to host both moderately and highly misaligned planets (since the two mechanisms by which excited hot Jupiters can be produced would be more likely to occur). Indeed, as the catalog of planets for which spin-orbit alignment measurements have been made has grown, this prediction appears to have been borne out. The fraction of highly misaligned planets appears to be a strong function of stellar effective temperature (and to a lesser degree stellar mass), with hotter and more massive stars far more likely to host such planets than those that are cooler and less massive \citep[e.g.,][]{2010ApJ...718L.145W,2010A&A...524A..25T,2021AJ....161...68L}.

In this work, we expand the catalog of planets orbiting hot main-sequence stars with the discovery and characterization of TOI-1431~b/MASCARA-5b --- an inflated, ultra-hot Jupiter orbiting a massive, hot, and metal-peculiar Am star.

In Section~\ref{data}, we describe the observations and data reduction. The host star properties derived from the analysis of the broadband spectral energy distribution and spectroscopy are outlined in Section~\ref{host_star}. The data analysis, modeling, and results are presented in Section~\ref{model_results}. We then discuss the phase curve and secondary eclipse analysis and measurements of the planet's dayside and nightside temperatures in Section~\ref{atmo}. Lastly, the discussion and conclusions are provided in Sections~\ref{discussion} and \ref{conclusions}, respectively.

%-----------------------------------------------------------------------
\section{Observations and Data Reduction}
\label{data}
TOI-1431/MASCARA-5 (HD 201033) was observed with space-based photometry by \textit{TESS} (Section~\ref{TESS_photometry}) and ground-based photometry from several facilities (Section~\ref{ground_photometry}). We also obtained follow-up spectroscopic observations from several ground-based facilities (Section~\ref{spectro}) as well as high-contrast imaging on the Keck II telescope (Section~\ref{direct_imaging}) to establish the planetary nature of the transit signals detected by \textit{TESS} and the Multi-site All-Sky CAmeRA. Here we describe the observations collected and the data reduction process.

%-------------------------------------------------------------------------------------

\subsection{TESS Photometry}\label{TESS_photometry}
The star TOI-1431 (HD 201033; TIC 375506058, \citealt{2019AJ....158..138S}) was observed in Sectors 15 (on Camera 2 and CCD chip number 4) and 16 (on Camera 2 and CCD chip number 3) by \textit{TESS} in 2-minute cadence mode nearly continuously between 2019 August 15 and 2019 October 7. The photometric data were processed by the \textit{TESS} Science Processing Operations Center (SPOC) pipeline \citep[see,][for a description of the SPOC pipeline]{jenkins2016}, resulting in two versions of the light curves: Simple Aperture Photometry \citep[SAP, see,][]{2010SPIE.7740E..23T,2020ksci.rept....6M} and Multi-scale Maximum A Posteriori (MAP) Presearch Data Conditioning \citep[PDC, see,][]{2012PASP..124..985S,2014PASP..126..100S,2012PASP..124.1000S}. Multi-scale MAP PDC divides each light curve into three wavelet band passes and then performs a MAP-like (Maximum A Posteriori) correction in each using separate cotrending basis vectors derived within each band-pass. A fit to neighboring targets is used as the Bayesian prior during the a-posteriori fit. We downloaded both versions of the light curves from the NASA's Mikulski Archive for Space Telescopes (MAST). Additionally, we produced a third version of the \textit{TESS} light curve, single-scale MAP PDC \citep[where no band-splitting is performed,][]{2012PASP..124.1000S}, for the analysis described in Section~\ref{model_results}.

\begin{figure*}
  \hspace*{-0.5cm}
  \begin{tabular}{cc}
  \includegraphics[width=0.48\textwidth,height=0.33\textwidth]{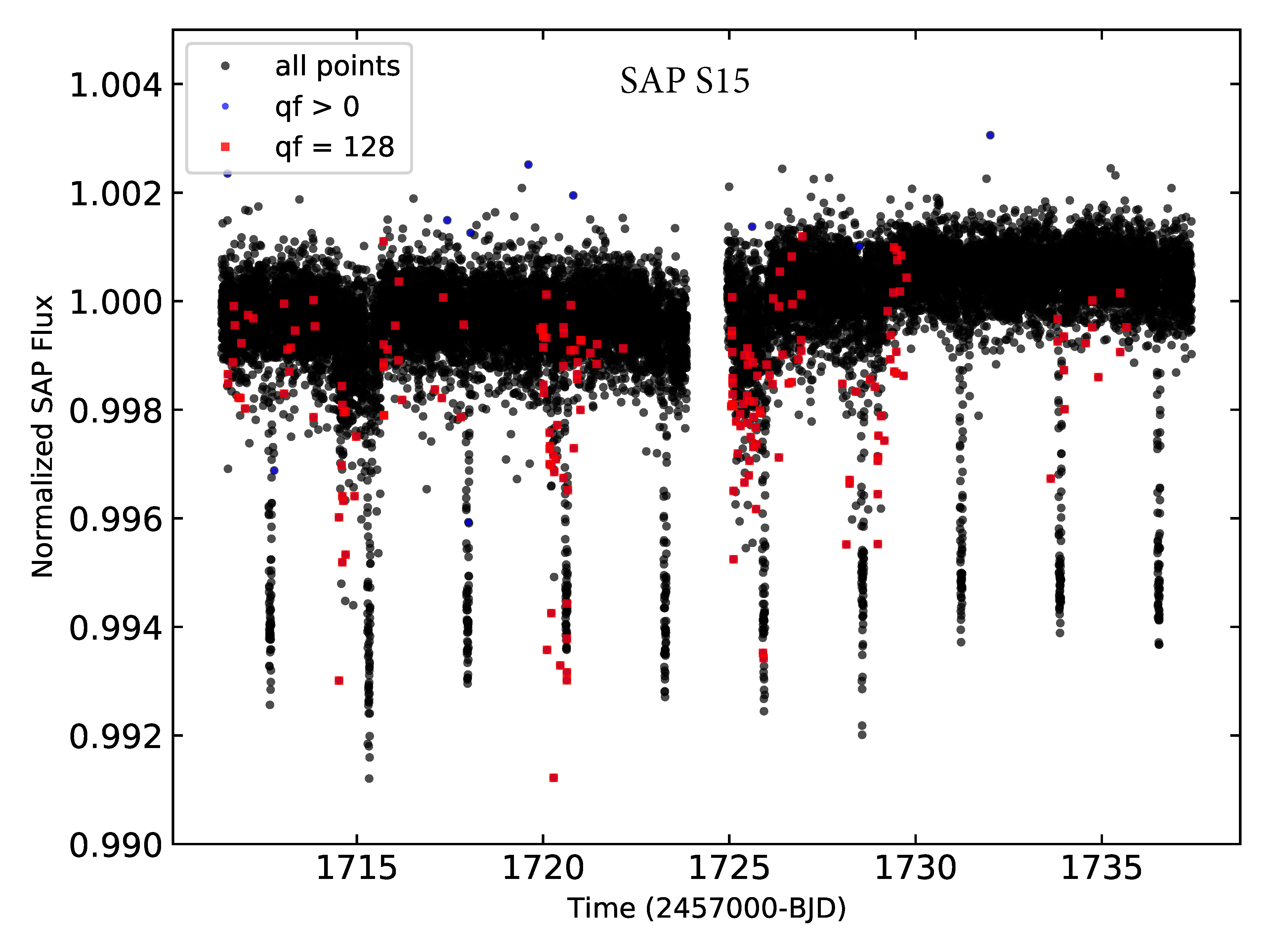} &
  \includegraphics[width=0.48\textwidth,height=0.33\textwidth]{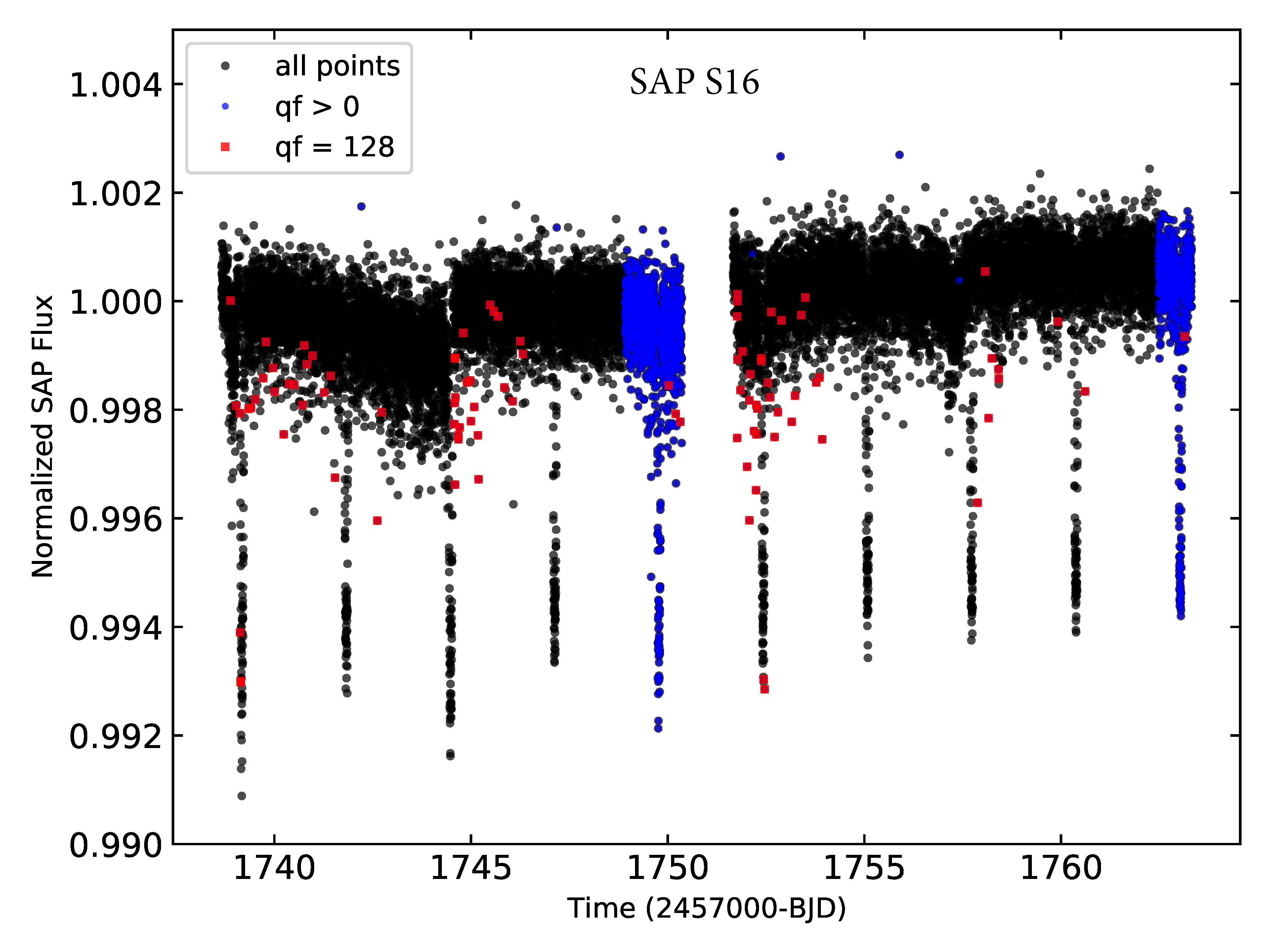} \\
  \includegraphics[width=0.48\textwidth,height=0.33\textwidth]{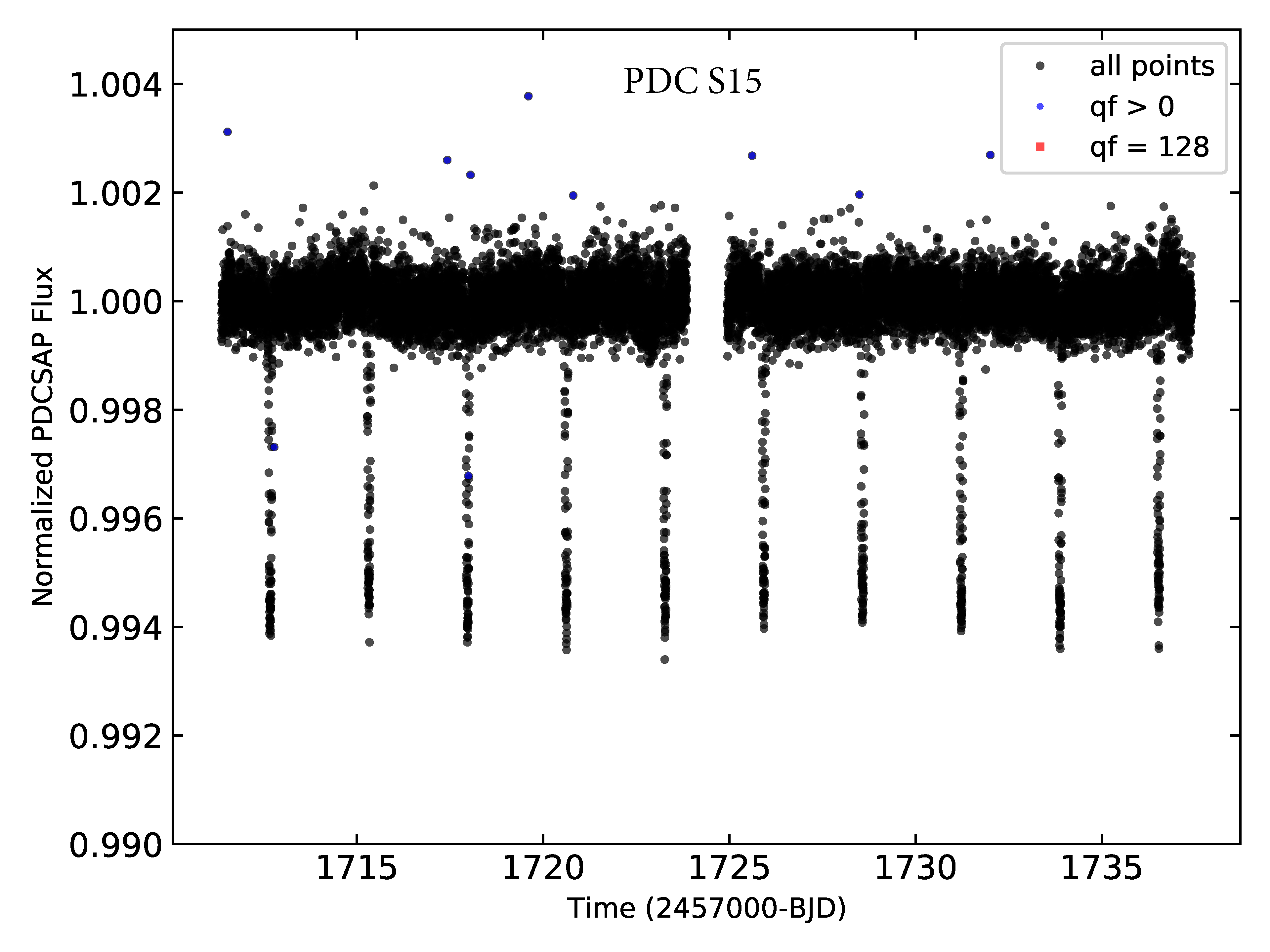} &
  \includegraphics[width=0.48\textwidth,height=0.33\textwidth]{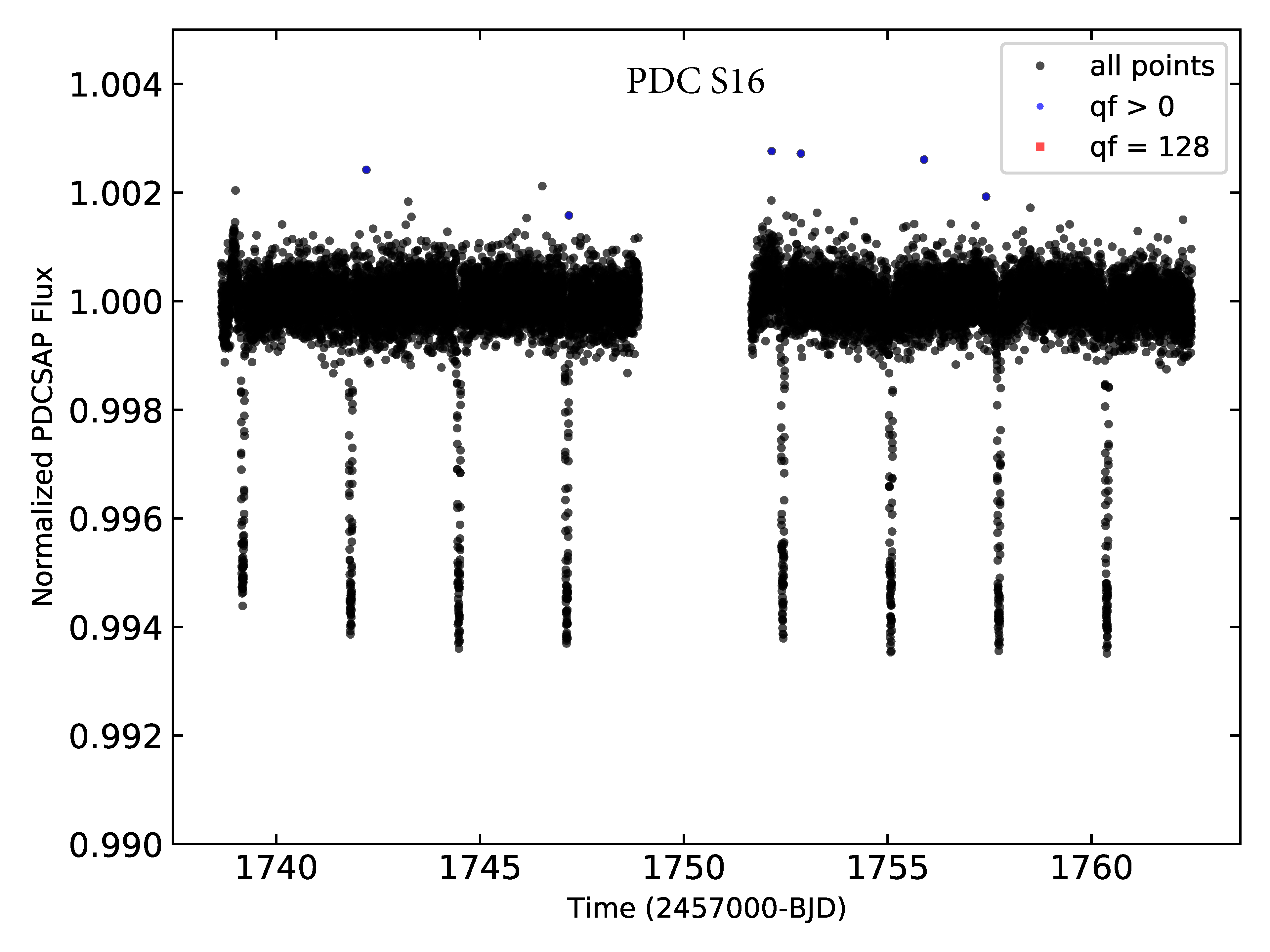}
  \end{tabular}
  \caption{The \textit{TESS} light curves of TOI-1431 from Sectors 15 (left panel) and 16 (right panel). The top panels show the Simple Aperture Photometry (SAP) and the bottom panels show the Presearch Data Conditioning (PDC) versions of the light curves. The circles plotted in blue represent data that have a quality flag greater than 0 (but not equal to 128) and the red squares are data with quality flag equal to 128 (flagged as manual exclude, i.e., the cadence was excluded because of an anomaly). The \textit{TESS} quality flags are described in the \textit{TESS} Data Products Description Document (see, \url{https://archive.stsci.edu/missions/tess/doc/EXP-TESS-ARC-ICD-TM-0014-Rev-F.pdf}.}
  \label{tessphotometry}
\end{figure*}

The transiting planet candidate was promoted to TOI status \citep[see TOI catalog,][]{2021ApJS..254...39G} and designated TOI-1431.01 by the \textit{TESS} Science Office based on model fit results and a passing grade on all diagnostic tests in the SPOC Data Validation (DV) report \citep{2018PASP..130f4502T,2019PASP..131b4506L} for Sector 15. Twenty transits (10 per sector), were observed by \textit{TESS}.
Each transit had a duration of $\sim$2.5\,hr, a depth of $\sim$6000 parts per million (ppm), and recurred with a period of $2.65$\,d. The first detected transit occurred on BJD$_{\mathrm{TDB}}$ 2458712 in Sector 15. The raw SAP and multi-scale MAP PDC light curves are shown in Figure~\ref{tessphotometry}.

Before detrending and fitting the SAP and the two versions of the PDC light curves, we first removed all quality-flagged data. The light curves were then split into a total of 10 segments, with each segment split at the spacecraft momentum dumps\footnote{As provided in the data release notes found at \url{https://archive.stsci.edu/tess/tess_drn.html}} at 4.25\,day and 5.83\,day intervals in Sectors 15 and 16, respectively, to minimize the offset effects on the light curves during the detrending and fitting analysis. The transits were then masked and a $5\sigma$ median filter was applied to the out-of-transit data to remove remaining outliers in each light curve segment. Lastly, the light curve segments were normalized with the mean of the out-of-transit flux before performing the detrending and fitting analysis as described in Section~\ref{model_results}.

\subsection{Ground-Based Transit Photometry}\label{ground_photometry}
\begin{figure*}
  \includegraphics[width=\linewidth]{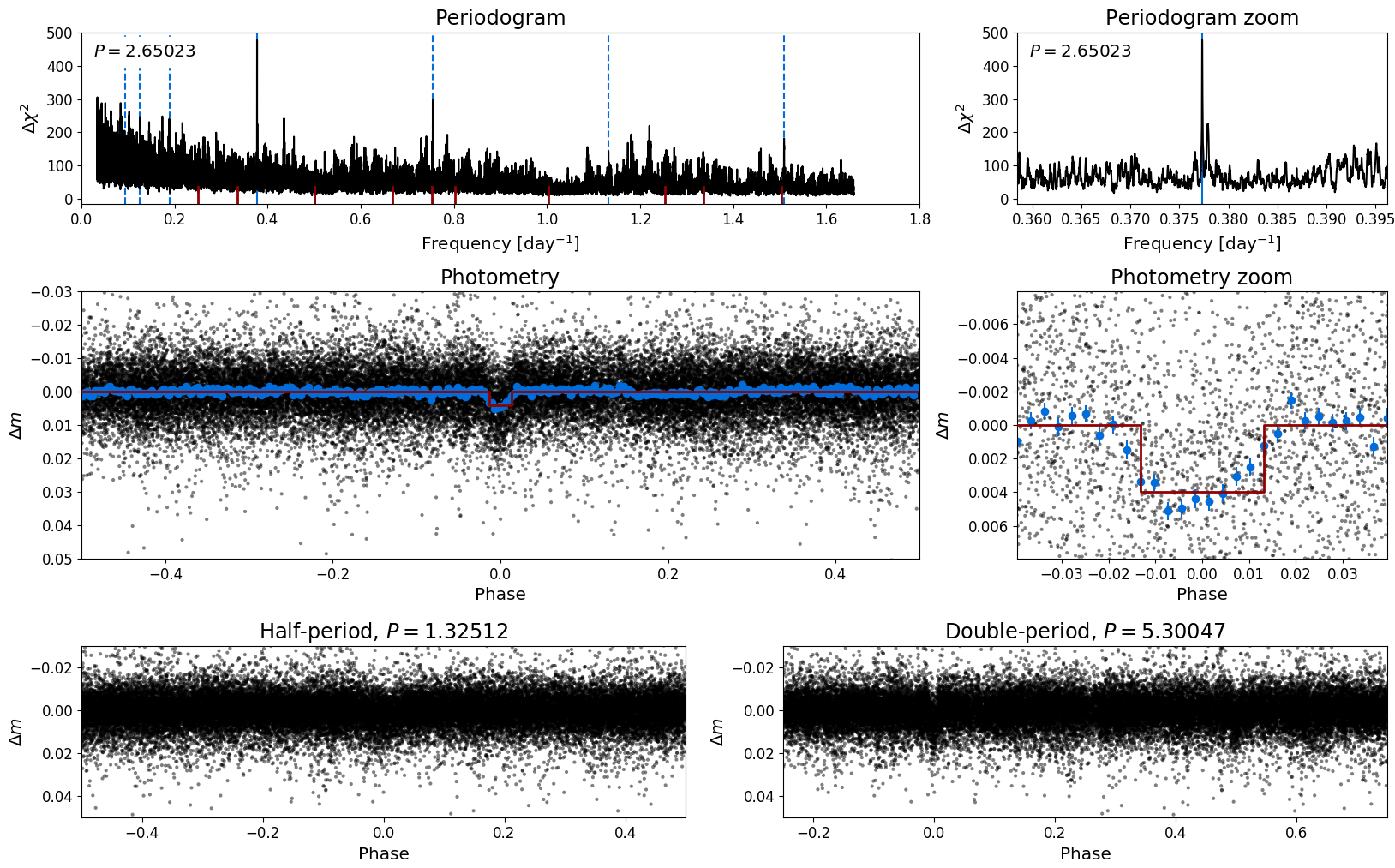}
  \caption{Photometry of TOI-1431 taken from the Multi-site All-Sky CAmeRA (MASCARA) telescope at the Roque de los Muchachos Observatory, La Palma. The top panels are the box least squares periodograms from the MASCARA light curves showing the strongest peak period at 2.65023\,d. The middle panels show the phase folded light curve from MASCARA, full-phase on the left and zoomed in on the transit on the right. The blue points represent binned data in 0.003 phase steps, the black points are the original observations in bins of 50 images (320\,s), and the red line is the best-fit box least squares model \citep[see,][]{2002A&A...391..369K} used in detecting the transit. The bottom panels show the photometry phase-folded at half and double the period (bottom left and right, respectively) found at the strongest peak in the periodogram. No obvious dip features are seen in the half-period phase-folded plot but two dip features are apparent in double-period phase-folded plot at phase 0 and 0.5, as expected for the transiting planet with a period of 2.65023\,d. The MASCARA light curves were not used in the global modeling in \texttt{Allesfitter}.}
  \label{fig:MASCARA}
\end{figure*}

Transit signals of TOI-1431\,b were first detected by the ground-based Multi-site All-Sky CAmeRA (MASCARA) telescope \citep{2017A&A...601A..11T} at the Roque de los Muchachos Observatory, La Palma, starting in February 2015. Transits were observed by MASCARA between February 2015 to March 2018 and were discovered before {\textit {TESS}} observed them. Figure~\ref{fig:MASCARA} shows the phase-folded light curve of TOI-1431\,b from MASCARA with the transit signal clearly detected at the same $\sim2.65$\,d period and approximate transit depth as measured by {\textit {TESS}}. The MASCARA observations were taken with exposure times of 6.4\,s without using any filters and achieve a photometric precision of $\sim5000$\,ppm, per 5\,m binned observation.

Both MASCARA and {\textit {TESS}} have large on-sky pixel sizes of 1\,\arcmin\,pixel$^{-1}$ \citep{2017A&A...601A..11T} and 0.35\,\arcmin\,pixel$^{-1}$ \citep{ricker2015}, respectively. As such, further photometric follow-up is required to rule out potential false positive scenarios such as nearby eclipsing binaries, where an eclipsing binary pair falls on or very close to the same pixel as the target star in the MASCARA and {\textit {TESS}} images.

To confirm the transit signals detected by {\textit {TESS}} and MASCARA were coming from TOI-1431 and to check for any changes in transit depth between multiple photometric bands (chromaticity), we acquired additional photometric follow-up observations from several ground-based facilities through the {\textit {TESS}} Follow-up Observing Program (TFOP) Working Group as summarized in Table~\ref{tab:SG1_summary} in the Appendix. We used the \textit{TESS} Transit Finder which is a customized version of the Tapir software package \cite[][]{Jensen:2013}, to schedule the observations. These facilities include the 0.36\,m CDK14 telescope at Howard Community College, the MuSCAT2 imager on the 1.52\,m Carlos Sanchez Telescope (TCS) at the Teide Observatory (OT), the 0.8\,m telescope at the Ankara University Kreiken Observatory (AUKR), the 0.3\,m telescope at the Kotizarovci Observatory (SCT) in Rijeka, Croatia, the 0.6\,m University of Louisville Manner Telescope (ULMT) at Mt. Lemmon, and the Las Cumbres Observatory Global Telescope \citep[LCOGT;][]{Brown:2013} 1-m node at the McDonald Observatory. For the final global analysis with \texttt{Allesfitter}, we included only the high-quality and high-precision (RMS\,$\lesssim3$\,ppt) ground-based observations with full transit coverage that are free of significant systematics to provide precise transit depth measurements and accurate transit ephemeris. As such, we have not included the light curve taken on 24 May 2020 from MuSCAT2 (but we do include the 16 May 2020 observation) due to significant systematics (see Section~\ref{MuSCAT2}), ULMT taken on the 28 September 2020 due to systematics and moderately low precision photometry obtained ($\sim3.4$\,ppt), nor the photometry from MASCARA due to its relatively low photometric precision of $\sim5$\,ppt.

\subsubsection{CDK14}\label{CDK14}
One full transit of TOI-1431\,b was observed on 24-25 December 2019 with the CDK14 telescope at Howard Community College. The observations were taken using alternating 10\,second exposures in the Sloan $g'$ and $z_s$ filters, starting at 23:02\,UT on the 24th December at an airmass of 1.2 and finishing at 02:56\,UT on the 25th December at an airmass of 2.4. We reduced the data and extracted the light curves for each filter using the {\tt AstroImageJ} ({\tt AIJ}) software package \citep{2017AJ....153...77C} with 7\,pixel apertures. Figure~\ref{ground_lcs} shows the transit signatures in the two bands with fitted transit depths of $5.1^{+0.6}_{-0.7}$\,ppt and $5.1^{+0.5}_{-0.6}$\,ppt for the Sloan $g'$ and $z_s$ filters, respectively. These transit depths are consistent with the SPOC pipeline transit depth of $5.68\pm0.03$ ppt to within $\sim1\sigma$.

\subsubsection{MuSCAT2}\label{MuSCAT2}
Two full transits of TOI-1431\,b were observed with MuSCAT2, one on 16 May 2020 and a second one on 24 May 2020, using simultaneous multi-color photometry in $g'$, $r'$, $i'$, and $z_s$ bands \citep{2019JATIS...5a5001N}. For the joint fitting, we only used the transit observed on 16 May 2020 since the second transit observation was affected by systematics due to weather, resulting in different transit depths between the filter bands and residuals larger than 1.0\,ppt. On the night of 16 May 2020, de-focused observations were taken from 01:50\,UT to 05:22\,UT, with exposure times of 8\,s for all channels, except for the $r'$, where an exposure time of 5\,s was used for auto-guiding. We reduced the data and extracted the light curves using the standard MuSCAT2 pipeline \citep{2019A&A...630A..89P}, recovering the expected transit signal as shown in Figure~\ref{ground_lcs}. Figure~\ref{fig:Muscat2} in the Appendix shows the transit light curve in all four bands with fits to the transits and systematics, fits to the transits with systematics removed, and the resulting residuals.

\subsubsection{AUKR}\label{AUKR}
A full transit of TOI-1431\,b was observed with the 0.8\,m Prof. Dr. Berahitdin Albayrak Telescope (T80) at the Ankara University Kreiken Observatory (AUKR) on 16 June 2020 in the Sloan $z_s$ band. Observations started approximately 30\,minutes before transit ingress at 21:15\,UT and lasted until 01:13\,UT the following day, with exposure times of 40\,seconds and the target was observed at an airmass between 1.0 and 1.3. We extracted the photometry using the {\tt AIJ} software package with 12\,pixel ($8.16\arcsec$) apertures and the resulting fit to the light curve is shown in Figure~\ref{ground_lcs2}. The measured transit depth of $4.5^{+0.5}_{-0.6}$\,ppt is moderately shallower compared to the transit depths found from the \textit{TESS} and other ground-based light curves, likely due to unaccounted for systematics.

\subsubsection{SCT}\label{SCT}
We used the 0.3\,m telescope at the Kotizarovci Observatory (SCT) near Viskovo, Croatia, to observe a full transit of TOI-1431\,b on 8 August 2020. The transit was observed in the \textit{TESS}-like band with 20\,second exposures continuously for nearly 6.5\,hours starting at 20:23\,UT and ending at 02:45\,UT in low airmass that ranged from 1.1\,\arcsec at the start to 1.3\,\arcsec at the end of the observations. The photometry was processed using the {\tt AIJ} software package with an aperture of 11\,pixels ($13.2\arcsec$) and Figure~\ref{ground_lcs2} shows the resulting fit to the light curve. We measure a transit depth of $5.6^{+0.2}_{-0.3}$\,ppt, consistent with the depth measured from the \textit{TESS} light curves.

\subsubsection{ULMT}\label{ULMT}
On 20 September 2020, we observed a full transit of TOI-1431\,b from the 0.6\,m University of Louisville Manner Telescope (ULMT) at Mt. Lemmon using the Sloan $z'$ filter. We set the exposure length to 32\,seconds and started the observations at 04:29\,UT. The observing was interrupted between 06:57\,UT and 07:31\,UT, around the start of transit ingress, due to a guiding error and then resumed afterwards until 10:26\,UT. The target was observed at an airmass of 1.1 at the start and airmass of 1.8 at the end of the observations. The images were bias and dark corrected and the light curve was extracted using an aperture of 13\,pixels ($5.1\arcsec$) with the {\tt AIJ} software package. We measure a transit depth of $5.8\pm0.3$\,ppt (see Figure~\ref{ground_lcs2}), in excellent agreement with the depth measured from the \textit{TESS} photometry.

\subsubsection{LCOGT}
A full transit of TOI-1431\,b was observed on 14 October 2020 from the LCOGT 1.0\,m network node at McDonald Observatory near Fort Davis, Texas (LCO-McD) in PANSTARRS Y-band. The SINISTRO camera images were calibrated using the standard LCOGT {\tt BANZAI} pipeline \citep{McCully:2018}, and the photometry was extracted using the {\tt AIJ} software package with 20\,pixel ($7.8\arcsec$) apertures. Figure~\ref{ground_lcs2} shows a clear transit detection with an individually fitted transit depth of $5.7\pm0.3$\,ppt, consistent with the SPOC pipeline transit depth of $5.68\pm0.03$ ppt to within $1\sigma$.

\begin{figure*}[hbt!]
    \includegraphics[width=\columnwidth]{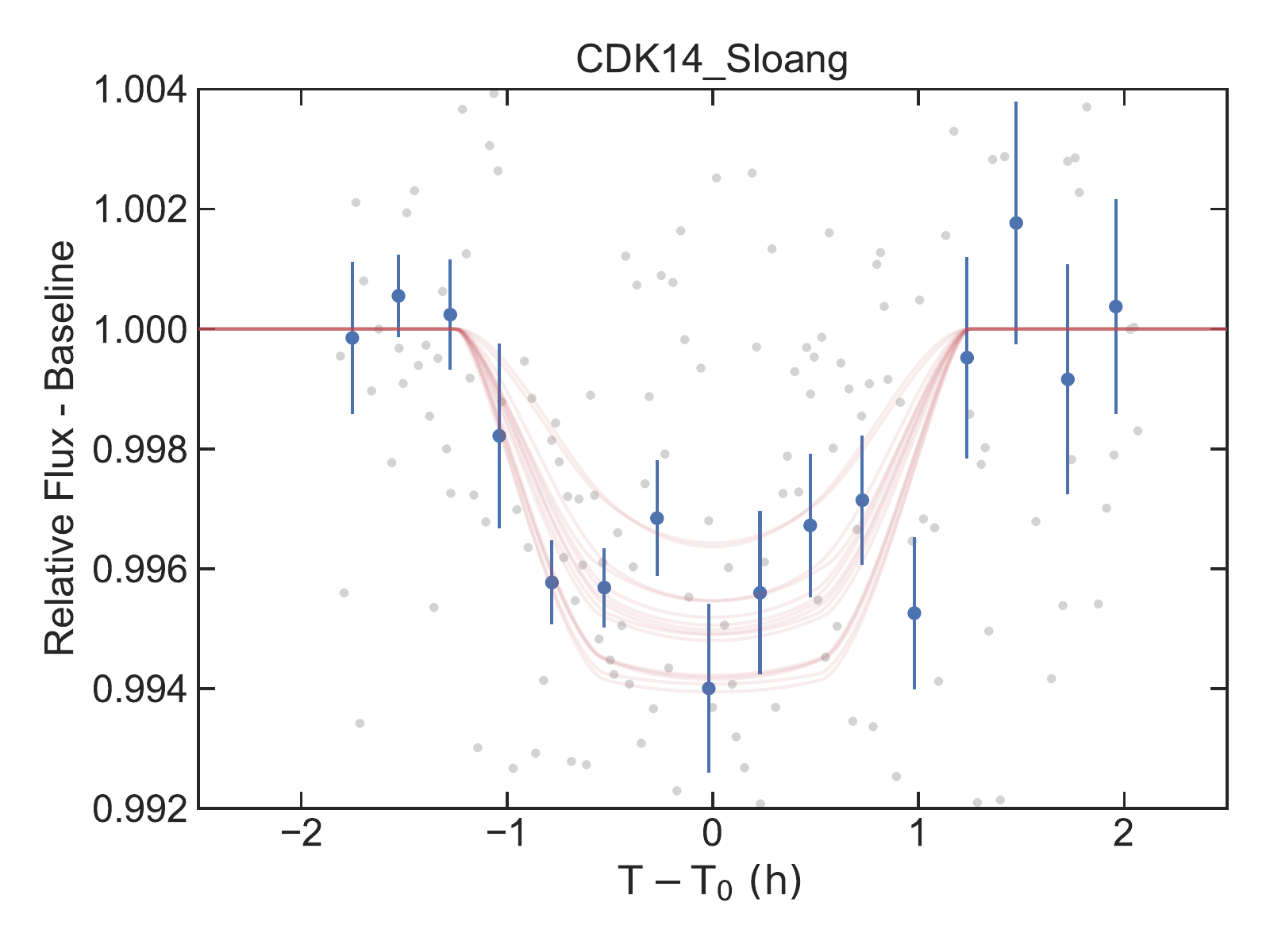}
    \includegraphics[width=\columnwidth]{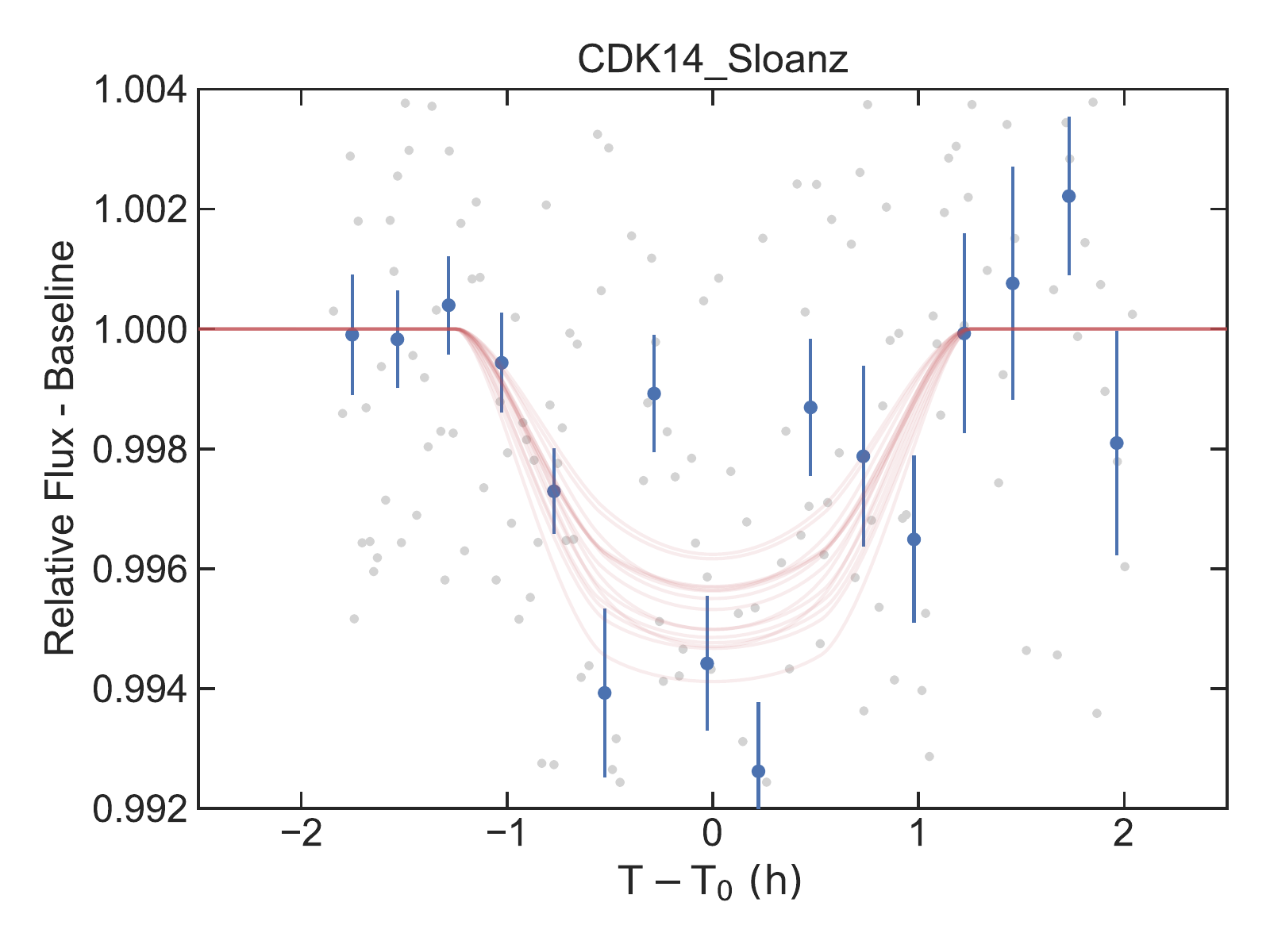}
  
    \medskip
  
    \includegraphics[width=\columnwidth]{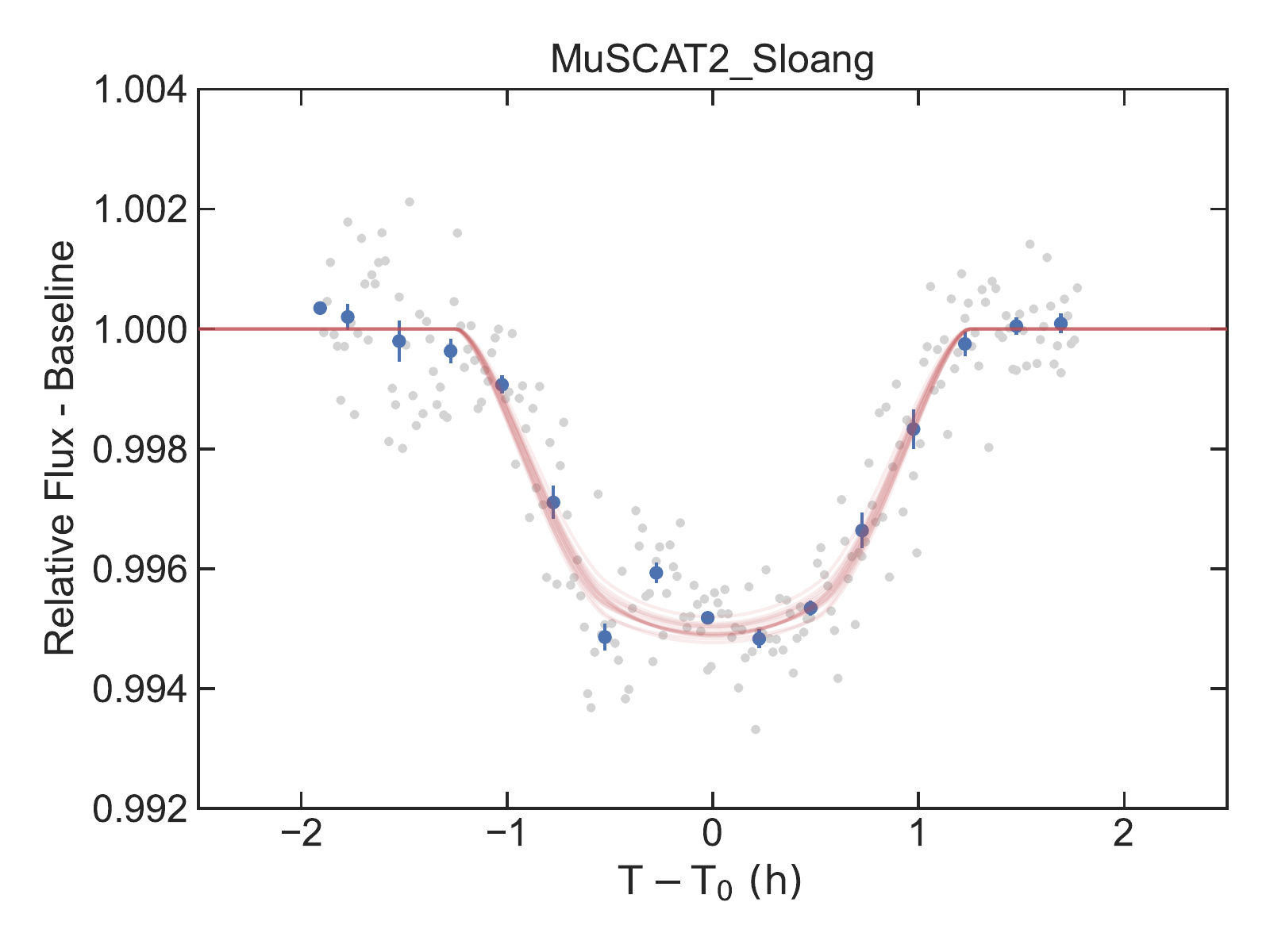}
    \includegraphics[width=\columnwidth]{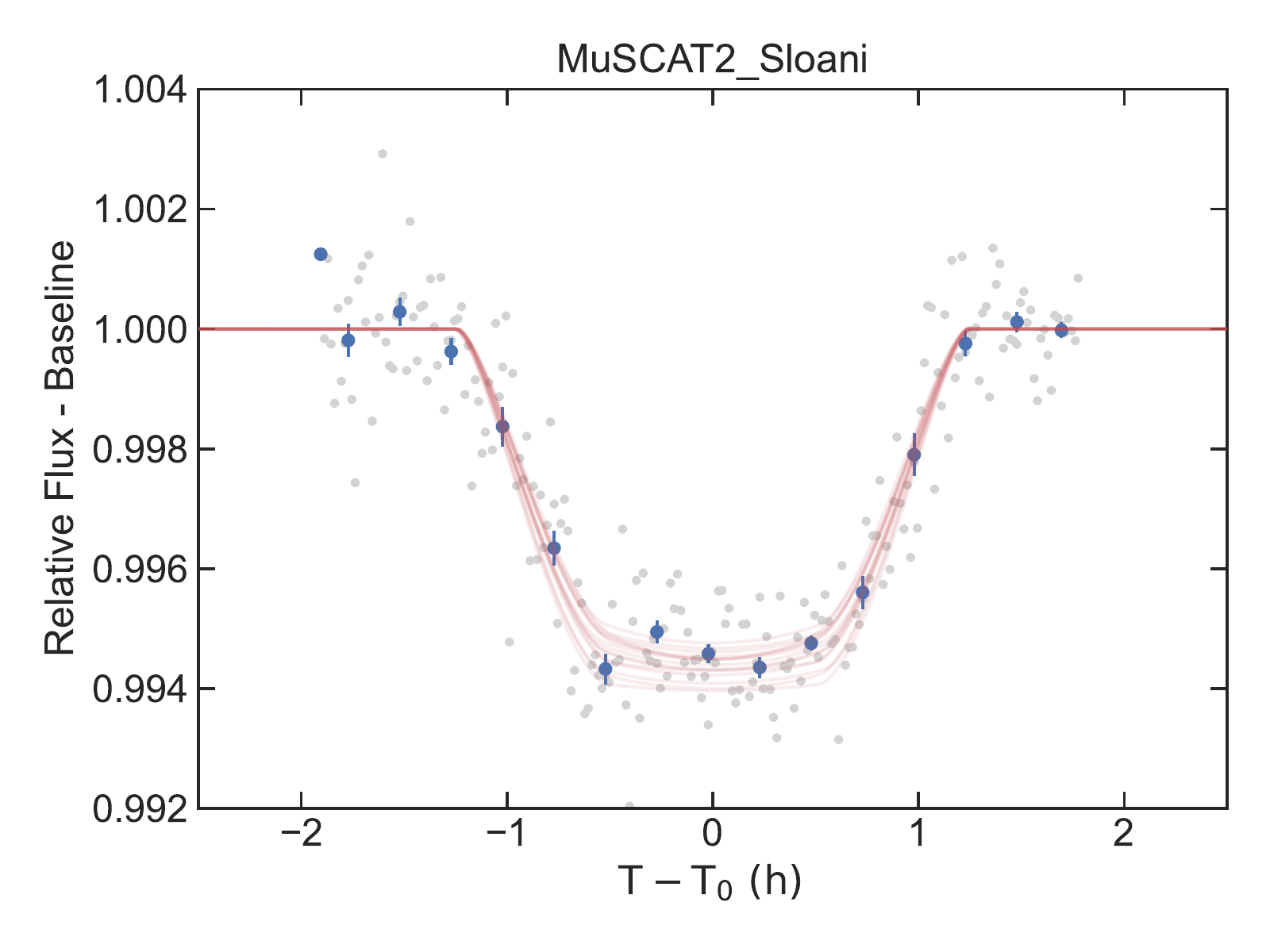}
  
    \medskip
  
    \includegraphics[width=\columnwidth]{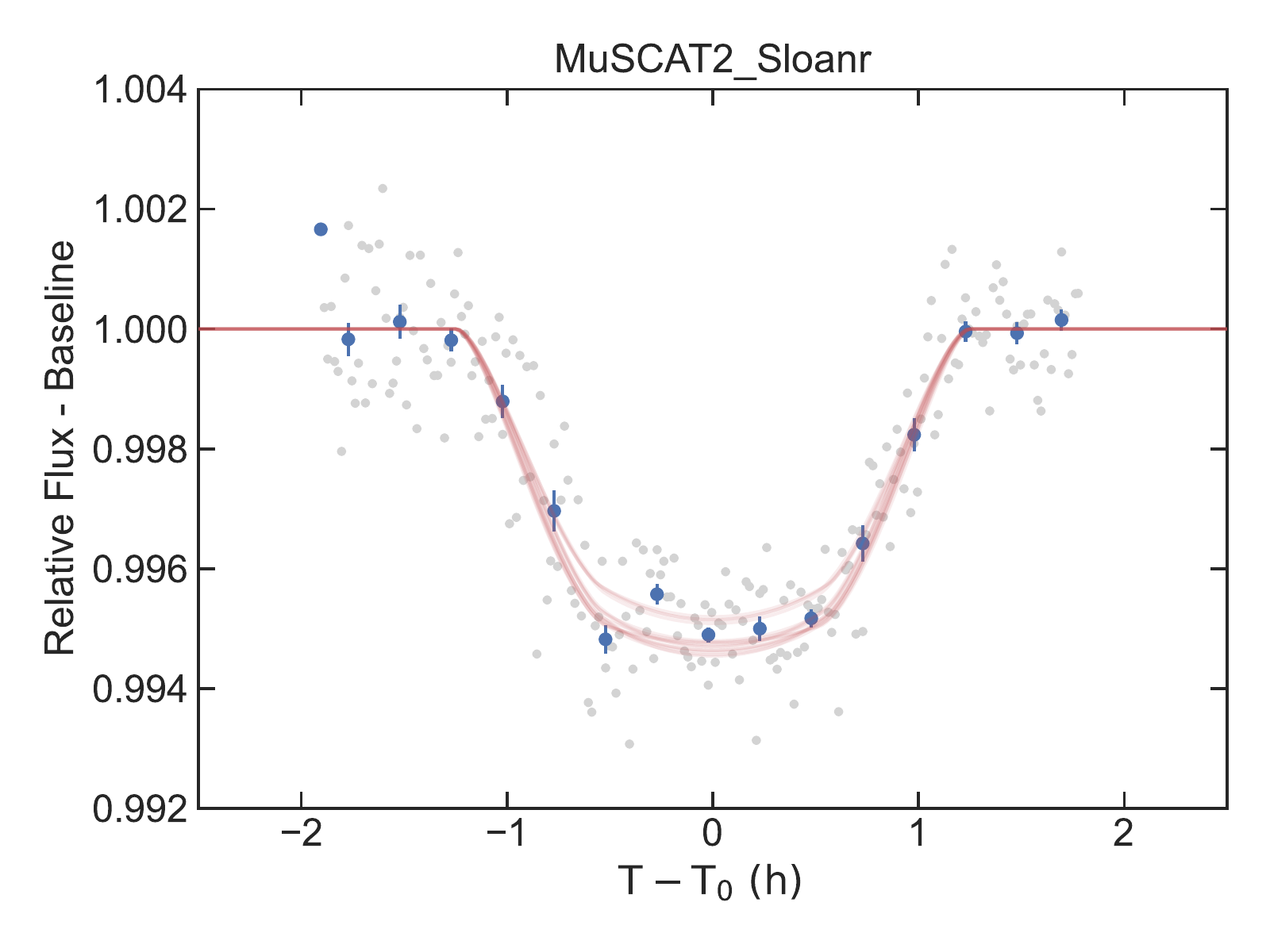}
    \includegraphics[width=\columnwidth]{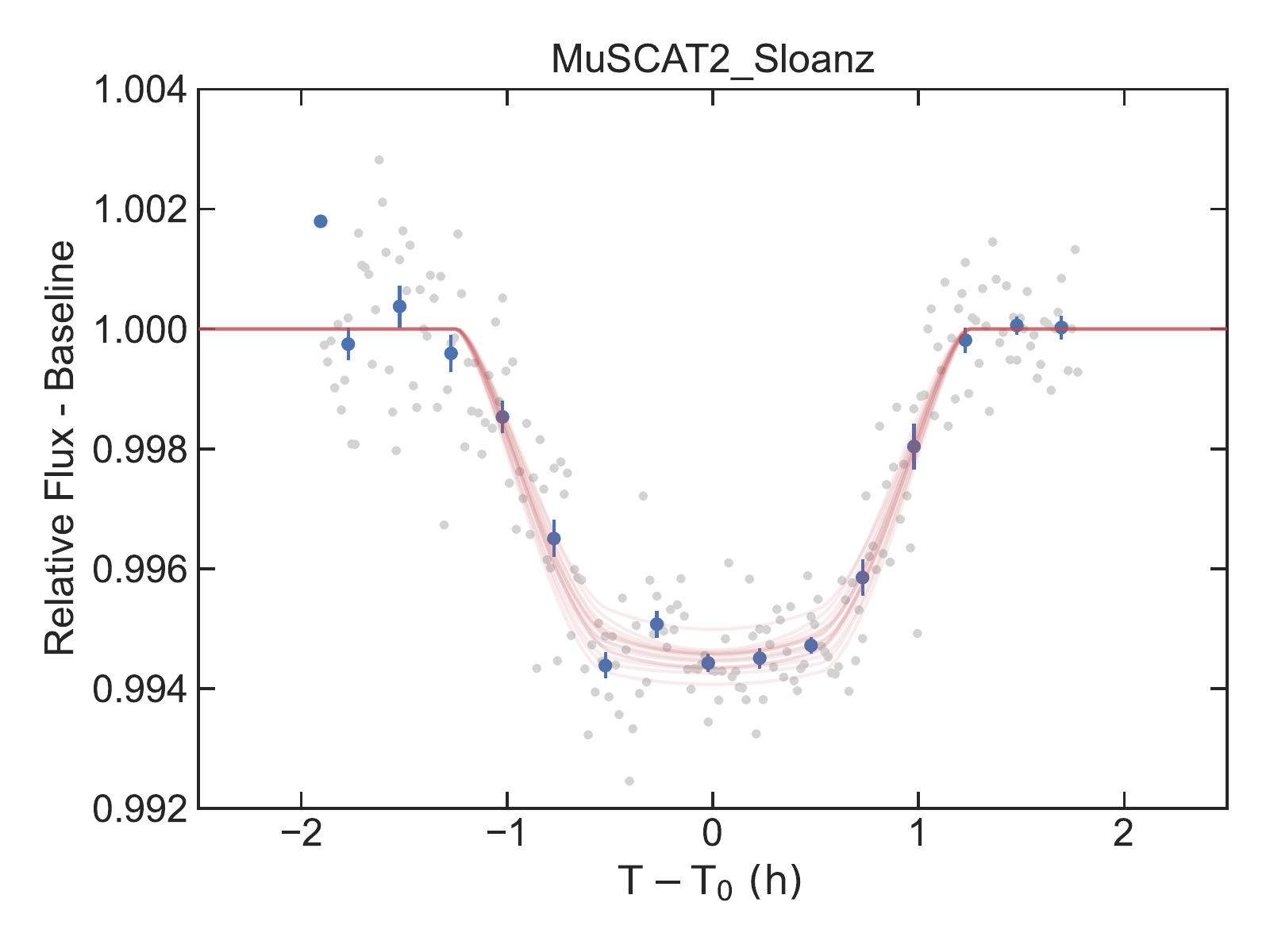}
    
    \caption{The ground-based transit light curves of TOI-1431\,b/MASCARA-5\,b as a function of time from the observations taken with the two CDK14 photometric bands and the four MuSCAT2 photometric bands. The normalized photometric measurements are represented by the gray points while the blue points are the binned photometry. The 15 red lines are the randomly drawn transit light curve models from the Nested Sampling posterior.}
    \label{ground_lcs}
\end{figure*}

\begin{figure*}[hbt!]
    \includegraphics[width=\columnwidth]{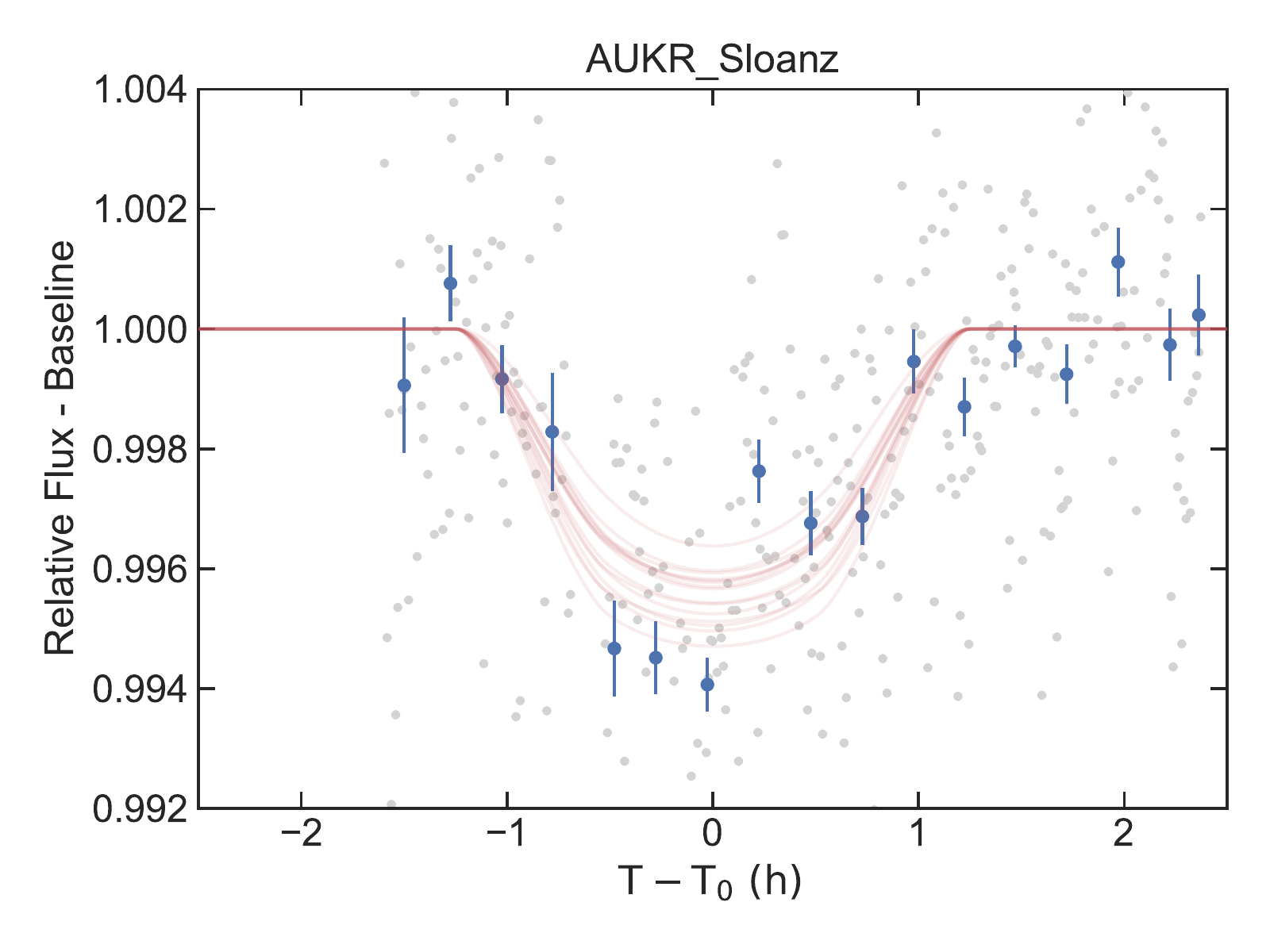}
    \includegraphics[width=\columnwidth]{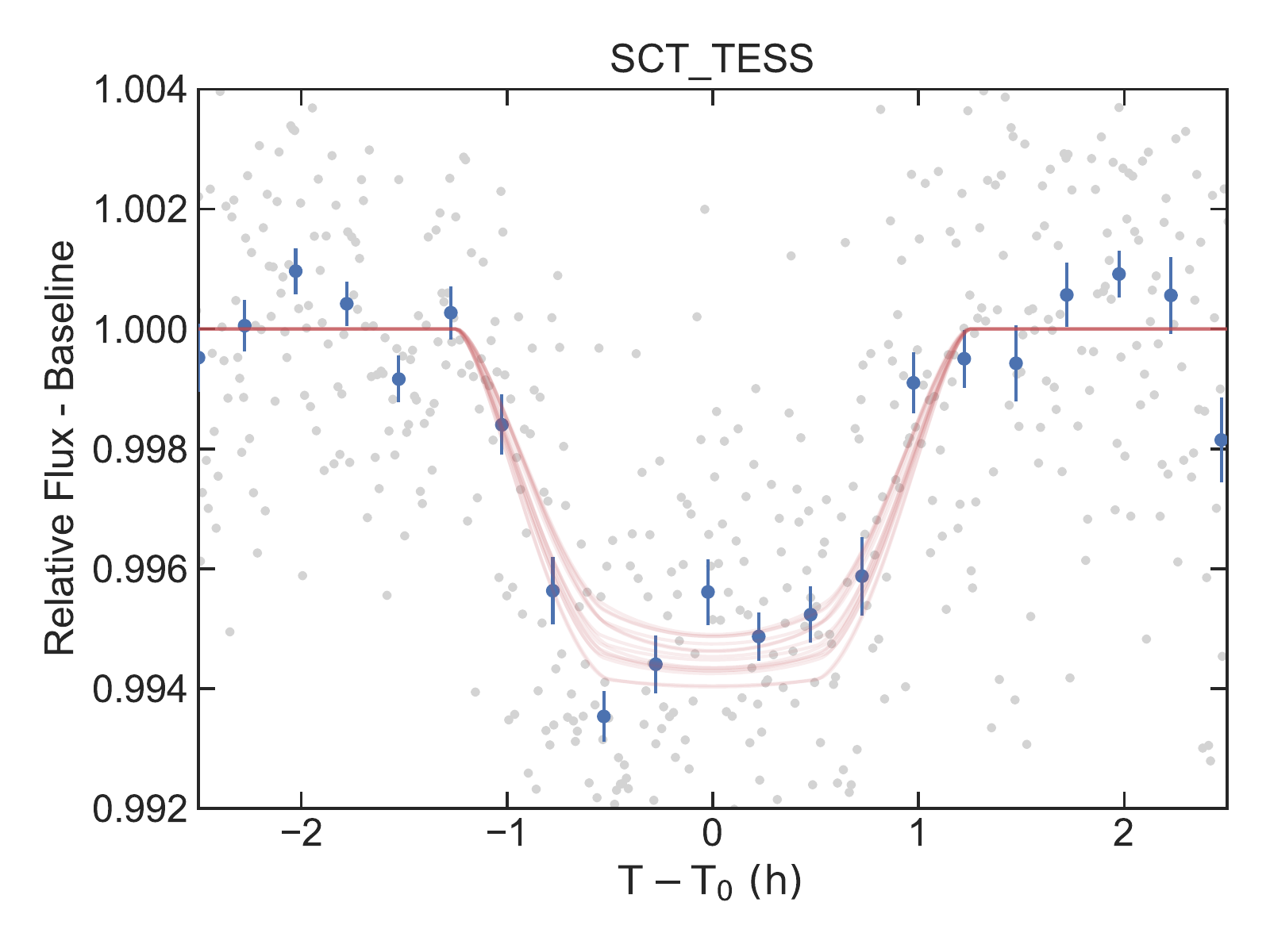}
  
    \medskip
  
    \includegraphics[width=\columnwidth]{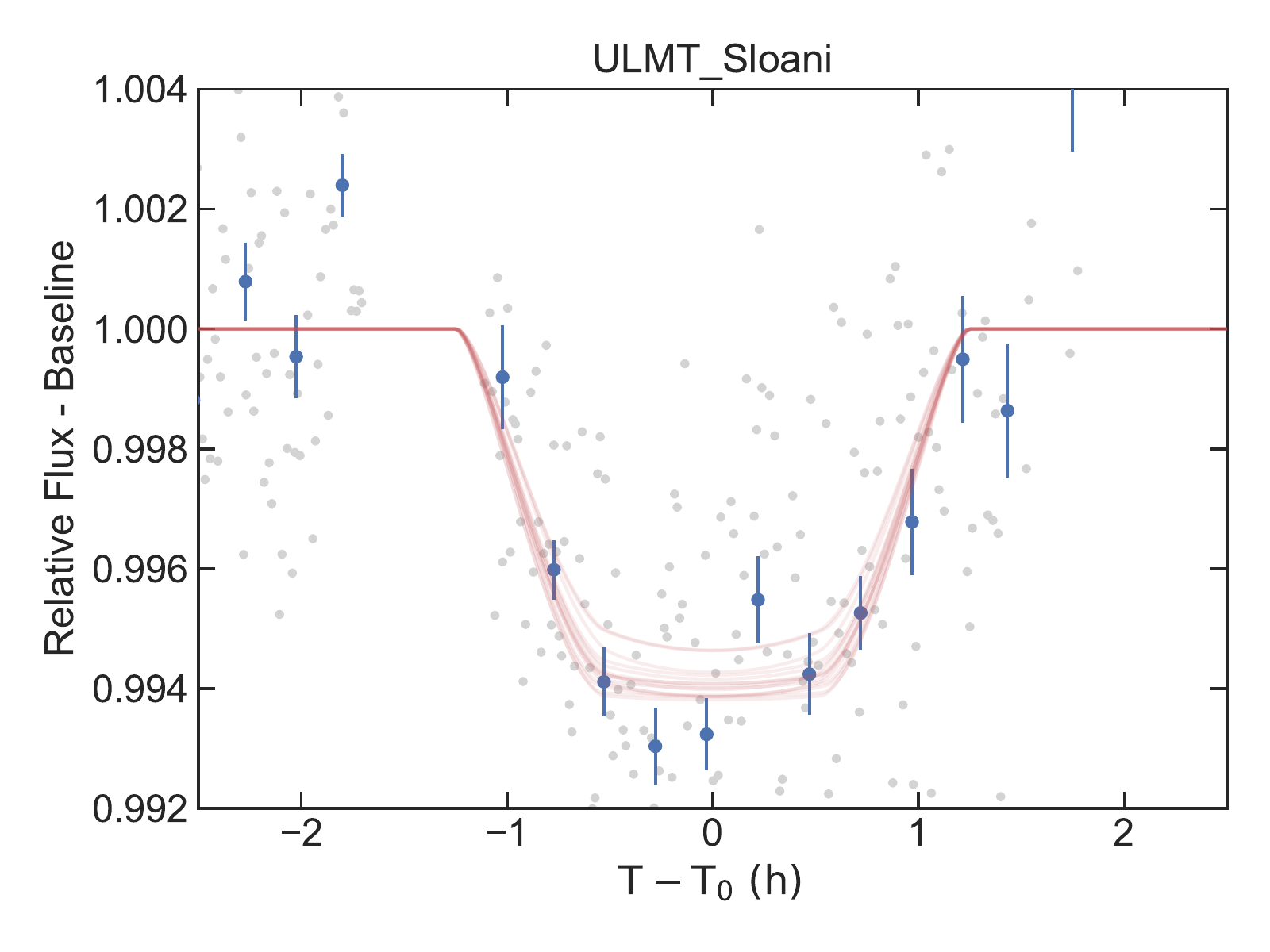}
    \includegraphics[width=\columnwidth]{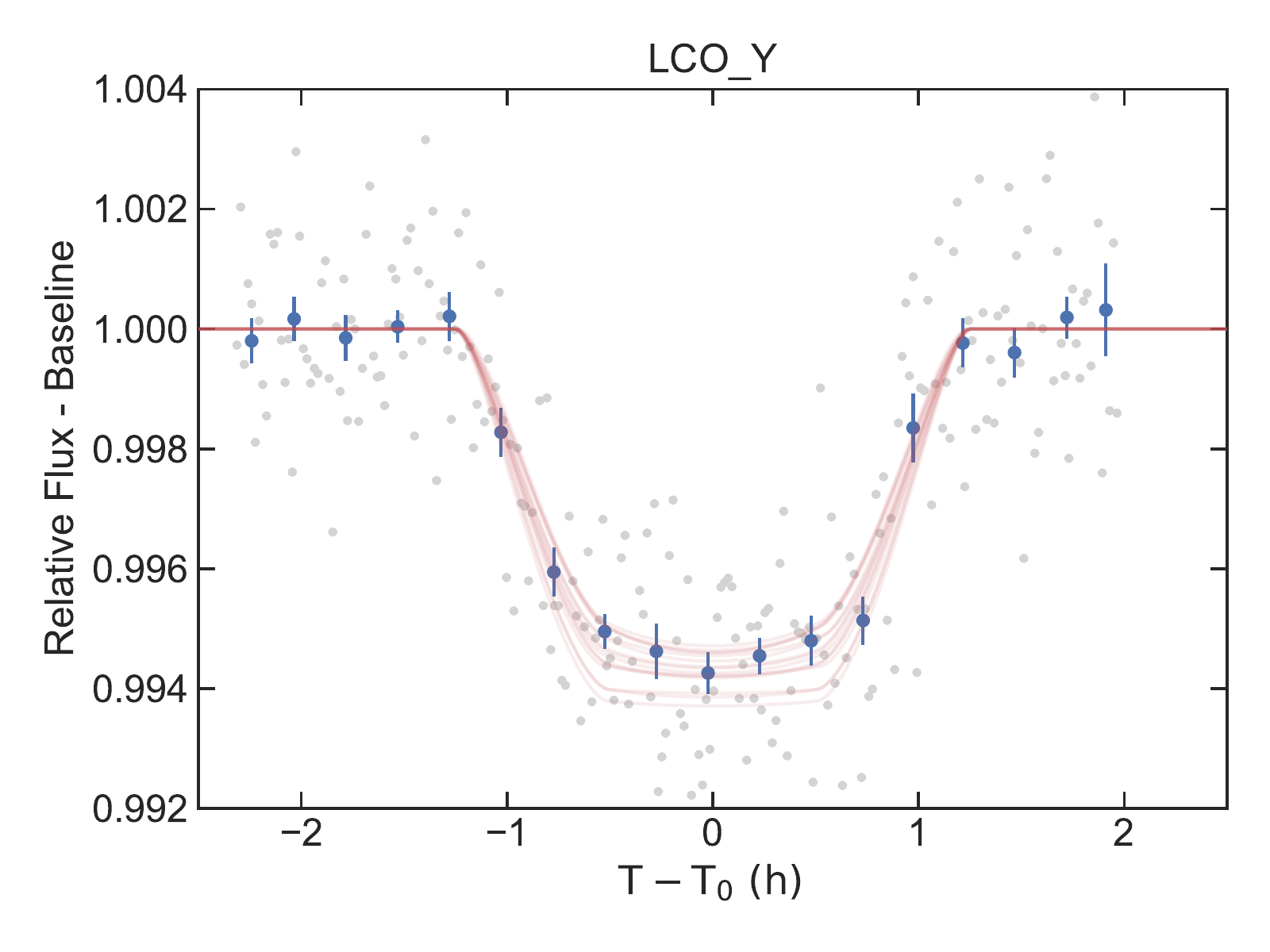}
    
    \caption{Same as Figure~\ref{ground_lcs}, but showing the observations taken with AUKR, SCT, ULMT, and LCO.}
    \label{ground_lcs2}
\end{figure*}

\subsection{Spectroscopic Observations}\label{spectro}
We obtained high-resolution spectroscopic observations of TOI-1431 with SONG, SOPHIE, FIES, NRES (TLV and ELP), and EXPRES to establish the planetary nature of the \textit{TESS} transiting candidate and measure its mass. Here we describe the observations from each spectrograph and list the derived radial velocities in Table~\ref{tab:vels} in the Appendix. 

\subsubsection{SONG}
High-resolution spectroscopic observations of TOI-1431 were obtained using the robotic Stellar Observations Network Group (SONG) 1\,m Hertzsprung telescope at the Teide Observatory in Tenerife \citep{2014RMxAC..45...83A,2019PASP..131d5003F}. In total, 19 spectra were taken with SONG between 06 March 2020 and 29 June 2020. SONG is equipped with a high-resolution coud\'{e} \'{e}chelle spectrograph with wavelength coverage between $4400-6900$\,\AA\, across 51 spectral orders. The observations were taken using the iodine cell to allow for precise wavelength calibration. For each observation, we took a 2400\,s exposure in slit mode 6 (slit width $1.2\arcsec$), providing a spectral resolution of $R=90,000$, a median count per pixel at 5560\,\AA\, of 1795.2\,ADU (signal-to-noise ratio of at least 50 per resolution element for all observations), and median radial velocity precision of 29\,\mos.

Before each target observation, calibration frames (including bias frames, flat fields, and ThAr spectra) were taken on the same night and applied to target images. The target observations were then reduced and the spectra extracted with a pipeline written in Python. This pipeline uses a C++ implementation of the optimal extraction method from \citet{2014PASP..126..170R}. We also acquired observations of the bright fast-rotating star HR 5191 that was used as an intrinsic stellar template for determining the spectral-line-spread function of the spectrograph and to deconvolve a high signal-to-noise spectrum of TOI-1431 taken without the iodine cell. We then analyzed the extracted spectra using the iSONG pipeline \citep{2013MNRAS.435.1563A} to produce radial velocities (given in Table~\ref{tab:vels} and shown in Figure~\ref{RVs}), following the procedure of \citet{2017ApJ...836..142G}. 

\subsubsection{SOPHIE}
TOI-1431 was observed with the fiber-fed SOPHIE high-resolution \'{e}chelle spectrograph on the 1.93\,m telescope at the Haute-Provence Observatory \citep{2008SPIE.7014E..0JP,2011SPIE.8151E..15P} between 18 December 2019 and 12 January 2020. We collected a total of eight spectra in the high-resolution (HR) mode ($R=75,000$ at $5500$\,\AA). The second SOPHIE aperture allowed us to check that there was no background sky pollution.

Thorium-argon calibration spectra were taken at the start of each night of the observations to determine the wavelength calibration zero point, as well as Fabry-P\'{e}rot \'{e}talon spectra regularly during the nights to monitor any possible instrumental drift. The spectra cover the wavelength range from $3872-6943$\,\AA\, across 41 spectral orders, of which 39 were extracted and used for computing the radial velocities. The exposure times ranged from 214 and 490\,s, resulting in a uniform signal-to-noise ratio for each observation of 50 per resolution element at 5500\,\AA\ and median Doppler velocity precision of 7\,\mos.

The spectroscopic data were reduced and then extracted using the standard SOPHIE pipeline \citep{2009A&A...505..853B}, which performs optimal order extraction, cosmic-ray rejection, and wavelength calibration. Radial velocities were computed using the cross-correlation technique and a numerical binary mask corresponding to a F0 spectral type star \citep[e.g., see,][]{2015A&A...581A..38C}, the closest available binary mask to TOI-1431's spectral type. Table~\ref{tab:vels} list the radial velocities from SOPHIE and they are shown in Figure~\ref{RVs}. The width of the cross-correlation function allows us to measure the rotational velocity of the star as $v \sin i = 6.0\pm0.2$\,\kms.

\subsubsection{FIES}\label{fies_spectra}
We acquired 52 spectra of TOI-1431 using the FIbre-fed {\'E}chelle Spectrograph \citep[FIES;][]{1999anot.conf...71F,2014AN....335...41T} at the 2.56\,m Nordic Optical Telescope \citep[NOT;][]{Djupvik2010} of Roque de los Muchachos Observatory (La Palma, Spain). The observations were carried out between 27 November 2019 and 20 June 2020 (UT). We used the FIES high-resolution mode, which provides a resolving power of $R$\,=\,67\,000 in the spectral range 3760--8220~{\AA}. We traced the RV drift of the instrument by acquiring long-exposure ThAr spectra ($T_\mathrm{exp}$\,$\approx$\,90\,s) immediately before and after each science exposure. The exposure time was set to 300--900\,s, depending on the sky conditions and scheduling constraints. 

The data reduction follows the steps of \citet{Buchhave2010} and includes optimal extraction of the spectrum and interpolation of the ThAr wavelength solutions. The SNR per pixel at 5500~\AA\ of the extracted spectra is in the range 42--127. We cross-correlate each observation order-by-order in velocity space against the strongest exposure of the star (BJD=2\,458\,995.703\,476). Finally, the radial velocity of each observation is measured from the peak of the summed cross-correlation function and transformed to the barycentric frame. The uncertainties for the derived radial velocities range 7.2--19.1\,\mos\, with a mean value of 12.2\,\mos.

\subsubsection{NRES}
We triggered observations of TOI-1431 on the Network of Robotic Echelle Spectrographs (NRES) \citep{NRES}, operated by the LCOGT \citep{LCOGT}. NRES consists of four similar spectrographs with two in the Northern Hemisphere and two in the Southern. The spectrographs cover the wavelength range from 380 to 860 nm with a resolution of R$\sim$53,000. They are fibre-fed, where one fibre observes the science target and the second fibre is used for the wavelength calibration source. In the case of NRES the calibration source is a ThAr lamp.

Due to the target's Northern declination, TOI-1431 was accessible and observed by the units at the Wise Observatory (TLV) and the McDonald Observatory (ELP). We obtained a total of 34 spectra. For spectral classification we used the standard NRES pipeline in combination with SpecMatch-Synth code\footnote{\url{https://github.com/petigura/specmatch-syn}}. In order to obtain high precision RV measurements, we processed the spectra using an adapted version of the CERES pipeline \citep{CERES}.

\subsubsection{EXPRES}
We observed TOI-1431 with the Extreme PREcision Spectrometer (EXPRES) \citep{Jurgenson2016} which was recently commissioned at the 4.3\,m Lowell Discovery Telescope (DCT) \citep{Levine2012}. EXPRES is a vacuum-stabilized and fiber-fed optical spectrograph, operating between 380 and 780\,nm at a resolution of $R\sim 137,500$. It achieves a single-measurement radial velocity precision of approximately 30 cm s$^{-1}$ at SNR of 250 per pixel \citep{Petersburg2020}. The wavelength solution is determined from both a ThAr lamp and a Menlo Systems laser frequency comb (LFC). Recent performance benchmarks and science results are shown by \citet{Blackman2020} and \citet{Brewer2020}, and the radial velocity pipeline is discussed by \citet{Petersburg2020}. For TOI-1431, we acquired 15 observations with EXPRES and achieved a mean radial velocity precision of 2.1\,\mos.

\begin{figure*}
  \centering
  \includegraphics[width=\textwidth]{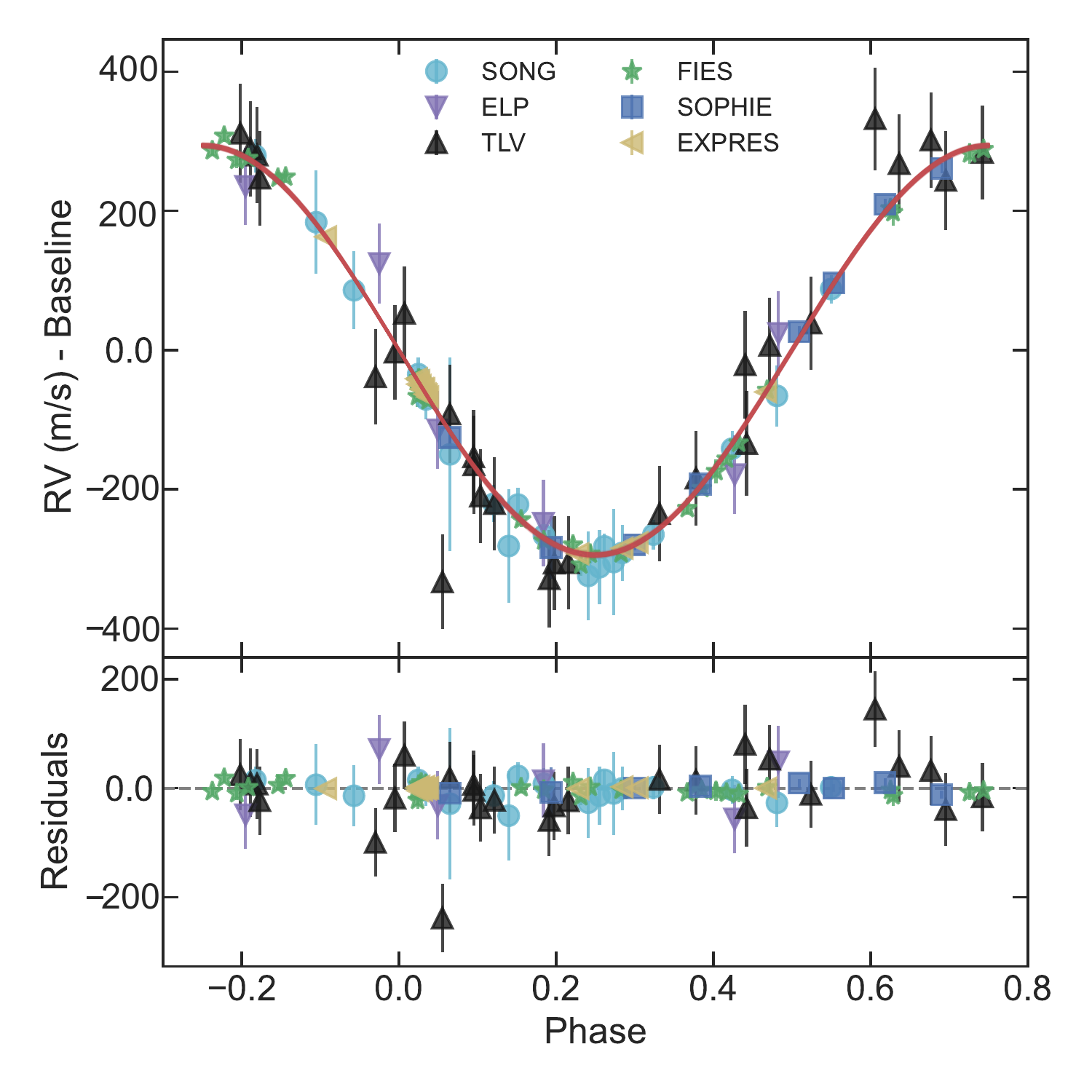}
  \caption{Radial velocity measurements of TOI-1431\,b/MASCARA-5\,b as a function of orbital phase. The radial velocities from SONG, NRES (ELP and TLV), FIES, SOPHIE, and EXPRES are represented by the cyan colored circles, purple triangles pointed down, black triangles pointed up, green stars, blue squares, and gold triangles pointed to the left, respectively. A random selection of 20 radial velocity curves drawn from the posterior of the Nested Sample modeling are plotted as the red lines. The bottom panel shows the residuals between the data and the best-fit (posterior median) model.}
  \label{RVs}
\end{figure*}

\subsection{High-Contrast Imaging Observations}\label{direct_imaging}

\begin{figure}
  \includegraphics[width=\linewidth]{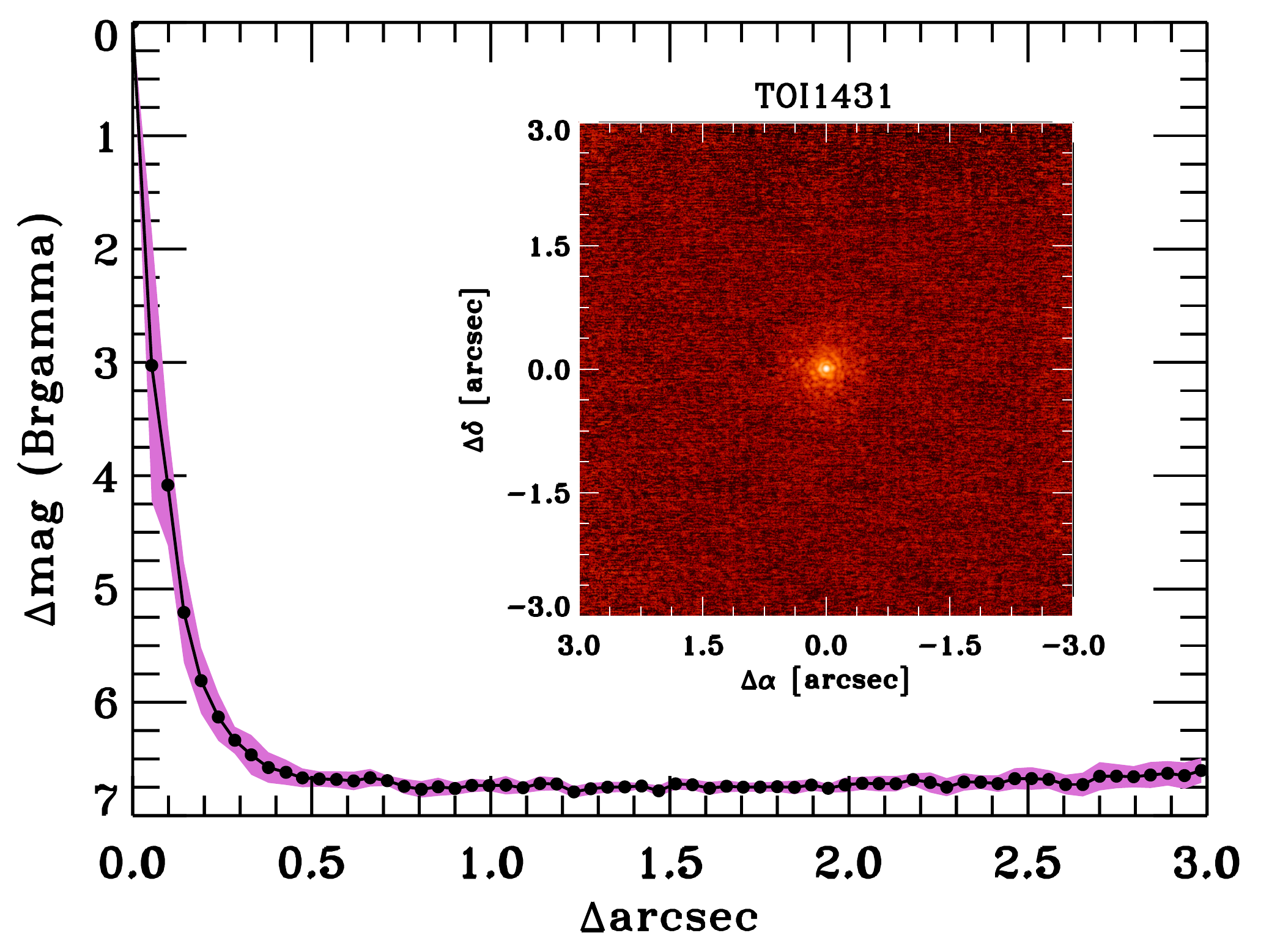}
  \caption{The $5\sigma$ detection sensitivity and AO image from Keck II NIRC2 high angular resolution imaging of TOI-1431 in the Brgamma-band. The orientation of the AO image has North pointed up and East to the left. No stars were detected within $3\arcsec$ of TOI-1431.}
  \label{fig:direct_image}
\end{figure}

We obtained high angular resolution imaging of TOI-1431 using the NIRC2 instrument on the Keck II. The observations were taken on the night of 09 September 2020 UTC under clear skies, very good seeing conditions of $\sim0.4\arcsec$, and at an airmass of 1.25. The resulting $5\sigma$ detection sensitivity and adaptive optics (AO) image from the NIRC2 observations are plotted in Figure~\ref{fig:direct_image}. No nearby companions or background sources were detected within $3\arcsec$ of TOI-1431.

\section{Host Star Properties}\label{host_star}
TOI-1431/MASCARA-5 (HD 201033) is a bright ($\mathrm{V}=8.049$) Am (chemically peculiar, kA5mF2) type star \citep{1991AAS...87..319F,1991AAS...89..429R,2009AA...498..961R} with a \textit{Gaia} DR2 parallax of $6.686 \pm 0.046$\,mas \citep{2018A&A...616A...1G} and distance of $149.6^{+1.1}_{-1.0}$\,pc. The star has a radius of $1.92\pm0.07$\,\Rsun, mass of $1.90^{+0.10}_{-0.08}$\,\Msun, surface gravity of $\log g=4.15\pm0.04$\,dex, effective temperature of $7690^{+400}_{-250}$\,K, metallicity of $[$Fe/H$]=0.43^{+0.20}_{-0.28}$\,dex, and luminosity of $11.7^{+2.3}_{-1.2}$\,\Lsun, derived from an analysis of the broadband spectral energy distribution (SED) and the Yonsei Yale (YY) stellar evolutionary models \citep{2001ApJS..136..417Y} using \texttt{EXOFASTv2} \citep{2019arXiv190709480E} as discussed in Section~\ref{SED}. We also derive independent values for the effective temperature and metallicity from the analysis of SONG, SOPHIE, and FIES spectra in Section~\ref{spec}.

\startlongtable
\begin{deluxetable}{lcc}
\tabletypesize{\scriptsize}
\tablewidth{\columnwidth}
%\tablecolumns{3}
\tablecaption{Stellar parameters for HD\,201033 as used in this work. \textbf{Notes.}--$^{\dagger}$Priors used in the Allesfitter analysis. $^{\ddagger}$Preferred solution for the stellar and spectroscopic parameters. $^{\star}$Upper limit on the V-band extinction from Schlegel Dust maps. \label{tab:star}}
\tablehead{
\colhead{Parameter}  & \colhead{Value}  & \colhead{Source}
}
\startdata
R.A. (hh:mm:ss)         & 21:04:48.89                       & (1) \textit{Gaia} DR2 \\
Decl. (dd:mm:ss)        & 55:35:16.88                       & (1) \textit{Gaia} DR2 \\
$\mu_{\alpha}$ (mas~yr$^{-1}$)  & $11.74 \pm 0.08$        & (1) \textit{Gaia} DR2 \\
$\mu_{\delta}$ (mas~yr$^{-1}$)  & $23.87 \pm 0.06$        & (1) \textit{Gaia} DR2 \\
Parallax (mas)          & $6.686 \pm 0.046$       & (1) \textit{Gaia} DR2 \\
Luminosity (\Lsun)      & $10.69 \pm 0.10$\,$^{\ddagger}$        & (1) \textit{Gaia} DR2 \\
$A_{V}$ (mag)           & $1.099$ $(\leq3.491)$\,$^{\star}$   & (2) Schlegel Dust maps \\
Distance (pc)           & $149.6^{+1.1}_{-1.0}$                  & (1) \textit{Gaia} DR2 \\
Spectral type           & Am C (kA5mF2)                     & (3,4,5) A-type star catalogs \vspace{2pt}\\ \hline
\vspace{-8pt}
\\\multicolumn{3}{l}{Broadband Magnitudes:} \vspace{4pt} \\
$B_{T}$ (mag)           & $8.368 \pm 0.016$          & (6) Tycho \\
$V_{T}$ (mag)           & $8.049 \pm 0.011$          & (6) Tycho \\
$TESS$ (mag)            & $7.798 \pm 0.006$        & (7) \textit{TESS} TIC v8 \\
$J$ (mag)               & $7.541 \pm 0.030$          & (8) 2MASS \\
$H$ (mag)               & $7.452 \pm 0.040$          & (8) 2MASS \\
$K_{s}$ (mag)           & $7.439 \pm 0.030$          & (8) 2MASS \\
$WISE1$ (mag)           & $7.335 \pm 0.034$          & (9) WISE \\
$WISE2$ (mag)           & $7.430 \pm 0.019$          & (9) WISE \\
$WISE3$ (mag)           & $7.446 \pm 0.015$          & (9) WISE \\
$WISE4$ (mag)           & $7.371 \pm 0.096$          & (9) WISE \\
$Gaia$ (mag)            & $7.976 \pm 0.001$          & (1) \textit{Gaia} DR2 \\
$Gaia_{BP}$ (mag)       & $8.123 \pm 0.001$          & (1) \textit{Gaia} DR2 \\
$Gaia_{RP}$ (mag)       & $7.772 \pm 0.001$          & (1) \textit{Gaia} DR2
\vspace{2pt} \\ \hline
\vspace{-8pt}
\\\multicolumn{3}{l}{Stellar Properties from SED \& YY Tracks:} \vspace{4pt} \\
$T_\mathrm{eff}$ (K)    & $7690^{+400}_{-250}$\,$^{\dagger,\ddagger}$ & (10) \texttt{EXOFASTv2}; this paper \\
$\log g$ (dex)          & $4.15\pm0.04$\,$^{\ddagger}$     & (10) \texttt{EXOFASTv2}; this paper \\
$[$Fe/H$]$ (dex)        & $0.43^{+0.20}_{-0.28}$        & (10) \texttt{EXOFASTv2}; this paper \\
$R_\star$ ($R_\odot$)   & $1.92\pm0.07$\,$^{\dagger,\ddagger}$           & (10) \texttt{EXOFASTv2}; this paper \\
$M_\star$ ($M_\odot$)   & $1.90^{+0.10}_{-0.08}$\,$^{\dagger,\ddagger}$           & (10) \texttt{EXOFASTv2}; this paper \\
$\rho_\star$ (g~cm$^{-3}$)     & $0.38\pm0.05$                  & (10) \texttt{EXOFASTv2}; this paper \\
$L_\star$ ($L_\odot$)   & $11.7^{+2.3}_{-1.2}$          & (10) \texttt{EXOFASTv2}; this paper \\
Age (Gyr)               & $0.29^{+0.32}_{-0.19}$\,$^{\ddagger}$        & (10) \texttt{EXOFASTv2}; this paper
\vspace{2pt} \\ \hline
\vspace{-8pt}
\\\multicolumn{3}{l}{Spectroscopic Properties from SONG spectra:} \vspace{4pt} \\
$T_\mathrm{eff}$ (K)    & $6764 \pm 120$                & (11,12) iSpec; this paper \\
$\log g$ (dex)          & $2.76 \pm 0.26$               & (11,12) iSpec; this paper \\
$[$M/H$]$ (dex)         & $-0.15 \pm 0.10$              & (11,12) iSpec; this paper \\
$[\alpha$/H$]$ (dex)     & $-0.27 \pm 0.10$              & (11,12) iSpec; this paper \\
$v \sin i$ (km\,s$^{-1}$)    & $7.1 \pm 1.0$            & (11,12) iSpec; this paper
\vspace{2pt} \\ \hline
\vspace{-8pt}
\\\multicolumn{3}{l}{Spectroscopic Properties from SOPHIE spectra:} \vspace{4pt} \\
$T_\mathrm{eff}$ (K)    & $6950 \pm 60$                 & (13) FASMA; this paper \\
$\log g$ (dex)          & $4.72 \pm 0.08$               & (13) FASMA; this paper \\
$[$Fe/H$]$ (dex)        & $0.09 \pm 0.03$ \,$^{\ddagger}$              & (13) FASMA; this paper \\
$v \sin i$ (km\,s$^{-1}$)    & $6.0 \pm 0.2$\,$^{\ddagger}$            & CCF; this paper
\vspace{2pt} \\ \hline
\vspace{-8pt}
\\\multicolumn{3}{l}{Spectroscopic Properties from FIES spectra:} \vspace{4pt} \\
$T_\mathrm{eff}$ (K)    & $6910 \pm 50$                 & (14) SPC; this paper \\
$\log g$ (dex)          & $3.29 \pm 0.10$               & (14) SPC; this paper \\
$[$M/H$]$ (dex)        & $0.03 \pm 0.08$               & (14) SPC; this paper \\
$v \sin i$ (km\,s$^{-1}$)    & $9.2 \pm 0.5$            & (14) SPC; this paper
\enddata
%\raggedright
% The next line causes 116 errors...WTF
\tablerefs{1. \cite{2018AA...616A...1G}; 2. \cite{2011ApJ...737..103S}; 3. \cite{1991AAS...87..319F}; 4. \cite{1991AAS...89..429R}; 5. \cite{2009AA...498..961R}; 6. \cite{2000AA...355L..27H}; 7. \cite{2019AJ....158..138S}; 8. \cite{2003yCat.2246....0C}; 9. \cite{2013yCat.2328....0C}; 10. \cite{2019arXiv190709480E}; 11. \cite{ispec2014}; 12. \cite{ispec2019}; 13. \cite{2018MNRAS.473.5066T}; 14. \cite{2012Natur.486..375B}}
\end{deluxetable}

\subsection{SED Analysis}\label{SED}
\begin{figure}
  \includegraphics[width=\linewidth]{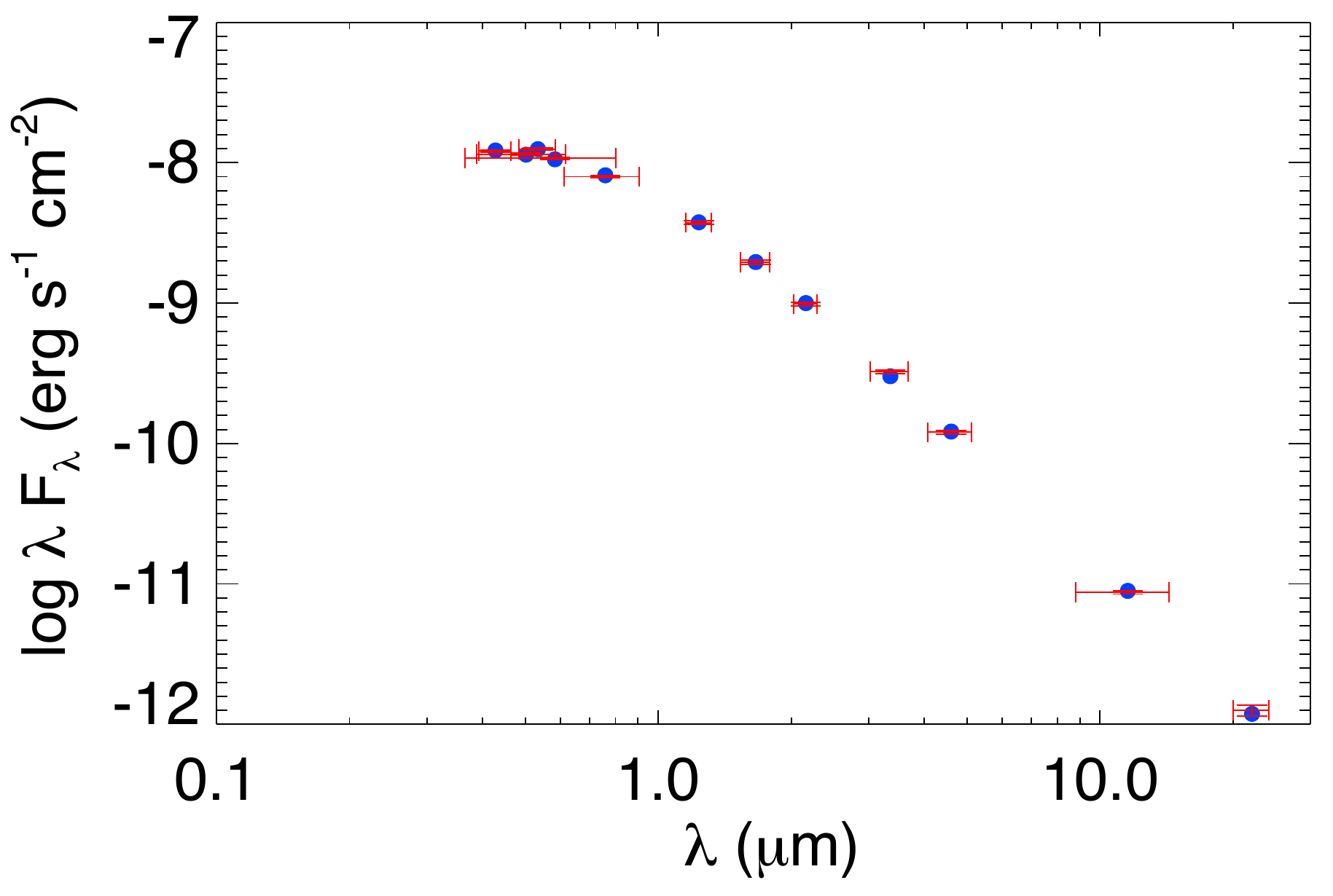}
  \caption{The SED for TOI-1431. The red symbols are the broadband photometric measurements used in the SED analysis (provided in Table~\ref{tab:star}) with the horizontal uncertainty bars representing the effective width of the passband. The blue symbols are the model fluxes from the best-fit Kurucz atmosphere model.}
  \label{fig:sed}
\end{figure}

We performed a SED analysis of TOI-1431 using \texttt{EXOFASTv2} to derive stellar parameters independent of those from spectroscopy. We used broadband photometry obtained from the Tycho \citep{2000AA...355L..27H}, 2MASS \citep{2003yCat.2246....0C}, WISE \citep{2013yCat.2328....0C}, and {\it Gaia\/} \citep{2018AA...616A...1G} catalogs for the fitting of the SED (Figure~\ref{fig:sed}). Gaussian priors were placed on the parallax from {\it Gaia\/} DR2, adding 82\,$\mathrm{\mu as}$ to correct for the systematic offset found by \citet{2018ApJ...862...61S} and adding the 33\,$\mathrm{\mu as}$\, uncertainty in their offset in quadrature to the {\it Gaia\/}-reported uncertainty. We applied an upper limit on the V-band extinction of 3.4905 from the \citet{2011ApJ...737..103S} dust maps at the location of TOI-1431 as well as an upper limit on the dilution effects of nearby stars of 0.00484 as reported in the \textit{TESS} SPOC light curve file. The SED analysis ran until convergence, i.e., the Gelman-Rubin statistic ($R_{z}$) and the number of independent chain draws ($T_{z}$) were less than 1.01 and greater than 1000, respectively (see \citealt{2019arXiv190709480E}). The YY stellar evolutionary models were used instead of the default MESA Isochrones and Stellar Tracks (MIST) to determine the stellar radius ($R_\star$), mass ($M_\star$), effective temperature ($T_\mathrm{eff}$), luminosity ($L_\star$), metallicity ($[$Fe/H$]$), surface gravity ($\log g$), and age of TOI-1431, which are given in Table~\ref{tab:star}. We found that using the MIST models resulted in many of the Markov Chain Monte Carlo (MCMC) steps reaching the $[$Fe/H$]$ grid's upper limit of 0.5\,dex while the MCMC chains never reached the $[$Fe/H$]$ upper limit of 0.78\,dex using the YY tracks. The YY evolutionary models are better suited for hot metal rich stars such as TOI-1431.

\begin{figure}
  \includegraphics[width=\linewidth]{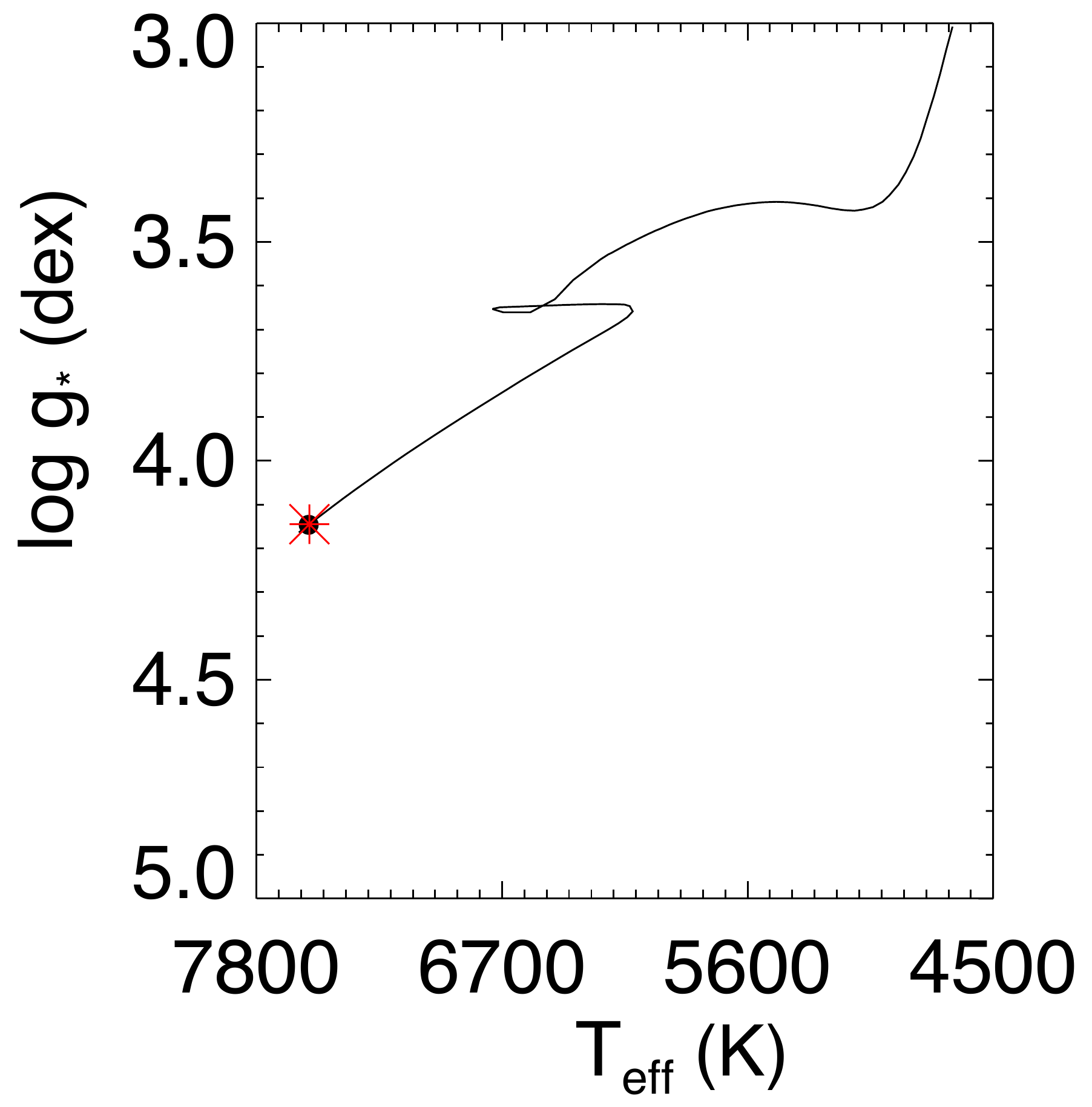}
  \caption{The YY mass track for TOI-1431. The black line represents the mass track interpolated at the model values for $M_\star$, [Fe/H], and age. The black circle is the best-fit model value for $T_\mathrm{eff}$ and $\log g$ while the red asterisk is the best-fit model age along the track. The black circle and red asterisk almost perfectly overlap indicating excellent consistency among all the components of the global model.}
  \label{fig:yy}
\end{figure}

From the best-fit Kurucz stellar atmosphere model from the SED and the best-fitting YY stellar evolutionary model (Figure~\ref{fig:yy}), we find that TOI-1431 is a very hot and metal rich A-type star with $R_{\star}=1.92\pm0.07$\,\Rsun, $M_{\star}=1.90^{+0.10}_{-0.08}$\,\Msun, $T_{\rm eff}=7690^{+400}_{-250}$\,K, $\log{g}=4.15\pm0.04$ (where $g$ is in units of \logg), and $[$Fe/H$]=0.43^{+0.20}_{-0.28}$\,dex. The effective temperature found for this star makes it one of the hottest known exoplanet hosts. The only known host stars hotter than TOI-1431 are KELT-9 \citep[$T_{\rm eff}=10170\pm450$\,K,][]{2017Natur.546..514G}, WASP-178 \citep[$T_{\rm eff}=9360\pm150$\,K,][]{2019MNRAS.490.1479H}, KELT-20 \citep[$T_{\rm eff}=8720^{+250}_{-260}$\,K,][]{2017AJ....154..194L}, Kepler-1115 \citep[$T_{\rm eff}=8480^{+300}_{-220}$\,K,][]{2016ApJ...822...86M}, HAT-P-70 \citep[$T_{\rm eff}=8450^{+540}_{-690}$\,K,][]{2019AJ....158..141Z}, HATS-70 \citep[$T_{\rm eff}=7930^{+630}_{-820}$\,K,][]{2019AJ....157...31Z}, and MASCARA-4 \citep[$T_{\rm eff}=7800\pm200$\,K,][]{2020A&A...635A..60D}. These stellar parameters (in particular $T_\mathrm{eff}$, $\log g$, and metallicity), however, are in strong disagreement with the spectroscopic parameters derived from the analysis of the SONG, SOPHIE, and FIES data. The source of this disagreement is likely due to the ``anomalous luminosity effect'' observed in Am stars \citep[e.g.,][]{1971A&A....14..233B} such as TOI-1431, which we discuss further in Section~\ref{Am_star}. The $\log g$ derived from the SED fit is in good agreement with the value obtained from the stellar density ($\log g\sim4.17$\,dex) from fitting the transits, and we have therefore adopted this value along with the values for $T_\mathrm{eff}$, $R_{\star}$, and $M_{\star}$ as the preferred values for the star.

\subsection{Spectroscopy}\label{spec}
We used the \textsc{iSpec} spectral analysis tool \citep{ispec2014,ispec2019} to derive the effective temperature, surface gravity, rotational velocity, alpha elemental abundance ([$\alpha$/H]), and overall metallicity ([M/H]) of TOI-1431 from the reduced and stacked SONG spectra. The \textsc{iSpec} synthetic grid was configured to incorporate a MARCS atmospheric model \citep{MARCSmodel} and the \textsc{spectrum} \citep{spectrumcode} radiative transfer code. [M/H] and [$\alpha$/H] were derived using version 5.0 of {\it Gaia\/}-ESO Survey's (GES) line-list \citep{GESlinelist} normalized by solar values obtained by \citet{Asplund09}. Initial values for $T_\mathrm{eff}$, $\log{g}$, and [M/H] as determined by the SED analysis with \texttt{EXOFASTv2} (see Section~\ref{SED}) were used to construct the synthetic spectra fit. From the \textsc{iSpec} analysis of the SONG spectra, we find that TOI-1431 has a $T_\mathrm{eff}=6764 \pm 120$\,K, $\log g =2.76 \pm 0.26$\,dex, $[$M/H$]=-0.15 \pm 0.10$\,dex, $[\alpha$/H$]=-0.27 \pm 0.10$\,dex, and $v \sin i=7.1 \pm 1.0$\,\kms. These results would seem to suggest that the star is quite evolved, is slightly metal poor, and is deficient in $\alpha$ elements. However, $T_\mathrm{eff}$ and $\log g$, in particular, are in very strong disagreement with the results obtained from the SED analysis as a consequence of the nature of the host being an Am star (see discussion in Section~\ref{Am_star}).

The atmospheric stellar parameters $T_\mathrm{eff}$, $\log{g}$, and [Fe/H] were derived from the SOPHIE spectra for TOI-1431 using the \textit{FASMA} spectral synthesis package \citep{2018MNRAS.473.5066T}. We provided initial stellar parameters for $T_\mathrm{eff}$, $\log{g}$, and [Fe/H] from the SED analysis with \texttt{EXOFASTv2} as a starting point of the spectral synthesis in \textit{FASMA}. The synthetic spectra were created using the radiative transfer code, MOOG \citep{1973ApJ...184..839S}, and the model atmospheres grid was generated by the ATLAS-APOGEE \citep{1993sssp.book.....K}, MARCS \citep{MARCSmodel}, and ATLAS9 \citep{2012AJ....144..120M} models. The line list used to determine the stellar parameters comes from the Vienna Atomic Line Database \citep[VALD,][]{1995A&AS..112..525P,2015PhyS...90e4005R} and includes all of the iron lines in the regions $5399-5619$\,\AA\, and $6347-6790$\,\AA\, as well as all the atomic and molecular lines within intervals of $\pm2$\,\AA\, of the iron lines. From this analysis with the SOPHIE spectra, we find the star has $T_\mathrm{eff}=6950 \pm 60$\,K, $\log g =4.72 \pm 0.08$\,dex, and $[$Fe/H$]=0.09 \pm 0.03$\,dex, consistent with a late A-type main-sequence star that is modestly metal rich. We have adopted the $[$Fe/H$]$ value derived here as the preferred value for the star's iron abundance.

The stellar parameters for this star were also derived from the FIES spectra using the stellar parameter classification (SPC) technique \citep{2012Natur.486..375B,2012ApJ...757..161T,Buchhave2014}. To derive $T_\mathrm{eff}$, $\log{g}$, $v \sin i$, and [M/H] for this star using SPC, we cross-correlated the reduced and the stacked FIES spectra against a library of synthetic spectra calculated from the Kurucz model atmospheres \citep{2003IAUS..210P.A20C}. The synthetic spectra covers the wavelength range from $5050-5360$\,\AA, of which five of the FIES echelle orders span that region and could be used for the spectral analysis. The best-fit values were determined by finding the maximum of the cross-correlation coefficient as a function of the stellar parameters. The results of the SPC analysis on the FIES spectra gives $T_\mathrm{eff}=6910 \pm 50$\,K, $\log g =3.29 \pm 0.10$\,dex, $[$M/H$]=0.03 \pm 0.08$\,dex, and $v \sin i=9.2 \pm 0.5$\,\kms, which suggest an A-type sub-giant star with solar metallicity. It should be noted that SPC was not designed for hot stars such as TOI-1431 and that surface gravities are often poorly constrained by spectral analyses \citep[e.g., see,][]{2007ApJ...664.1190S,2008ApJ...683.1076W}.

The spectral analyses carried out on the SONG, SOPHIE, and FIES spectra give significantly different results for the host star properties, and in particular for the surface gravity. However, the stellar density (and surface gravity) determined from the SED and transit light curve analysis are in good agreement with each other. \citet{2008ApJ...683.1076W} has demonstrated that stellar densities can be more accurately determined through transit light curve analysis and should be preferred over those derived through spectroscopy if there are strong disagreements between them. In this light, we therefore have adopted the $T_\mathrm{eff}$, $\log g$, $R_{\star}$, and $M_{\star}$ from the SED analysis as the preferred values for the star.

%-------------------------------------------------------------------------------------
\section{Modelling and Results}
\label{model_results}
To determine the system parameters for TOI-1431 and its planet, including the planet's thermal emissions and day-night temperature contrast, we used \texttt{Allesfitter} \citep{allesfitter-code,allesfitter-paper} to perform a joint analysis of the \textit{TESS} light curve segments (from the multi-scale MAP PDC, single-scale MAP PDC, and SAP photometry), the photometric ground-based light curves, and the radial velocity measurements. The ground-based light curves included in the analysis are from CDK14 (Sloan $g'$ and $z_s$ bands) taken on the 24 December 2019, MuSCAT2 (four Sloan bands $g'$, $r'$, $i'$, and $z_s$) obtained on the 16 May 2020, AUKR (Sloan $z_s$ band) observed on the 16 June 2020, SCT (\textit{TESS}-equivalent band) acquired on the 8 August 2020, ULMT in the Sloan $z'$ band collected on the 20 September 2020, and LCOGT captured on the 14 October 2020. We used the radial velocity measurements from the SONG, SOPHIE, FIES, NRES, and EXPRES instruments in the joint analysis.

\subsection{Joint Transit, Phase-curve, \& Radial Velocity Analysis}
\label{global_analysis}
For the joint analysis, Gaussian priors were placed on the stellar parameters $R_{\star}$, $M_{\star}$, and $T_{\rm eff}$ from the SED and YY isochrones fitting using \texttt{EXOFASTv2}. We also applied a Gaussian prior on the dilution parameter ($D_\mathrm{0}$) from the \textit{TESS} SPOC light curve parameter CROWDSAP, after transforming the CROWDSAP value to conform to the definition of the dilution parameter used in \texttt{Allesfitter} ($D_{0}=1-CROWDSAP$). For the other physical model parameters, we used uniform priors with reasonable boundaries and starting values. These parameters include the planet-to-star radius ratio ($R_{p}/R_{\star}$), the ratio of the sum of the planet and star radii to the semi-major axis ($(R_{\star} +R_{p})/a_{p}$), cosine of the inclination angle ($\cos{i_{p}}$), mid-transit time ($T_{0;b}$), orbital period ($P_{p}$), the radial velocity semi-amplitude ($K$), eccentricity ($\sqrt{e_{p}} \cos{\omega_{p}}$ and $\sqrt{e_{p}} \sin{\omega_{p}}$), surface brightness ratio ($J_{p}$), set of quadratic limb-darkening coefficients ($q_{1}$ and $q_{2}$) and flux error scaling ($\ln{\sigma_{F_{inst}}}$) for each of the light curves, and a radial velocity baseline offset ($\Delta RV_{inst}$) and jitter term ($\log{\sigma_{RV_{inst}}}$) for each radial velocity instrument.

We experimented with several different detrending techniques in our analysis of each of the three versions of the \textit{TESS} photometry, including: (1) baseline flux offset, (2) rspline detrending prior to the light curve fitting with a baseline flux offset, (3) hybrid cubic spline detrending run simultaneously with the light curve fitting, and (4) a red noise detrending model using a Mat\'{e}rn 3/2 Gaussian Processes (GP) kernel run simultaneously with the light curve fitting. For the GP parameters, we applied uniform priors with reasonable boundaries and starting values on the flux offset ($GP_{offset}$, between -0.01 and 0.01), characteristic amplitude ($\ln{\sigma_{flux}}$, between -15 and 0), and characteristic length scale ($\ln{\rho_{flux}}$, between 2.07 and 12). These parameters were also coupled across the \textit{TESS} light curve segments so that a single GP model is sampled at every sampling step. The lower boundary on the characteristic length scale term was set to three times the planet's orbital period to ensure that the GP preserves any phase curve modulations while still removing most of the remaining systematics present in the light curve segments.

Close inspection of the \textit{TESS} light curves reveals periodic photometric modulations that are synchronized with the orbital period of the planet. These phase curve variations are driven by the changing viewing angle of the planet and the planet--star gravitational interaction. There are three main components to the phase curve that can be detected in visible-light photometry (see detailed review in \citealt{shporer2017}): (1) the atmospheric brightness modulation, which arises due to the variations in atmospheric temperature, thermal emission, and reflected starlight across the surface of a tidally-locked planet, (2) ellipsoidal distortion of the host star due to the tidal bulge raised by the orbiting planet, and (3) beaming, i.e., modulations in the star's flux within the band-pass due primarily to the radial velocity driven Doppler shifting of the stellar spectrum. Setting the zero-point of orbital phase at mid-transit, these three phase curve modulations appear as, to the leading order, the cosine of the fundamental frequency, the cosine of the first harmonic, and the sine of the fundamental. 

In this work, we also fit for these phase curve terms by including free parameters for the respective full amplitudes: $A_{\mathrm{atmospheric}}$, $A_{\mathrm{ellipsoidal}}$, and $A_{\mathrm{beaming}}$. Similar methodologies have been used in extensive previous phase curve analyses of \textit{TESS}-observed exoplanet systems \citep[e.g.,][]{2019AJ....157..178S,2020AJ....159..104W,2020AJ....160...88W,2020AJ....160..155W,daylan2021}. The full list of priors used in the analysis are given in Table~\ref{tab:joint_fit_results}.

\subsubsection{Multi-scale MAP PDC Photometry}\label{multi_PDC}
We started the joint analysis using the \textit{TESS} multi-scale MAP corrected PDC light curve (after removing all quality-flagged data and outliers as described in Section~\ref{TESS_photometry}) since most of the instrumental systematics are removed by the SPOC PDC pipeline from the use of co-trending basis vectors and flux corrections due to crowding from other stars \citep[e.g., see][]{2014PASP..126..100S}. This results in a cleaner light curve compared to the SAP version while still preserving astrophysical signals on timescales shorter than the orbital period of TOI-1431\,b, important for phase curve and secondary eclipse analysis.

\begin{figure}
  \includegraphics[width=\linewidth]{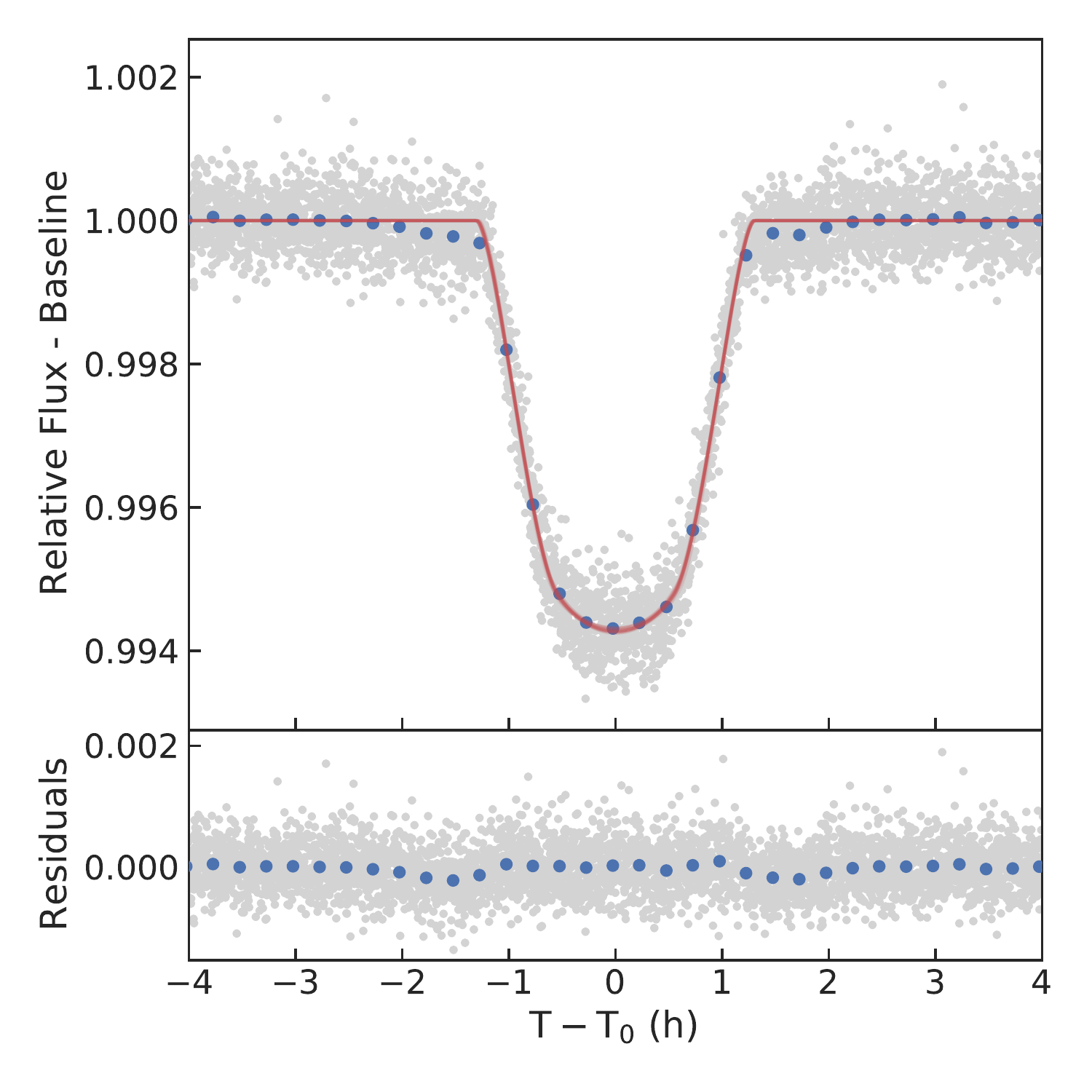}
  \caption{Phase-folded and detrended \textit{TESS} multi-scale MAP PDC light curve of TOI-1431 (Sectors 15 and 16) from the simultaneous hybrid cubic spline detrending model, focusing on the transit. Systematics and other quality-flagged data (as shown in Figure~\ref{tessphotometry}) that are the result of, for example, spacecraft pointing anomalies have been removed prior to performing the analysis on the light curve. The blue points are binned at a cadence of 15\,m and the red solid lines are 20 light curve models drawn from the posteriors of the Markov Chain Monte Carlo (MCMC) analysis in \texttt{Allesfitter}. Dip features starting $\sim1$\,hr before and after transit ingress and egress, respectively, are apparent in the light curve and were introduced during the PDC multi-scale MAP cotrending process. These dip features are not present in the single-scale MAP corrected PDC and SAP versions of the light curve (see Figure~\ref{fig:TESS_LC_multi_PDC}).}
  \label{fig:TESS_LC_PDC}
\end{figure}

The analysis was carried out using a MCMC that ran until convergence (all chains are $>42$ times their autocorrelation lengths) for the four different detrending techniques mentioned in Section~\ref{global_analysis}. Figure~\ref{fig:TESS_LC_PDC} is the resulting phase-folded and detrended PDC light curve of TOI-1431 from the hybrid cubic spline detrending, overplotted with 20 light curve models drawn from the MCMC posteriors. It is apparent in the residuals to the best-fit transit model that there are two significant dip features in the light curve ($\sim200$\,ppm depth), one starting $\sim1$\,hr before transit ingress and continuing until just after the start of ingress and another starting just prior to the end of egress with a duration of $\sim1$\,hr. The residuals from the fit using the other three detrending techniques also show similar dip features. As additional checks, we ran the \texttt{Allesfitter} analysis without any detrending separately on the Sector 15 and 16 PDC light curves and observed similar dip signatures, ruling out the possibility that the light curve detrending was responsible for these features. These signatures are also present in the Data Validation (DV) light curve provided in the Data Validation Report Summary available on exo.MAST.

Further investigations revealed the source of the dip features to be a wavelet artifact caused by the multi-scale MAP correction applied by PDC (see Section~\ref{TESS_photometry}). Normally, multi-scale MAP performs better than single-scale MAP for removing long period systematics while preserving signals at transit scales. However, there is a mechanism where multi-scale MAP can slightly corrupt a very deep transit in a light curve with little long period systematics, such as we see in TOI-1431. The single-scale MAP corrected light curve was found to not contain the dip features and as such, we chose to report the results of the analysis performed on that light curve.

\subsubsection{Single-scale MAP PDC Photometry}\label{PDC_analysis}

The joint analysis with the pre-cleaned (as described in Section~\ref{TESS_photometry}) single-scale MAP corrected PDC light curve segments in \texttt{Allesfitter} follows the same procedure as with the multi-scale MAP corrected PDC light curves. We ran MCMC fits until convergence for the four different detrending techniques. We then used the best-fit parameter values found from the MCMC as starting values in subsequent Nested Sampling fits to estimate the Bayesian evidence for use in a robust comparison of the four detrending models. We used the default \texttt{Allesfitter} settings for the nested sampling fits \citep[see][]{allesfitter-paper} that ran until they reached the convergence criterion threshold of $\Delta\ln{Z}\leq0.01$.

The results from the Nested Sampling fits are in good agreement with the MCMC results to within $1\sigma$. Comparing the Bayesian evidence of the Nested Sampling models, we find that the hybrid cubic spline detrending run simultaneously with the light curve fitting is decisively favored over the second most favored model, the red noise GP detrending model, with a Bayes model comparison factor of $\ln{R}=119$. We also applied the hybrid cubic spline detrending to each of the ground-based transit light curves used in the joint fit.

\begin{figure}
  \includegraphics[width=\linewidth]{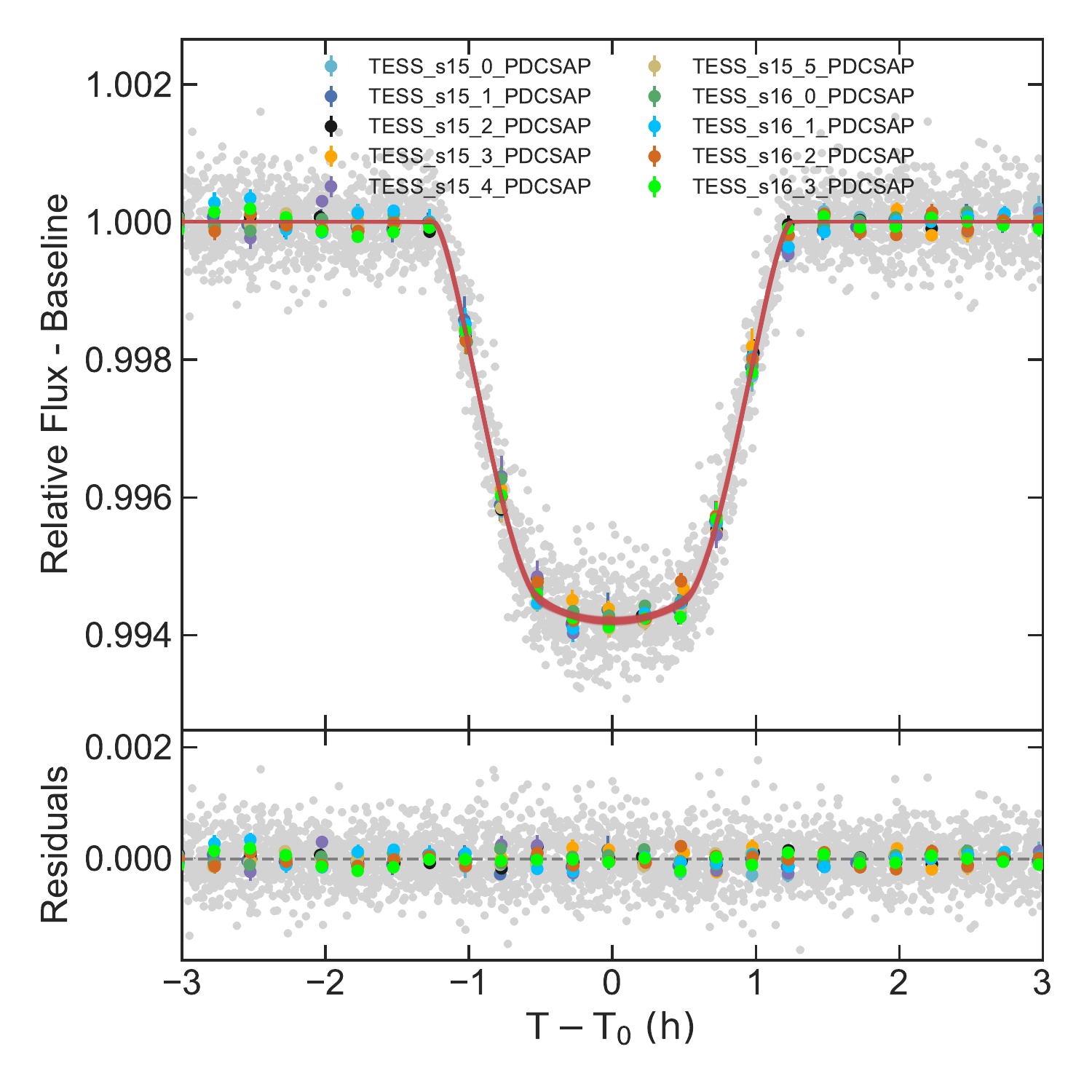}
  \caption{Phase-folded and detrended light curve of TOI-1431 from the simultaneous hybrid cubic spline detrending model, similar to Figure~\ref{fig:TESS_LC_PDC}, but with the single-scale MAP corrected PDC data instead. The colored points represent the individual phased light curve segments across Sectors 15 and 16 that have been binned at a cadence of 15\,m. The red solid lines are 20 light curve models drawn from the posteriors of the Nested Sampling analysis in \texttt{Allesfitter}.}
  \label{fig:TESS_LC_multi_PDC}
\end{figure}

Figure~\ref{fig:TESS_LC_multi_PDC} shows the resulting phase folded and detrended single-scale MAP PDC light curve of TOI-1431 from the simultaneous cubic spline detrending Nested Sampling model. Unlike the multi-scale MAP PDC light curve shown in Figure~\ref{fig:TESS_LC_PDC}, there is no evidence of the strange dip features near the transit, motivating our decision to use the results derived from the single-scale MAP version of light curve analysis. Both the MCMC and Nested Sampling give consistent results, and we report both the fitted and derived parameter results of the Nested Sampling model in Table~\ref{tab:joint_fit_results}. In Figures~\ref{ground_lcs}, \ref{ground_lcs2}, and \ref{RVs}, we show the  ground-based transit light curves and radial velocity measurements with the corresponding models drawn from the Nested Sampling posterior, respectively.

Our preferred solution from the joint analysis using the single-scale MAP PDC light curve shows that TOI-1431 hosts an inflated and highly irradiated Jovian planet with a radius of $R_{p}=1.49\pm0.05$\,$\rm{R_J}$\ ($16.7\pm0.6$\,\Rearth), mass of $3.12\pm0.18$\,\Mjup, high equilibrium temperature of $T_\mathrm{eq}=2370\pm70$\,K (assuming zero Bond albedo), and a circular orbit with a period of $P=2.650237\pm0.000003$\,d. The mean density of the planet is $1.17_{-0.16}^{+0.18}$\,\densitycgs, slightly less dense but consistent with the density of Jupiter (1.326\,\densitycgs) and similar in density to other Jovian worlds. The large number and high quality of radial velocity measurements (as shown in Figure~\ref{RVs}), coupled with the host star's low $v \sin i$ of $6.0 \pm 0.2$\,\kms, has allowed us to measure the radial velocity semi-amplitude of the orbit ($294.1\pm1.1$\,\mos) with unusually high precision for a planet orbiting such a hot star ($7690^{+400}_{-250}$\,K). In fact, this is the highest precision orbit measured for any planet that has a host star hotter than 6600\,K. The largest contributor to the uncertainty in the planet mass ($17.4\sigma$ detection) is from the uncertainty on the host star mass.

\begin{figure*}
  \includegraphics[width=\linewidth]{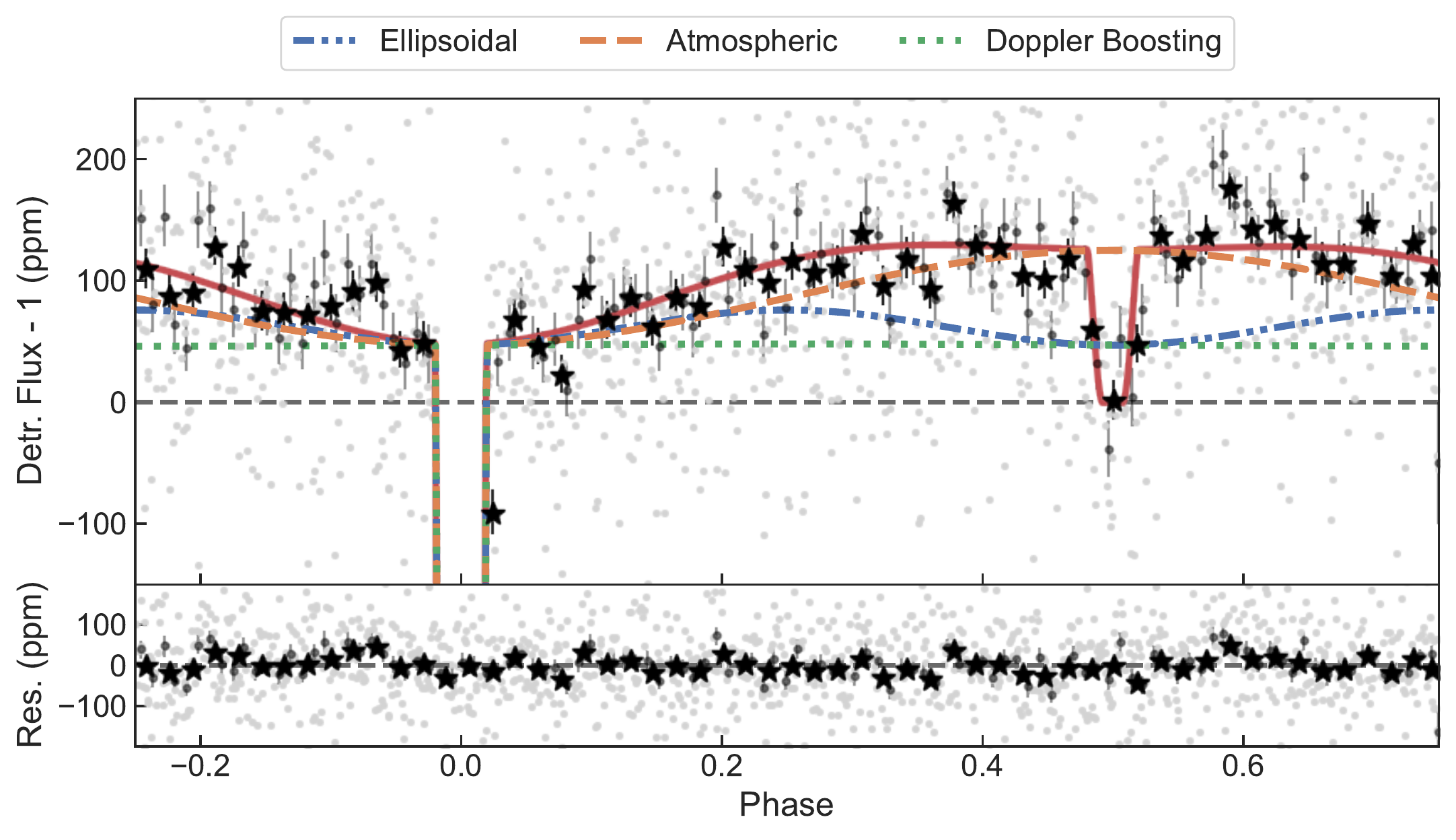}
  \caption{The phase-folded \textit{TESS} single-scale MAP PDC light curve for TOI-1431\,b, zoomed in to show the phase curve and secondary eclipse. The grey points are the detrended photometry, the dark grey points are the binned photometry with a cadence of $P_{\mathrm{p}}/300$ ($\sim12.7$\,m), and the black stars are the binned photometry with a cadence of $P_{\mathrm{p}}/150$ ($\sim25.5$\,m). Included in the phase curve model fit are the ellipsoidal (blue dash-dotted line), the atmospheric (orange dashed line), and the beaming modulations (green dotted line). The combined median model fit is plotted as the solid red line. The dark grey dashed line represents no emission from TOI-1431\,b. The bottom panel shows the residuals to the combined median model.}
  \label{fig:TESS_phase_curve}
\end{figure*}

There is also a clear signature of the secondary eclipse (occultation, i.e., when the planet passes behind the star) with a depth of $\delta_\mathrm{occ; dil}=0.127_{-0.005}^{+0.004}$\,ppt. We also robustly detect the longitudinal modulation of the planet's light in the \textit{TESS} band-pass ($A_\mathrm{atmospheric}=0.078_{-0.008}^{+0.007}$\,ppt) from the full phase curve, as shown in Figure~\ref{fig:TESS_phase_curve}, providing us with a rare opportunity to measure the planet's day/night brightness temperature contrast (see Section~\ref{atmo}).

From the light curve, we see a very weak beaming signal with a measured amplitude of $A_\mathrm{beaming}=0.0010_{-0.0006}^{+0.0012}$\,ppt and a marginal detection of ellipsoidal modulation with an amplitude of $A_\mathrm{ellipsoidal}=0.029\pm0.006$\,ppt. The predicted full amplitudes of both of these phase curve components can be derived from theory and depend on the planet--star mass ratio $q_{p}=M_{p}/M_{*}$ (see, for example, the review in \citealt{shporer2017}):
\begin{align}
\label{ellip}
A_{\mathrm{ellipsoidal}} & = 2\alpha_{\mathrm{ellip}}q_{p}\left(\frac{R_{*}}{a}\right)^{3}\sin^2 i_{p}, \\
\label{dopp}
A_{\mathrm{beaming}} & =2\left\lbrack\frac{2\pi G} {P_{p}c^{3}}\frac{q_{p}^2M_{p}\sin^3 i_{p}}{(1+q_{p})^2}\right\rbrack^{\frac{1}{3}}  \left\langle \frac{xe^{x}}{e^{x}-1}\right\rangle_{\mathrm{TESS}}.
\end{align}
Here, we have assumed for simplicity that the stellar spectrum is a blackbody, and $x\equiv hc/k\lambda T_{*}$; the angled brackets indicate averaging over the \textit{TESS} band-pass. The prefactor $\alpha_{\mathrm{ellip}}$ in the ellipsoidal distortion is an expression that includes the limb- and gravity-darkening coefficients for the host star.

To calculate the predicted amplitudes, we use the stellar parameters derived from the SED fit (Table~\ref{tab:star}) and the best-fit system parameters from the joint analysis. For the limb- and gravity-darkening coefficients, we take tabulated values from \citet{claret} and interpolate to the measured stellar parameters. Uncertainties are propagated to the estimates using Monte Carlo sampling. We obtain predicted amplitudes of $A_{\mathrm{ellipsoidal}}=0.028\pm0.004$\,ppt and $A_{\mathrm{beaming}}=0.0052\pm0.0004$\,ppt. These theoretical values are in excellent agreement with our measured values to within the $1\sigma$ level.

Our phase-curve analysis did not consider a possible phase shift in the time of maximum brightness, which would correspond to an offset in the dayside hotspot. Previous studies of ultra-hot Jupiter have indicated very small (if any) phase-curve offsets (see, for example, the case of KELT-9\,b; \citealt{2020AJ....160...88W}). Any offset would manifest itself in our phase-curve fits through its effect on the measured beaming amplitude. We have shown that the measured and predicted values of $A_{\mathrm{beaming}}$ are consistent to within $1\sigma$. Therefore, we conclude that there is no evidence for a significant phase offset in the planet's atmospheric brightness modulation.

\subsubsection{SAP Photometry}\label{SAP_analysis}

We also performed the analysis of the pre-cleaned SAP light curve segments following the same procedures as done in Sections~\ref{multi_PDC} \& \ref{PDC_analysis}. The SAP light curve (see Figure~\ref{fig:TESS_LC_SAP} in the Appendix) also shows no evidence of the dip features near the transit. We do find some discrepancies in the results between the joint fit using the SAP light curve and the single-scale MAP PDC light curve that are likely due to spacecraft point excursions (see Figure~\ref{fig:TESS_SAP_vs_PDC} in the Appendix). However, most of our derived parameters were consistent within $1\sigma$. We adopt the results from the joint analysis using the single-scale MAP PDC light curve based on Bayesian model comparison for the discussion of this paper.

%In particular, the deeper transit depth measured with the SAP light curve compared to the single-scale MAP PDC light curve is likely the result of `whisker'-like flux dips that are caused by brief spacecraft pointing excursions. These excursions result in the target pixel response function momentarily moving off center from the optimal aperture, causing slightly less flux to be measured for a fraction of a cadence that occasionally overlap with some of the transits (see Figure~\ref{tessphotometry} and Figure~\ref{fig:TESS_SAP_vs_PDC} in the Appendix). While most of the `whiskers' were removed in the SAP light curve (either from being quality flagged or from the median filter), the ones that occurred during a transit that were not quality flagged by the \textit{TESS} pipeline, remained in the light curve, resulting in a deeper transit depth. The correction applied to the single-scale MAP PDC light curve from the use of the co-trending basis vectors has removed high frequency noise, including the `whiskers' from the pointing shifts.

\startlongtable
\begin{deluxetable*}{lccc}
\tabletypesize{\small}
\tablecaption{Median values and 68\% confidence interval of the fitted and derived parameters for TOI-1431b from the joint Nested Sampling \texttt{Allesfitter} analysis of the \textit{TESS} single-scale MAP corrected PDC photometry, ground-based light curves, and radial velocity measurements.\label{tab:joint_fit_results}}
\tablehead{
\colhead{Parameter} & \colhead{Description} & \colhead{Prior\tablenotemark{a}} & \colhead{Value}
}
\startdata
\hline 
\multicolumn{4}{c}{\textit{Fitted parameters}} \smallskip \\ 
\hline
$R_{b} / R_\star$\dotfill & Radius ratio\dotfill & $\mathcal{U}(0.082;0.010;0.300)$ & $0.07955_{-0.00053}^{+0.00063}$ \\ 
$(R_\star + R_b) / a$\dotfill & Radii sum to semi-major axis\dotfill & $\mathcal{U}(0.21;0.05;0.50)$ & $0.2096\pm0.0017$ \\ 
$\cos{i_b}$\dotfill & Cosine of inclination\dotfill & $\mathcal{U}(0.169;0;1)$ & $0.1715\pm0.0022$ \\ 
$T_{0}$\dotfill & Mid transit time (2458700-BJD$_{\mathrm{TDB}}$)\dotfill & $\mathcal{U}(39.177;38.177;40.177)$ & $39.17737\pm0.00007$ \\ 
$P_b$\dotfill & Orbital period (d)\dotfill & $\mathcal{U}(2.65;2.35;2.95)$ & $2.650237\pm0.000003$ \\
$\sqrt{e} \cos{\omega}$\dotfill & Eccentricity term\dotfill & $\mathcal{U}(0.0;-0.3;0.3)$ & $-0.009_{-0.021}^{+0.023}$ \\ 
$\sqrt{e} \sin{\omega}$\dotfill & Eccentricity term\dotfill & $\mathcal{U}(0.0;-0.3;0.3)$ & $0.036_{-0.041}^{+0.033}$ \\ 
$K_b$\dotfill & RV semi-amplitude (\kms)\dotfill & $\mathcal{U}(0.300;0;1.0)$ & $0.2941\pm0.0011$ \\
$D_\mathrm{0_{TESS}}$\dotfill & Dilution\dotfill & $\mathcal{N}(0.004904;0.000588)$ & $0.00486\pm0.00051$ \\ 
$q_{1_\mathrm{TESS}}$\dotfill & Transformed limb darkening\dotfill & $\mathcal{U}(0.40;0;1)$ & $0.22\pm0.05$ \\ 
$q_{2_\mathrm{TESS}}$\dotfill & Transformed limb darkening\dotfill & $\mathcal{U}(0.21;0;1)$ & $0.18_{-0.09}^{+0.13}$ \\ 
$q_{1_\mathrm{LCO Y}}$\dotfill & Transformed limb darkening\dotfill & $\mathcal{U}(0.24;0;1)$ & $0.19_{-0.10}^{+0.13}$ \\ 
$q_{2_\mathrm{LCO Y}}$\dotfill & Transformed limb darkening\dotfill & $\mathcal{U}(0.44;0;1)$ & $0.50\pm0.29$ \\ 
$q_{1_\mathrm{MuSCAT2 z'}}$\dotfill & Transformed limb darkening\dotfill & $\mathcal{U}(0.35;0;1)$ & $0.26_{-0.10}^{+0.12}$ \\ 
$q_{2_\mathrm{MuSCAT2 z'}}$\dotfill & Transformed limb darkening\dotfill & $\mathcal{U}(0.42;0;1)$ & $0.52\pm0.28$ \\ 
$q_{1_\mathrm{MuSCAT2 r'}}$\dotfill & Transformed limb darkening\dotfill & $\mathcal{U}(0.50;0;1)$ & $0.41_{-0.12}^{+0.13}$ \\ 
$q_{2_\mathrm{MuSCAT2 r'}}$\dotfill & Transformed limb darkening\dotfill & $\mathcal{U}(0.47;0;1)$ & $0.51\pm0.25$ \\ 
$q_{1_\mathrm{MuSCAT2 i'}}$\dotfill & Transformed limb darkening\dotfill & $\mathcal{U}(0.26;0;1)$ & $0.19_{-0.08}^{+0.09}$ \\ 
$q_{2_\mathrm{MuSCAT2 i'}}$\dotfill & Transformed limb darkening\dotfill & $\mathcal{U}(0.53;0;1)$ & $0.58_{-0.28}^{+0.24}$ \\ 
$q_{1_\mathrm{MuSCAT2 g'}}$\dotfill & Transformed limb darkening\dotfill & $\mathcal{U}(0.62;0;1)$ & $0.54_{-0.11}^{+0.13}$ \\ 
$q_{2_\mathrm{MuSCAT2 g'}}$\dotfill & Transformed limb darkening\dotfill & $\mathcal{U}(0.61;0;1)$ & $0.63_{-0.21}^{+0.19}$ \\
$q_{1_\mathrm{ULMT i'}}$\dotfill & Transformed limb darkening\dotfill & $\mathcal{U}(0.26;0;1)$ & $0.14_{-0.10}^{+0.18}$ \\ 
$q_{2_\mathrm{ULMT i'}}$\dotfill & Transformed limb darkening\dotfill & $\mathcal{U}(0.53;0;1)$ & $0.35_{-0.23}^{+0.32}$ \\ 
$q_{1_\mathrm{CDK14 z'}}$\dotfill & Transformed limb darkening\dotfill & $\mathcal{U}(0.35;0;1)$ & $0.54\pm0.28$ \\ 
$q_{2_\mathrm{CDK14 z'}}$\dotfill & Transformed limb darkening\dotfill & $\mathcal{U}(0.42;0;1)$ & $0.51\pm0.30$ \\ 
$q_{1_\mathrm{CDK14 g'}}$\dotfill & Transformed limb darkening\dotfill & $\mathcal{U}(0.62;0;1)$ & $0.45_{-0.27}^{+0.29}$ \\ 
$q_{2_\mathrm{CDK14 g'}}$\dotfill & Transformed limb darkening\dotfill & $\mathcal{U}(0.61;0;1)$ & $0.63_{-0.31}^{+0.23}$ \\ 
$q_{1_\mathrm{AUKR z'}}$\dotfill & Transformed limb darkening\dotfill & $\mathcal{U}(0.35;0;1)$ & $0.83_{-0.18}^{+0.11}$ \\ 
$q_{2_\mathrm{AUKR z'}}$\dotfill & Transformed limb darkening\dotfill & $\mathcal{U}(0.42;0;1)$ & $0.60_{-0.29}^{+0.24}$ \\ 
$q_{1_{\mathrm{SCT} TESS}}$\dotfill & Transformed limb darkening\dotfill & $\mathcal{U}(0.40;0;1)$ & $0.22_{-0.11}^{+0.15}$ \\ 
$q_{2_{\mathrm{SCT} TESS}}$\dotfill & Transformed limb darkening\dotfill & $\mathcal{U}(0.21;0;1)$ & $0.57_{-0.30}^{+0.26}$ \\
$J_{p}$\dotfill & Surface brightness ratio\dotfill & $\mathcal{U}(0.0074;0;0.1)$ & $0.0076\pm0.0023$ \\ 
$A_\mathrm{beaming}$\dotfill & Phase curve term (ppt)\dotfill & $\mathcal{U}(0.0099;0;1)$ & $0.0010_{-0.0007}^{+0.0012}$ \\ 
$A_\mathrm{atmospheric}$\dotfill & Phase curve term (ppt)\dotfill & $\mathcal{U}(0.0976;0;1)$ & $0.0785_{-0.0077}^{+0.0073}$ \\ 
$A_\mathrm{ellipsoidal}$\dotfill & Phase curve term (ppt)\dotfill & $\mathcal{U}(0.0409;0;1)$ & $0.0291\pm0.0064$ \\ 
$\ln{\sigma_\mathrm{TESS}}$\dotfill & Flux error scaling ($\ln{ \mathrm{rel. flux.} }$)\dotfill & $\mathcal{U}(-7.545;-10;-5)$ & $-7.779_{-0.004}^{+0.003}$ \smallskip \\
$\ln{\sigma_\mathrm{LCO Y}}$\dotfill & Flux error scaling ($\ln{ \mathrm{rel. flux.} }$)\dotfill & $\mathcal{U}(-6.53;-10;-3)$ & $-6.528\pm0.040$ \smallskip \\
$\ln{\sigma_\mathrm{MuSCAT2 z'}}$\dotfill & Flux error scaling ($\ln{ \mathrm{rel. flux.} }$)\dotfill & $\mathcal{U}(-6.92;-10;-3)$ & $-6.926\pm0.041$ \smallskip \\
$\ln{\sigma_\mathrm{MuSCAT2 r'}}$\dotfill & Flux error scaling ($\ln{ \mathrm{rel. flux.} }$)\dotfill & $\mathcal{U}(-7.00;-10;-3)$ & $-7.007\pm0.041$ \smallskip \\
$\ln{\sigma_\mathrm{MuSCAT2 i'}}$\dotfill & Flux error scaling ($\ln{ \mathrm{rel. flux.} }$)\dotfill & $\mathcal{U}(-7.02;-10;-3)$ & $-7.019\pm0.041$ \smallskip \\
$\ln{\sigma_\mathrm{MuSCAT2 g'}}$\dotfill & Flux error scaling ($\ln{ \mathrm{rel. flux.} }$)\dotfill & $\mathcal{U}(-7.05;-10;-3)$ & $-7.045\pm0.042$ \smallskip \\
$\ln{\sigma_\mathrm{ULMT i'}}$\dotfill & Flux error scaling ($\ln{ \mathrm{rel. flux.} }$)\dotfill & $\mathcal{U}(-6.5;-10;-3)$ & $-5.826\pm0.030$ \\ 
$\ln{\sigma_\mathrm{CDK14 z'}}$\dotfill & Flux error scaling ($\ln{ \mathrm{rel. flux.} }$)\dotfill & $\mathcal{U}(-5.9;-10;-3)$ & $-5.330\pm0.043$ \\ 
$\ln{\sigma_\mathrm{CDK14 g'}}$\dotfill & Flux error scaling ($\ln{ \mathrm{rel. flux.} }$)\dotfill & $\mathcal{U}(-6.1;-10;-3)$ & $-5.358\pm0.044$ \\ 
$\ln{\sigma_\mathrm{AUKR z'}}$\dotfill & Flux error scaling ($\ln{ \mathrm{rel. flux.} }$)\dotfill & $\mathcal{U}(-5.8;-10;-3)$ & $5.808\pm0.031$ \\ 
$\ln{\sigma_{\mathrm{SCT} TESS}}$\dotfill & Flux error scaling ($\ln{ \mathrm{rel. flux.} }$)\dotfill & $\mathcal{U}(-6.5;-10;-3)$ & $-5.845\pm0.023$ \\ 
$\log{\sigma_\mathrm{RV_{SONG}}}$\dotfill & RV jitter (\kms)\dotfill & $\mathcal{U}(-7.35;-10;-1)$ & $-7.2\pm1.6$ \\ 
$\log{\sigma_\mathrm{RV_{ELP}}}$\dotfill & RV jitter (\kms)\dotfill & $\mathcal{U}(-2.93;-10;-1)$ & $-2.97_{-0.31}^{+0.34}$ \\ 
$\log{\sigma_\mathrm{RV_{TLV}}}$\dotfill & RV jitter (\kms)\dotfill & $\mathcal{U}(-2.79;-10;-1)$ & $-2.79_{-0.14}^{+0.14}$ \\ 
$\log{\sigma_\mathrm{RV_{FIES}}}$\dotfill & RV jitter (\kms)\dotfill & $\mathcal{U}(-4.43;-10;-1)$ & $-7.3_{-1.5}^{+1.3}$ \\ 
$\log{\sigma_\mathrm{RV_{SOPHIE}}}$\dotfill & RV jitter (\kms)\dotfill & $\mathcal{U}(-5.09;-10;-1)$ & $-5.54_{-1.3}^{+0.73}$ \\ 
$\log{\sigma_\mathrm{RV_{EXPRES}}}$\dotfill & RV jitter (\kms)\dotfill & $\mathcal{U}(-2.50;-10;-1)$ & $-7.15_{-1.1}^{+0.70}$ \\
$\Delta \mathrm{RV_{SONG}}$\dotfill & RV offset (\kms)\dotfill & $\mathcal{U}(-25.15;-26.0;-24.4)$ & $-25.154\pm0.006$ \\ 
$\Delta \mathrm{RV_{ELP}}$\dotfill & RV offset (\kms)\dotfill & $\mathcal{U}(-25.57;-26.5;-24.8)$ & $-25.579\pm0.019$ \\ 
$\Delta \mathrm{RV_{TLV}}$\dotfill & RV offset (\kms)\dotfill & $\mathcal{U}(-25.56;-26.0;-25.0)$ & $-25.570\pm0.011$ \\ 
$\Delta \mathrm{RV_{FIES}}$\dotfill & RV offset (\kms)\dotfill & $\mathcal{U}(-0.28;-1.2;0.2)$ & $-0.274\pm0.002$ \\ 
$\Delta \mathrm{RV_{SOPHIE}}$\dotfill & RV offset (\kms)\dotfill & $\mathcal{U}(-25.24;-26.0;-24.4)$ & $-25.249\pm0.003$ \\ 
$\Delta \mathrm{RV_{EXPRES}}$\dotfill & RV offset (\kms)\dotfill & $\mathcal{U}(0.0;-1.0;1.0)$ & $0.0051\pm0.0007$ \\ 
\hline 
\multicolumn{4}{c}{\textit{Derived parameters}} \smallskip \\ 
\hline
$R_\star/a$\dotfill & Host radius to semi-major axis\dotfill & \dotfill & $0.194\pm0.002$ \\ 
$a/R_\star$\dotfill & Semi-major axis to host radius\dotfill & \dotfill & $5.15\pm0.05$ \\ 
$R_\mathrm{b}$\dotfill & Planet radius ($\mathrm{R_{\oplus}}$)\dotfill & \dotfill & $16.7\pm0.6$ \\ 
$R_\mathrm{b}$\dotfill & Planet radius ($\mathrm{R_{jup}}$)\dotfill & \dotfill & $1.49\pm0.05$ \\ 
$a$\dotfill & Semi-major axis (AU)\dotfill & \dotfill & $0.046\pm0.002$ \\ 
$i$\dotfill & Inclination (deg)\dotfill & \dotfill & $80.13\pm0.13$ \\ 
$b_\mathrm{tra}$\dotfill & Impact parameter\dotfill & \dotfill & $0.881\pm0.004$ \\
$e$\dotfill & Eccentricity\dotfill & \dotfill & $0.0022_{-0.0016}^{+0.0030}$ \\ 
$\omega$\dotfill & Argument of periastron (deg)\dotfill & \dotfill & $108_{-31}^{+92}$ \\  
$q_\mathrm{b}$\dotfill & Mass ratio\dotfill & \dotfill & $0.00157\pm0.00006$ \\ 
$M_\mathrm{b}$\dotfill & Companion mass ($\mathrm{M_{\oplus}}$)\dotfill & \dotfill & $990\pm60$ \\ 
$M_\mathrm{b}$\dotfill & Companion mass ($\mathrm{M_{jup}}$)\dotfill & \dotfill & $3.12\pm0.18$ \\
$T_\mathrm{tot}$\dotfill & Total transit duration (h)\dotfill & \dotfill & $2.489\pm0.009$ \\ 
$T_\mathrm{full}$\dotfill & Full-transit duration (h)\dotfill & \dotfill & $1.057\pm0.034$ \\ 
$T_\mathrm{0;occ}$\dotfill & Epoch occultation (2458700-BJD$_{\mathrm{TDB}}$)\dotfill & \dotfill & $40.502\pm0.002$ \\ 
$b_\mathrm{occ}$\dotfill & Impact parameter occultation\dotfill & \dotfill & $0.885_{-0.005}^{+0.006}$ \\ 
$\rho_\mathrm{\star}$\dotfill & Host density from orbit (cgs)\dotfill & \dotfill & $0.37\pm0.01$ \\ 
$\rho_\mathrm{b}$ \dotfill & Companion density (cgs)\dotfill & \dotfill & $1.17_{-0.16}^{+0.18}$ \\ 
$g_\mathrm{b}$\dotfill & Companion surface gravity (cgs)\dotfill & \dotfill & $3430_{-60}^{+50}$ \\ 
$T_\mathrm{eq}$\tablenotemark{b}\dotfill & Equilibrium temperature (K)\dotfill & \dotfill & $2370\pm70$ \\
$H_\mathrm{b}$\dotfill & Companion atmospheric scale height (km)\dotfill & \dotfill & $230\pm30$ \\ 
$u_\mathrm{1; TESS}$\dotfill & Limb darkening\dotfill & \dotfill & $0.13_{-0.08}^{+0.13}$ \\ 
$u_\mathrm{2; TESS}$\dotfill & Limb darkening\dotfill & \dotfill & $0.33_{-0.12}^{+0.10}$ \\ 
$u_\mathrm{1; LCO Y}$\dotfill & Limb darkening\dotfill & \dotfill & $0.40_{-0.24}^{+0.27}$ \\ 
$u_\mathrm{2; LCO Y}$\dotfill & Limb darkening\dotfill & \dotfill & $0.00_{-0.22}^{+0.25}$ \\ 
$u_\mathrm{1; MuSCAT2 z'}$\dotfill & Limb darkening\dotfill & \dotfill & $0.51\pm0.26$ \\ 
$u_\mathrm{2; MuSCAT2 z'}$\dotfill & Limb darkening\dotfill & \dotfill & $-0.02_{-0.24}^{+0.29}$ \\ 
$u_\mathrm{1; MuSCAT2 r'}$\dotfill & Limb darkening\dotfill & \dotfill & $0.64_{-0.31}^{+0.28}$ \\ 
$u_\mathrm{2; MuSCAT2 r'}$\dotfill & Limb darkening\dotfill & \dotfill & $-0.01_{-0.28}^{+0.33}$ \\ 
$u_\mathrm{1; MuSCAT2 i'}$\dotfill & Limb darkening\dotfill & \dotfill & $0.49\pm0.24$ \\ 
$u_\mathrm{2; MuSCAT2 i'}$\dotfill & Limb darkening\dotfill & \dotfill & $-0.07_{-0.20}^{+0.24}$ \\ 
$u_\mathrm{1; MuSCAT2 g'}$\dotfill & Limb darkening\dotfill & \dotfill & $0.91_{-0.28}^{+0.25}$ \\ 
$u_\mathrm{2; MuSCAT2 g'}$\dotfill & Limb darkening\dotfill & \dotfill & $-0.18_{-0.26}^{+0.31}$ \\ 
$u_\mathrm{1; ULMT i'}$\dotfill & Limb darkening\dotfill & \dotfill & $0.23_{-0.16}^{+0.25}$ \\ 
$u_\mathrm{2; ULMT i'}$\dotfill & Limb darkening\dotfill & \dotfill & $0.09_{-0.19}^{+0.23}$ \\ 
$u_\mathrm{1; CDK14 z'}$\dotfill & Limb darkening\dotfill & \dotfill & $0.67_{-0.40}^{+0.50}$ \\ 
$u_\mathrm{2; CDK14 z'}$\dotfill & Limb darkening\dotfill & \dotfill & $-0.02_{-0.39}^{+0.42}$ \\ 
$u_\mathrm{1; CDK14 g'}$\dotfill & Limb darkening\dotfill & \dotfill & $0.76_{-0.43}^{+0.46}$ \\ 
$u_\mathrm{2; CDK14 g'}$\dotfill & Limb darkening\dotfill & \dotfill & $-0.15_{-0.33}^{+0.35}$ \\ 
$u_\mathrm{1; AUKR z'}$\dotfill & Limb darkening\dotfill & \dotfill & $1.04_{-0.50}^{+0.44}$ \\ 
$u_\mathrm{2; AUKR z'}$\dotfill & Limb darkening\dotfill & \dotfill & $-0.17_{-0.42}^{+0.51}$ \\ 
$u_\mathrm{1; SCT TESS}$\dotfill & Limb darkening\dotfill & \dotfill & $0.48_{-0.24}^{+0.27}$ \\ 
$u_\mathrm{2; SCT TESS}$\dotfill & Limb darkening\dotfill & \dotfill & $-0.05_{-0.21}^{+0.29}$ \\ 
$\delta_\mathrm{TESS_{tr}; undil}$\dotfill & Transit depth (undil.) (ppt)\dotfill & \dotfill & $5.825\pm0.007$ \\ 
$\delta_\mathrm{TESS_{tr}; dil}$\dotfill & Transit depth (dil.) (ppt)\dotfill & \dotfill & $5.797\pm0.007$ \\
$\delta_\mathrm{TESS_{occ}; undil}$\dotfill & Occultation depth (undil.) (ppt)\dotfill & \dotfill & $0.127_{-0.005}^{+0.004}$ \\ 
$\delta_\mathrm{TESS_{occ}; dil}$\dotfill & Occultation depth (dil.) (ppt)\dotfill & \dotfill & $0.127_{-0.005}^{+0.004}$ \\ 
$F_\mathrm{TESS_{night}; undil}$\dotfill & Nightside flux (undil.) (ppt)\dotfill & \dotfill & $0.049_{-0.005}^{+0.004}$ \\ 
$F_\mathrm{TESS_{night}; dil}$\dotfill & Nightside flux (dil.) (ppt)\dotfill & \dotfill & $0.049_{-0.005}^{+0.004}$ \\
$\delta_\mathrm{LCO Y}$\dotfill & Transit depth (ppt)\dotfill & \dotfill & $5.67_{-0.27}^{+0.25}$ \\ 
$\delta_\mathrm{MuSCAT2 z'}$\dotfill & Transit depth (ppt)\dotfill & \dotfill & $5.54_{-0.24}^{+0.22}$ \\ 
$\delta_\mathrm{MuSCAT2 r'}$\dotfill & Transit depth (ppt)\dotfill & \dotfill & $5.30_{-0.27}^{+0.22}$ \\ 
$\delta_\mathrm{MuSCAT2 i'}$\dotfill & Transit depth (ppt)\dotfill & \dotfill & $5.62_{-0.23}^{+0.21}$ \\ 
$\delta_\mathrm{MuSCAT2 g'}$\dotfill & Transit depth (ppt)\dotfill & \dotfill & $4.90\pm0.26$ \\
$\delta_\mathrm{ULMT i'}$\dotfill & Transit depth (ppt)\dotfill & \dotfill & $5.84_{-0.31}^{+0.26}$ \\ 
$\delta_\mathrm{CDK14 z'}$\dotfill & Transit depth (ppt)\dotfill & \dotfill & $5.13_{-0.61}^{+0.47}$ \\ 
$\delta_\mathrm{CDK14 g'}$\dotfill & Transit depth (ppt)\dotfill & \dotfill & $5.09_{-0.68}^{+0.55}$ \\ 
$\delta_\mathrm{AUKR z'}$\dotfill & Transit depth (ppt)\dotfill & \dotfill & $4.51_{-0.56}^{+0.53}$ \\ 
$\delta_{\mathrm{SCT} TESS}$\dotfill & Transit depth (ppt)\dotfill & \dotfill & $5.58_{-0.27}^{+0.24}$
\enddata
\tablenotetext{a}{$\mathcal{N}(\mu;\sigma)$ is a normal distribution with mean $\mu$ and width $\sigma$, $\mathcal{U}(s;a;b)$ is a uniform prior with a starting value $s$ and lower and upper limits of $a$ and $b$, respectively.}
\tablenotetext{b}{Calculated assuming zero Bond albedo.}
\end{deluxetable*}

%\newpage
%-------------------------------------------------------------------------------------
%\clearpage

\subsection{The Atmosphere}\label{atmo}

We have used the \textit{TESS} single-scale MAP corrected PDC photometry and \texttt{Allesfitter} to characterize the red-optical ($6000-9500$\,\AA\,) phase curve of TOI-1431\,b, which includes contributions from beaming, ellipsoidal modulation, and atmospheric brightness modulation, as well as the secondary eclipse (see Section~\ref{PDC_analysis}). We also fit the phase curve and secondary eclipse using the \textit{TESS} SAP photometry (see Figure~\ref{fig:TESS_phase_curve_SAP} in the Appendix) as a comparison to our preferred solution using the single-scale MAP PDC photometry.

\begin{figure*}
  \includegraphics[width=\linewidth]{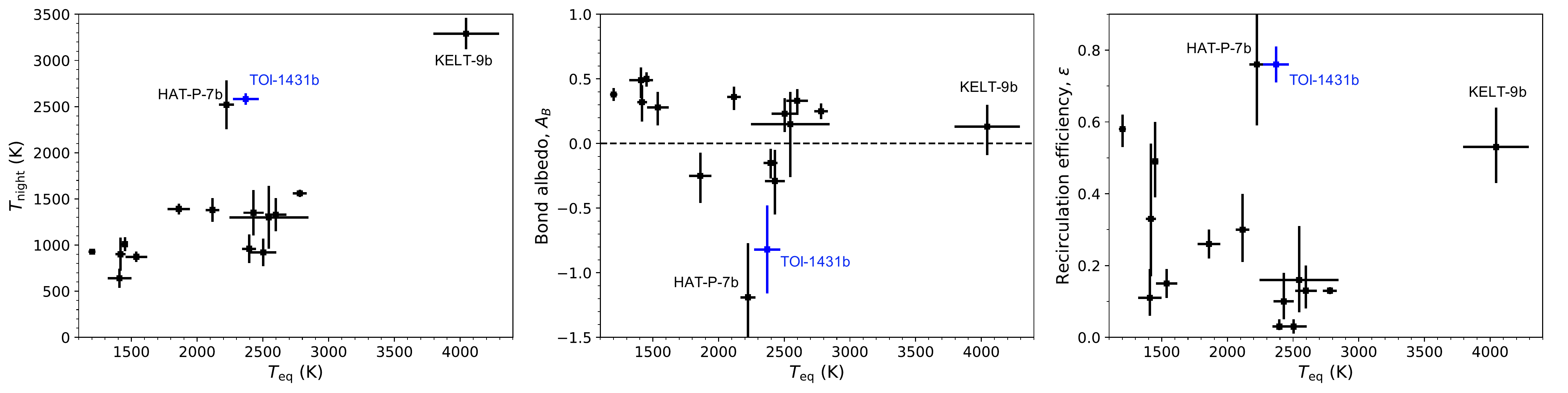}
  \caption{Plots of nightside brightness temperature $T_{\mathrm{night}}$, Bond albedo $A_{B}$, and day-night heat recirculation efficiency $\epsilon$ as a function of equilibrium temperature $T_{\mathrm{eq}}$ for a population of hot and ultra-hot Jupiters. TOI-1431\,b is indicated in blue; the black points are taken/derived from the results of \citet{bell2021}. KELT-16b has been omitted due to its very large nightside temperature uncertainty ($>400$~K) and poorly constrained $\epsilon$ value. With its high nightside temperature, low day/night temperature contrast, and strongly negative Bond albedo, TOI-1431\,b is a notable outlier among gas giants with similar levels of stellar irradiation, and is very similar to HAT-P-7b.}
  \label{fig:context}
\end{figure*}

From the measured secondary eclipse depth $\delta_{\mathrm{occ}}$ and nightside planetary flux $\delta_{\mathrm{night}}$, we can derive the dayside and nightside brightness temperatures of TOI-1431\,b using the relationship between the planet--star flux ratio $\delta$, planetary emission $F_{p}(\lambda,T_{p})$ (assumed to be a blackbody), and stellar spectrum $F_{*}(\lambda,T_{\mathrm{eff}})$: 
\begin{equation}\label{temp}
\delta = \left(\frac{R_{p}}{R_{*}}\right)^{2}\frac{\int F_{p}(\lambda,T_{p})\tau(\lambda) d\lambda}{\int F_{*}(\lambda,T_{\mathrm{eff}})\tau(\lambda) d\lambda}+A_{g}\left(\frac{R_{p}}{a}\right)^{2}.
\end{equation}
Any reflected light on the dayside atmosphere contributes to the measured flux ratio through the geometric albedo $A_{g}$; this second term is zero when computing the nightside temperature. The planetary and stellar spectra are integrated over the \textit{TESS} band-pass, which has a transmission function (in energy units) represented by $\tau(\lambda)$. 

We model the star's spectrum using PHOENIX models \citep{phoenix}. To properly interpolate the stellar models and propagate the uncertainties in stellar parameters to the planetary temperature estimates, we follow the methodology described in \citet{2020AJ....160...88W} and calculate the integrated stellar flux within the \textit{TESS} band-pass for a grid of PHOENIX models, before fitting a cubic polynomial in $(T_{\mathrm{eff}},\log\,g,\mathrm{[Fe/H]})$ to the full set of values. We then use a standard MCMC routine to compute the posterior distribution of the planetary brightness temperature $T_{p}$, with Gaussian priors on the stellar and system parameters, as well as the measured dayside or nightside flux ($\delta_{\mathrm{occ}}$ or $\delta_{\mathrm{night}}$; see Table~\ref{tab:joint_fit_results}).

For the dayside brightness temperature, we obtain $T_\mathrm{day}=3004\pm64$\,K when assuming zero geometric albedo; across an albedo range of 0--0.2, we find dayside brightness temperatures spanning 2700--3100~K. We measure a very high nightside brightness temperature of $T_\mathrm{night}=2583\pm63$\,K. The extremely hot dayside is expected to preclude the formation of condensates, resulting in a cloudfree dayside hemisphere \citep[e.g.,][]{wakeford2017}. Previous analyses of other ultra-hot Jupiters combining secondary eclipses at optical and thermal infrared wavelengths confirm the predictions of low reflectivity, yielding near-zero geometric albedos \citep[e.g., WASP-18b and WASP-33b;][]{2019AJ....157..178S,vonessen2020}.

We use the measured dayside and nightside brightness temperatures to jointly constrain the Bond albedo $A_{\mathrm{B}}$ and the efficiency of heat recirculation from the dayside to nightside of the planet $\epsilon$ (0 for no recirculation; 1 for full recirculation), following the thermal balance formalism outlined in \citet{2011ApJ...729...54C} and the methodology described in \citet{2020AJ....160...88W}. We find $A_{B}=-0.82^{+0.30}_{-0.38}$ and $\epsilon=0.76\pm0.05$. The very efficient day-night heat recirculation is borne out by the low day/night temperature contrast ($\sim$420\,K). However, the unusual negative Bond albedo suggests that the overall thermal emission level across the entire planet cannot be accounted for by the energy input from stellar insolation alone.

By placing TOI-1431\,b in context, we can appreciate the exceptional atmospheric properties of this planet. Figure~\ref{fig:context} shows the nightside brightness temperatures, Bond albedos, and recirculation efficiencies for a sample of hot and ultra-hot Jupiters, taken and/or analogously derived from the brightness temperatures listed in the comprehensive Spitzer 4.5~$\mu$m phase curve analysis by \citet{bell2021}. From the plots, it is evident that TOI-1431\,b has a much higher nightside temperature and day-night heat recirculation efficiency than most other hot Jupiters with comparable equilibrium temperatures. In particular, this planet has a higher heat recirculation efficiency and significantly lower Bond albedo when compared to KELT-9\,b ($\epsilon=0.39\pm0.05$ and $A_{B}=0.19^{+0.12}_{-0.11}$; \citealt{2020AJ....160...88W}), the hottest exoplanet ever discovered \citep{2017Natur.546..514G} with a dayside and nightside brightness temperatures of $T_\mathrm{day}=4450^{+220}_{-210}$\,K and $T_\mathrm{day}=3290\pm170$\,K \citep{bell2021}, respectively. Meanwhile, only HAT-P-7\,b has a similar strongly ($>2\sigma$) negative inferred Bond albedo; indeed, HAT-P-7\,b is very similar to TOI-1431\,b in all respects. Broadly speaking, KELT-9\,b, HAT-P-7\,b, and TOI-1431\,b are outliers amid the weak overall trends across gas giants with $T_{\mathrm{eq}}>1200$\,K, i.e., increasing nightside temperature, decreasing Bond albedo, and decreasing day-night heat recirculation efficiency with increasing equilibrium temperature. In the case of KELT-9\,b in particular, given its extreme dayside temperature, its anomalously high heat recirculation efficiency could be explained from the moderation of the day-night temperature contrast due to the transport of thermally dissociated atomic hydrogen from the dayside to the nightside and subsequent recombination into molecular hydrogen, which releases a significant amount of heat \citep[e.g.,][]{2018ApJ...857L..20B}.

The unexpected negative Bond albedo may indicate the limitations of the simple thermal balance considerations underpinning our estimates of $A_{B}$ and $\epsilon$. In particular, gradients in chemical composition between the dayside and nightside hemispheres can entail drastically different atmospheric opacities, meaning that the pressure levels probed by our broadband photometric measurements may vary significantly across the planet's surface. Another explanation for a negative Bond albedo is additional thermal emission from residual heat of formation. This scenario would increase the emitted flux at all longitudes, raising the measured brightness temperatures on both the dayside and the nightside hemispheres.

%($T_{\mathrm{night}}>T_{\mathrm{eq}}$)
This explanation may be especially applicable to TOI-1431\,b: the stellar age inferred from SED modeling is $0.29^{+0.32}_{-0.19}$\,Gyr, making the system among the youngest giant planet hosting systems hitherto discovered. However, the age of the planet is sufficient for it to have deflated through the initial Kelvin-Helmholtz contraction phase and for its atmosphere to have cooled and reached the equilibrium temperature. The time scale over which this initial cooling and deflation occurred is $\sim1$\,Myr (see Equation 17 from \citealt{2015ApJ...803..111G}), the photospheric cooling time to reach the regime where stellar irradiation acts to slow cooling. The current level of incident stellar flux is sufficient to slow the planet's interior cooling and allow it to remain hot and inflated at its present-day radius (see Figure~\ref{fig:radius_time} in the Appendix and \citealt{Lous:2020ug}, \citealt{2020ApJ...893...36K}). We do not require deposited heating in the deep interior to explain the present-day radius, however, a weak conversion of $< 0.05\%$ of the incident stellar flux to deposited heating is allowed. Importantly, we find that the effects of irradiation slowing cooling can only explain the present-day radius if TOI-1431\,b arrived at its current orbit within $\sim1$\,Myr after formation. This may imply that TOI-1431\,b either formed in-situ or that a nearby stellar or massive planetary companion rapidly scattered the planet inward very soon after formation. The high obliquity orbit as measured by the Rossiter-McLaughlin effect \citep[][see Section~\ref{obliquity}]{2021arXiv210412414S}, suggests that the planet experienced just such a scattering event early in its history.

Future spectrally-resolved emission spectra with the James Webb Space Telescope (JWST) can enable detailed analyses of the atmospheric composition and temperature--pressure profiles across the planet's surface, providing a full picture of the thermal energy budget for TOI-1431\,b.

\begin{figure*}[hbt!]
    \includegraphics[width=\columnwidth]{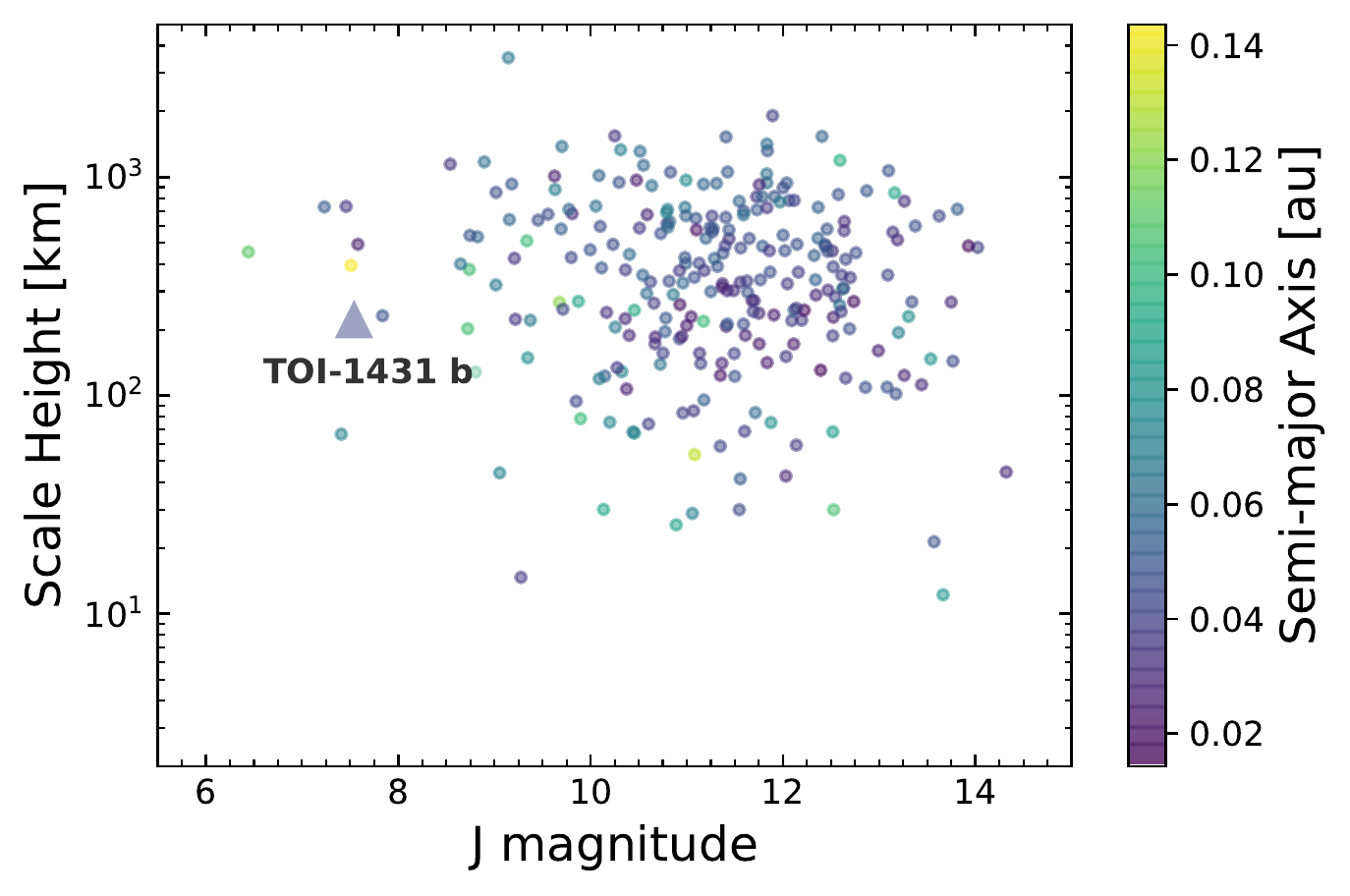}
    \includegraphics[width=\columnwidth]{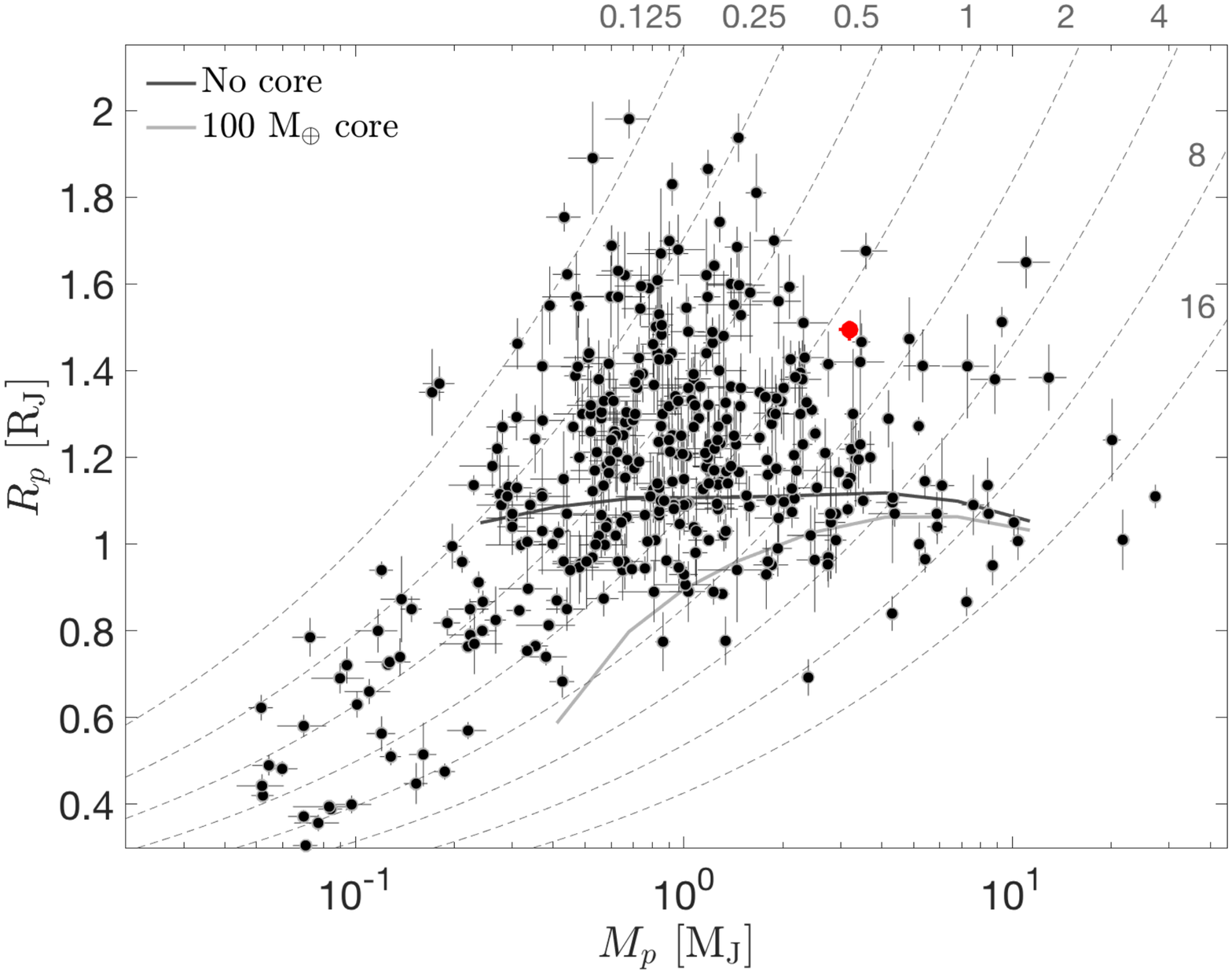}
  
    \medskip
  
    \includegraphics[width=\columnwidth]{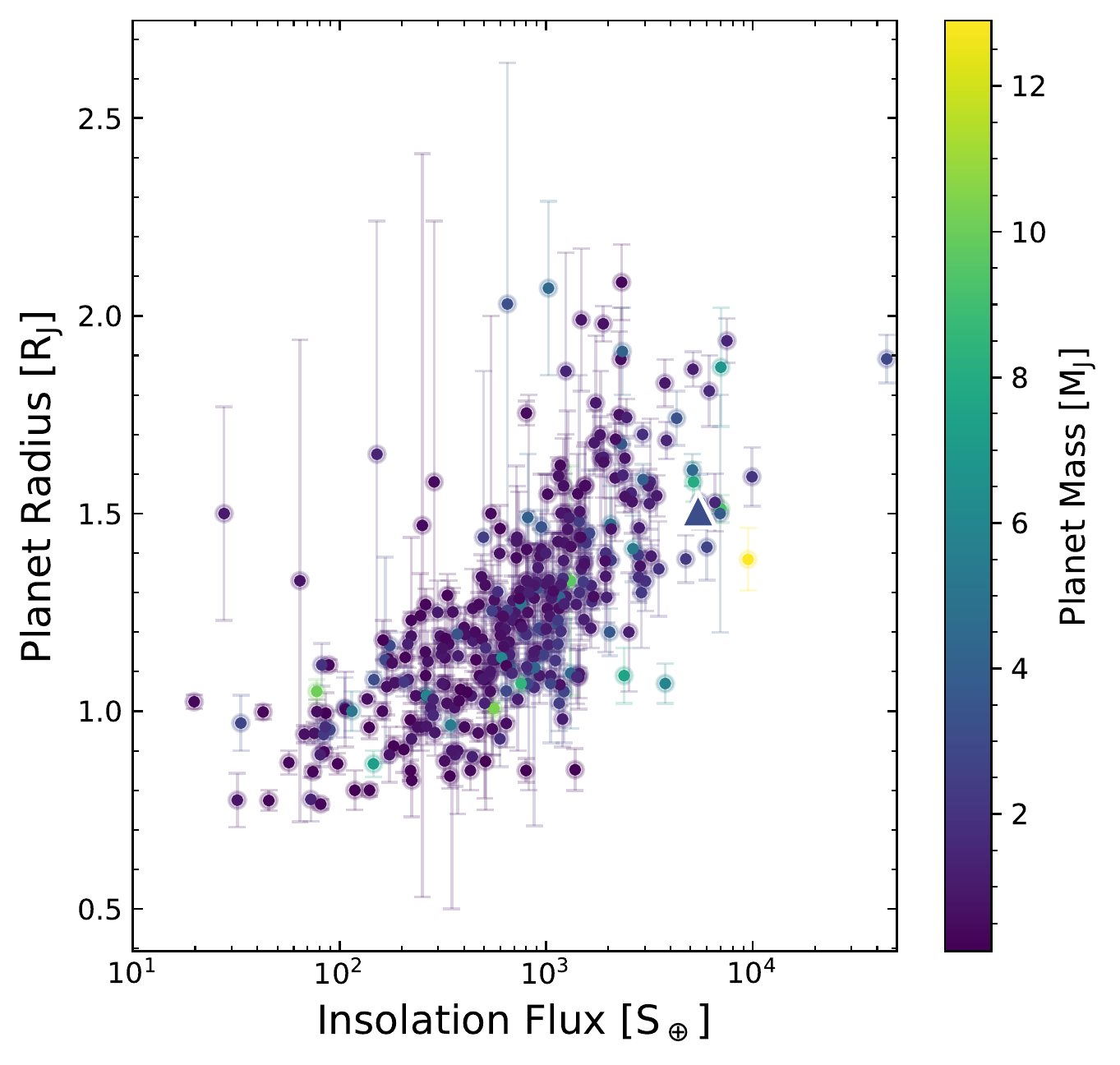}
    \includegraphics[width=\columnwidth]{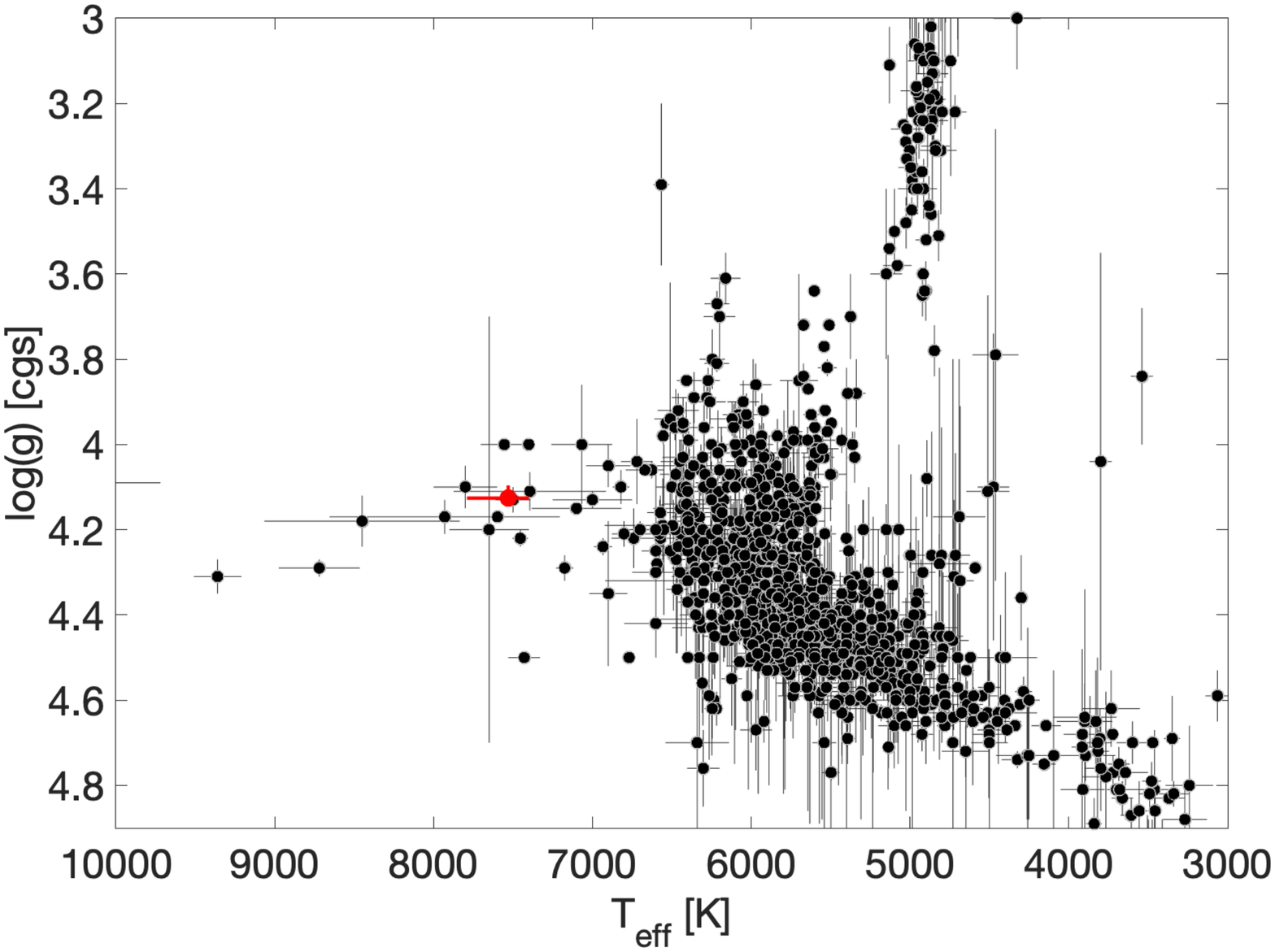}
    
    \caption{Comparison of TOI-1431\,b and its host star with other hot and ultra-hot Jupiter systems. \textbf{Top left:} The atmospheric scale heights of hot Jupiters as a function of host star brightness in the Jmag. The points are colored based on the distance from the host star, with TOI-1431\,b plotted as a triangle. This planet transits one of the brightest stars, making it an excellent candidate for future follow-up atmospheric studies. \textbf{Top right:} Planet mass versus radius of hot Jupiters. The red point is the location of TOI-1431\,b. Planets located above the dark grey solid line are predicted to have no solid core while those plotted below the light grey solid line are predicted to have a solid core of at least 100\,\Mearth\ \citep{2007ApJ...659.1661F}. The dotted lines are isodensity lines, with the density value given in the top-left, in cgs units. This plot shows that TOI-1431\,b is moderately inflated compared to other Jovians of similar mass. \textbf{Bottom left:} Incident stellar irradiation versus planet radius for all known Jovians and sub-Jovians that have measured masses and radii with the planet masses indicated by the color scale on the left. TOI-1431\,b is plotted as the colored triangle on the right and it is clear that this planet is one of the most highly irradiated planets. \textbf{Bottom right:} Plot of stellar effective temperature versus surface gravity for planet hosting stars. TOI-1431 is plotted as the red point and shows that it is one of the hottest (top 1\%) known planet hosting stars.}
    \label{fig:comparisons}
\end{figure*}

From the derived equilibrium temperature and surface gravity of the planet and assuming a H/He atmosphere with a mean molecular mass of $\mu=2.3$\,amu, the calculated atmospheric scale height (from $H_{\mathrm{p}}=kT_{eq}/(\mu g_{\mathrm{p}})$) is $H_{\mathrm{p}}=220\pm30$\,km. While not the largest scale height among the population of hot Jupiters, this planet orbits one of the brightest host stars ($J_{mag}=7.541\pm0.030$ and $K_{mag}=7.439\pm0.030$, see Figure~\ref{fig:comparisons}) making it a potential target for future atmospheric characterization through transmission spectroscopy.

%Maybe some of the below text can be trimmed.

The transmission spectroscopy metric \citep[TSM, see,][]{2018PASP..130k4401K}, used to assess the suitability of transmission spectroscopy observations with JWST, is $\sim110$ for TOI-1431\,b. Planets with TSM values greater than 90 (for Jovians and sub-Jovians), such as for this planet, are considered suitable for these observations with JWST. However, transmission spectroscopy carried out by \citet{2021arXiv210412414S} from two HARPS-N and one EXPRES transit observations finds no absorption signatures in the planet's atmosphere, likely due to its high surface gravity. Additionally, given the detection of the phase curve and secondary eclipse in the \textit{TESS} red-optical band photometry, this will be an excellent target for emission spectroscopy with JWST. In particular, phase curve observations carried out with the Near Infrared Imager and Slitless Spectrograph (NIRISS) over wavelengths between 0.6 and 5.0\,$\mathrm{\mu m}$\, should provide a high-precision global temperature map of this planet's atmosphere \citep{2018haex.bookE.116P}. Measurements of the abundances of molecular species as a function of longitude (chemical mapping) could also be probed through phase-resolved spectroscopic observations taken with JWST's NIRCam and NIRSpec instruments.

Previous atmospheric modeling of the extreme end-member ultra-hot Jupiter KELT-9\,b can serve as a guide for exploring what future intensive emission spectroscopy might reveal for TOI-1431\,b. The comparison is particularly apropos due to the similar planetary surface gravities of the two planets --- $\log\,g_p=3.30$ and 3.54 (in cgs units) for KELT-9\,b and TOI-1431\,b, respectively. In \citet{2020AJ....160...88W}, radiative transfer calculations of the dayside emission spectrum of KELT-9\,b produced largely featureless spectra across the visible and near-infrared wavelength range. In contrast, model nightside emission spectra spanning 2500--3500\,K showed large near-infrared absorption features due to H$_2$O and CO, as well as excess continuum opacity in the visible and near-infrared through $\sim$1.7~$\mu$m due to dissociated H$^-$; this latter spectral feature is a unique property of the hottest ultra-hot Jupiters and has been studied in detail by numerous earlier theoretical works \citep[e.g.,][]{arcangeli2018,lothringer2018,2018A&A...617A.110P}. Measuring the detailed shape of this H$^-$ feature can provide the abundance of H$^-$ and the temperature--pressure profile across the nightside, which will in turn probe the effect of hydrogen dissociation on the global atmospheric dynamics and thermal energy budget.

%and potentially a cloud map (assuming the atmosphere is not completely cloud free) of this planet's atmosphere if phase curve offsets with wavelength are detected \citep{2018haex.bookE.116P}

TOI-1431\,b will also be a great target for detailed atmospheric characterization by the European Space Agency's \textit{Atmospheric Remote-sensing Infrared Exoplanet Large-survey} \citep[ARIEL,][]{2018ExA....46..135T} telescope. ARIEL will operate in the infrared with a spectral range between 1.25--7.8\,$\mathrm{\mu m}$\, as well as multiple narrow-band photometry in the optical. As such, it will be well suited to potentially measure this planet's equilibrium chemistry, trace gases, vertical and horizontal thermal structures, and the detection of clouds and cloud composition (assuming the atmosphere is not cloud free) through transmission and emission spectroscopy and phase curve observations.

\section{Discussion}
\label{discussion}
TOI-1431\,b joins the growing list of $\sim16$ hot and ultra-hot Jovians with measured full phase curves and secondary eclipses \citep[e.g., see,][]{bell2021}. With a dayside and nightside brigntness temperatures of $T_\mathrm{day}=3004\pm64$\,K and $T_\mathrm{night}=2583\pm63$\,K, respectively, TOI-1431\,b is one of the hottest known exoplanets. 

%\textit{TESS} observations of this system in Sectors 15 and 16 (2019 August 15 to 2019 October 7) have provided us with measurements of the amplitude of the phase curve and the depth of the secondary eclipse, in turn allowing us to determine the day and night side temperatures of the planet's atmosphere. Assuming zero geometric albedo, we found the planet's dayside temperature to be $T_\mathrm{day}=3067^{+75}_{-81}$\,K and the nightside temperature as $T_\mathrm{night}=2554^{+75}_{-81}$\,K, giving a day/night temperature difference of around 500~K. 

We have measured the mass of TOI-1431\,b to be $M_\mathrm{p}=3.12\pm0.18$\,\Mjup\, from radial velocity observations obtained from the SONG, SOPHIE, FIES, NRES TLV and ELP, and EXPRES spectrographs. When combined with the planet radius of $R_\mathrm{p}=1.49\pm0.05$\,\Rjup, this mass gives a bulk density of $\rho_\mathrm{p}=1.17_{-0.16}^{+0.18}$\,\densitycgs, similar to Jupiter (1.326\,\densitycgs). The measured radial velocity semi-amplitude of $K=294.1\pm1.1$\,\mos\, is the most precise $K$ measurement for a planet-hosting star hotter than $6600$\,K. A large contributing factor for this is the unusually slow projected stellar rotational velocity of $v \sin i<10$\,\kms.

Figure~\ref{fig:comparisons} shows a plot of the planet radius as a function of planet mass as well as a plot of planet radius as a function of stellar irradiation for all known Jovians and sub-Jovians with measured masses and radii. The radius of TOI-1431\,b is inflated compared to other exoplanets of similar mass as shown in Figure~\ref{fig:comparisons}, but is not so inflated when accounting for the very high insolation flux ($5300^{+500}_{-470}\mathrm{S_{\oplus}}$), indicating that the planet inflation is likely the result of stellar irradiation and youthful age ($0.29^{+0.32}_{-0.19}$\,Gyr).

\subsection{Metal Peculiar Am Star}\label{Am_star}
TOI-1431\,b orbits a bright ($V_{T}\sim8.0$ and $J\sim7.5$) and very hot ($T_\mathrm{eff}\sim7700$\,K) host star. In fact, TOI-1431 is one of the hottest planet hosting stars as shown in Figure~\ref{fig:comparisons}. The star is also classified as being a non-magnetic metallic-line chemically peculiar Am star \citep[see,][]{1970PASP...82..781C, 1991AAS...87..319F,1991AAS...89..429R,2009AA...498..961R}. Am stars typically have slow rotation rates compared to typical `field' A stars of similar effective temperatures (A star $v\sin{i}$ range between $\sim100$ to $\sim150$\,\kms, see for example, \citealt{1995ApJS...99..135A}), which is certainly the case for TOI-1431 (assuming the star is not being observed nearly pole-on). The slow rotation of these stars is believed to be due to a close orbiting ($2.5\lesssim P\lesssim100$\,d) stellar companion (at least 64\% of Am stars are found to be binary systems, see e.g., \citealt{2006PASP..118..419B} and \citealt{2007MNRAS.380.1064C}) that raises tidal forces on the primary star, causing the primary to lose angular momentum and spin down \citep{1983ApJ...269..239M}. 

This in turn results in the onset of gravitational settling and radiative levitation due to the lack of mixing in the shallow convective envelopes of slow rotating A stars that produces a chemical peculiarity observed as a photospheric overabundance of iron-peak elements (such as Cr, Mn, Fe, and Ni and specifically of Ba, Y and Sr) but depleted abundances of light elements such as Ca, Sc and Mg \citep[e.g., see,][]{1974ARA&A..12..257P,1983ApJ...269..239M,1993ASPC...44..474C, 2006PASP..118..419B,2020ApJ...898...28X}. 

The spectroscopic analysis of TOI-1431 from spectra collected from SONG reveals that the star has an overall metallicity slightly lower than the Solar abundance ($[$M/H$]=-0.15 \pm 0.10$\,dex) and even less so in alpha elements (such as O, Mg, Si, S, Ca, and Ti) with $[\alpha$/H$]=-0.27 \pm 0.10$\,dex, while analysis of the SOPHIE spectra gives a somewhat higher than Solar abundance of iron ($[$Fe/H$]=0.09 \pm 0.03$). This seems to suggest that the star has an overabundance of iron group elements and an under-abundance of light elements (compared with normal A stars) that is characteristic of Am stars. However, a more detailed analysis of the stellar spectra is required to confirm the Am star classification.

Another interesting feature of Am stars is the ``anomalous luminosity effect'', which causes deviations in the strengths of specific spectral lines from what is expected from normal main-sequence A stars at a given effective temperature \citep{1971A&A....14..233B}. This means that depending on the spectral lines used when fitting the stellar spectrum model, one can obtain very different values for a star's effective temperature and surface gravity, hence the overall stellar classification. This appears to be the case for TOI-1431 (especially for $\log g$); the analysis of SONG spectra gives a $T_\mathrm{eff}=6764 \pm 120$\,K and $\log g=2.76 \pm 0.26$\,dex, the FIES spectral analysis gives $T_\mathrm{eff}=6910 \pm 50$\,K and $\log g=3.29 \pm 0.10$\,dex, and the SOPHIE spectral analysis gives $T_\mathrm{eff}=6950 \pm 60$\,K and $\log g=4.72 \pm 0.08$\,dex. 

%\BAcom{Might include some details regarding the lines and/or wavelength range used to extract stellar properties from SONG, SOPHIE, and FIES}.

Planets orbiting main-sequence Am stars appear to be quite rare--only five that transit have been discovered to date and these include TOI-1431\,b, WASP-33\,b \citep{2010MNRAS.407..507C}, KELT-17\,b \citep{2016AJ....152..136Z,2020A&A...641A.145S,2021arXiv210104416S}, KELT-19A\,b \citep{2018AJ....155...35S}, and WASP-178\,b \citep{2019MNRAS.490.1479H}/ KELT-26\,b \citep{2020AJ....160..111R}. This raises some interesting questions. First, what is the occurrence rate of planets around Am stars? Radial velocity surveys have historically avoided hot and early type stars ($T_{\mathrm{eff}}>6500$\,K and mid-F and earlier), instead have targeted `Solar analogs' \citep[i.e., late-F, G, and K type stars, see,][]{1999ApJ...526..890C,2001ApJ...551..507T,2004A&A...423..385P,2005ApJS..159..141V,2008ApJ...683L..63W} in the search for planets. As such, planets orbiting hot main-sequence A stars like TOI-1431 have gone nearly undiscovered, until now thanks to \textit{TESS} indiscriminately surveying the sky for transiting planets orbiting bright stars. Planets have been discovered orbiting former A stars, i.e., stars that have evolved off the main-sequence to become giant and sub-giant stars, in radial velocity surveys targeting such stars \citep[e.g.][]{Astar1,2014A&A...568A..64T,2015A&A...574A.116R,Astar2,Astar3,Astar4}. Recent results from surveys of evolved stars indicate that the occurrence rate of giant planets around giant stars is $\sim$10 percent (e.g., \citealt{Astar4}; Wolthoff et al., under review), indicating that giant planets are relatively common around hot stars and could also be common around Am stars as well.

If giant planet formation around A-type stars is indeed a relatively common occurrence, could the migration of a Jovian planet to a close-in orbit around its host (as presumed to be the case for TOI-1431\,b) play a significant role in its tidal spin-down and as such, contribute to its nature as being an Am star? For TOI-1431\,b, we do not see any conclusive evidence for this given that the system does not appear to be tidally synchronized ($P_\mathrm{p}\simeq P_\mathrm{rot}$) based on the host star's rotation period of between $\sim10$ to $\sim16$\,d (from spectroscopic measurements of $v\sin{i}$) and the planet's orbital period of $\sim2.65$\,d. However, the one caveat is that the orientation of the stellar spin axis ($I_\mathrm{\star}$) is unknown and the star could perhaps be spinning more rapidly if we are observing it nearly pole-on (leaving open the possibility of $P_\mathrm{p}\simeq P_\mathrm{rot}$). If indeed giant planets can contribute to the nature of their host's being an Am star, for Am stars without evidence of a stellar binary, do most or all host a hot sub-stellar or massive planetary companion? Radial velocity and transit searches targeting Am stars could resolve this question as well as provide valuable insights into the formation and evolution of Am stars, a process that is not fully understood.

\subsection{Projected Spin-Orbit Angle}\label{obliquity}
We attempted to measure the projected spin-orbit alignment ($\lambda$) of this system, using the planetary shadow technique \citep[e.g., see,][]{2010MNRAS.403..151C,2014ApJ...790...30J,2016MNRAS.460.3376Z}, by acquiring in-transit high-resolution spectroscopic observations with the FIES instrument. We observed a transit of TOI-1431\,b on the night of 23 May 2020 using FIES, obtaining a total of 30 exposures, each with an exposure time of 300\,s and with airmass decreasing from 1.88 to 1.16 throughout the observation. The wavelength calibrated 2-D reduced spectra were extracted as outlined in Section~\ref{fies_spectra}. We then followed a procedure similar to that of \citet{2020A&A...641A.123H} to attempt to extract the absorption line deformations during the transit of the planet. This involved first cross-correlating a continuum-normalized model template of the stellar spectrum of TOI-1431 with the telluric-corrected FIES spectra (in the stellar rest frame) to produce in-transit cross-correlation functions (CCFs). The in-transit CCFs were then divided by the mean out-of-transit CCF to retrieve the Doppler shadow map.

From our analysis, the planetary shadow is not clearly detected from the FIES data, likely because of the slow rotation of the star. However, observations taken with HARPS-N and EXPRES of three transits of TOI-1431\,b do successfully detect the Rossiter-McLaughlin effect, revealing that the planet is on a retrograde orbit with $\lambda={-155.3^{+16.1}_{-11.3}}^{\circ}$ \citep{2021arXiv210412414S}. The results of \citet{2021arXiv210412414S} suggest that TOI-1431\,b likely experienced high-eccentricity migration in the past that produced its high obliquity orbit and then later the orbit was tidally circularized to the $\sim2.65$\,day period we observe today \citep[for dissenting views of the formation of close-in gas giant planets in-situ via the core-accretion process and spin-orbit misalignments due to processes unrelated to planet migration, see e.g.,][]{2016ApJ...829..114B,2019A&A...629L...1H,2021AJ....161...68L,2021PNAS..11820174H}. This result follows the general pattern observed by other studies that the hottest stars tend to host planets on misaligned orbits \citep[e.g.,][]{winn09,albrecht12,2021AJ....161...68L}.

\section{Conclusions}
\label{conclusions}
We have presented the discovery of the transiting ultra-hot Jupiter, TOI-1431\,b. This planet orbits one of the hottest ($T_\mathrm{eff}=7690^{+400}_{-250}$\,K) and brightest ($V_{T}\sim8.0$) of the known host stars with a period of just $P_{p}=2.650237\pm0.000003$\,d, resulting in it receiving a high amount of insolation flux and being moderately inflated. A joint analysis of the \textit{TESS} light curve, ground-based light curves from MuSCAT2 and LCOGT, and radial velocities from SONG, SOPHIE, FIES, NRES, and EXPRES instruments results in a planet radius of $R_{p}=1.49\pm0.05$\,\Rjup\, ($16.7\pm0.6$\,\Rearth) and a planet mass of $3.12\pm0.18$\,\Mjup, corresponding to a bulk density of $1.17_{-0.16}^{+0.18}$\,\densitycgs.

The planet's phase curve and secondary eclipse have been detected from the \textit{TESS} photometry, providing us with the exciting opportunity to measure the planet's dayside and nightside temperatures as $T_\mathrm{day}=3004\pm64$\,K and $T_\mathrm{night}=2583\pm63$\,K, respectively, when assuming zero dayside geometric albedo. Among the population of hot/ultra-hot Jupiters, TOI-1431\,b has the second highest measured nightside temperature and day-night heat recirculation efficiency.

It is also an excellent candidate for future follow-up observations with JWST and ARIEL to measure its transmission and emission spectra as well as obtain a high-precision global temperature and cloud map of this planet's atmosphere. Furthermore, the discovery and characterization of planets orbiting Am stars, for which few planets have been found, provides good opportunities to probe the tidal interactions between Jovian planets and hot host stars and the potential mechanisms responsible for the creation and evolution of Am stars.

%-------------------------------------------------------------------------------------
\acknowledgments

% SONG
Based on observations made with the Hertzsprung SONG telescope operated on the Spanish Observatorio del Teide on the island of Tenerife by the Aarhus and Copenhagen Universities and by the Instituto de Astrofísica de Canarias.

Funding for the Stellar Astrophysics Centre is provided by The Danish National Research Foundation (Grant agreement no.: DNRF106).

%NOT/FIES
Based on observations made with the Nordic Optical Telescope, operated by the Nordic Optical Telescope Scientific Association at the Observatorio del Roque de los Muchachos, La Palma, Spain, of the Instituto de Astrofisica de Canarias under programs 59-210, 59-503, and 61-804.

%SOPHIE
Based on observations collected with the SOPHIE spectrograph on the 1.93\,m telescope at the Observatoire de Haute-Provence (CNRS), France.

%MuSCAT2
This article is based on observations made with the MuSCAT2 instrument, developed by ABC, at Telescopio Carlos Sánchez operated on the island of Tenerife by the IAC in the Spanish Observatorio del Teide.

%LCOGT/NOIRlab
This work makes use of observations from the LCOGT network.
LCOGT telescope time was granted by NOIRLab through the Mid-Scale Innovations Program (MSIP). MSIP is funded by NSF.

% All about the Benjamins
B.A. is supported by Australian Research Council Discovery Grant DP180100972.  The SONG telescope network is partially supported by the Australian Research Council Linkage, Infrastructure, Equipment and Facilities grant LE190100036.

%grant support for V.A. 
V.A. was supported by a research grant (00028173) from VILLUM FONDEN.

%postdoc funding for I.W.
I.W. is supported by a Heising-Simons \textit{51 Pegasi b} postdoctoral fellowship.

% from TFOP wiki publications page:
Funding for the TESS mission is provided by NASA's Science Mission directorate. We acknowledge the use of public TESS Alert data from pipelines at the TESS Science Office and at the TESS Science Processing Operations Center.  This research has made use of the Exoplanet Follow-up Observation Program website, which is operated by the California Institute of Technology, under contract with the National Aeronautics and Space Administration under the Exoplanet Exploration Program.  Resources supporting this work were provided by the NASA High-End Computing (HEC) Program through the NASA Advanced Supercomputing (NAS) Division at Ames Research Center for the production of the SPOC data products.  This paper includes data collected by the TESS mission, which are publicly available from the Mikulski Archive for Space Telescopes (MAST).
% http://www.not.iac.es/news/publications/
This paper is partially based on observations made with the Nordic Optical Telescope, operated by the Nordic Optical Telescope Scientific Association at the Observatorio del Roque de los Muchachos, La Palma, Spain, of the Instituto de Astrofisica de Canarias.

TD acknowledges support from MIT's Kavli Institute as a Kavli postdoctoral fellow.

%research support for AUKR
M.Y. and H.V.S.  acknowledge  the  support  by  the  research fund  of  Ankara  University (BAP)  through  the project 18A0759001.

%research support for Porto 
This work was supported by Funda\c{c}\~ao para a Ci\^encia e a Tecnologia (FCT) and Fundo Europeu de Desenvolvimento Regional (FEDER) via COMPETE2020 through the research grants UIDB/04434/2020, UIDP/04434/2020, PTDC/FIS-AST/32113/2017 \& POCI-01-0145-FEDER-032113, PTDC/FIS-AST/28953/2017 \& POCI-01-0145-FEDER-028953.

O.D.S.D. is supported in the form of work contract (DL 57/2016/CP1364/CT0004) funded by FCT.

%Nuno Santos acknowledgements.
This work was supported by FCT - Funda\c{c}\~ao para a Ci\^encia e a Tecnologia through national funds and by FEDER through COMPETE2020 - Programa Operacional Competitividade e Internacionaliza\c{c}\~ao by these grants: UID/FIS/04434/2019; UIDB/04434/2020; UIDP/04434/2020; PTDC/FIS-AST/32113/2017 \& POCI-01-0145-FEDER-032113; PTDC/FIS-AST/28953/2017 \& POCI-01-0145-FEDER-028953; PTDC/FIS-AST/28987/2017 \& POCI-01-0145-FEDER-028987.

SH acknowledges CNES funding through the grant 837319

G.H. and I.B. received funding from the French Programme National de Planétologie (PNP) of CNRS (INSU).

This work is partly supported by JSPS KAKENHI Grant Numbers JP18H01265 and JP18H05439, and JST PRESTO Grant Number JPMJPR1775, and a University Research Support Grant from the National Astronomical Observatory of Japan (NAOJ).

%Max Guenther acknowledgements
M.N.G. acknowledges support from MIT's Kavli Institute as a Juan Carlos Torres Fellow.

%Tad Komacek postdoc funding
T.D.K.\ acknowledges support from the 51 Pegasi b Fellowship in Planetary Astronomy sponsored by the Heising–Simons Foundation.

%% To help institutions obtain information on the effectiveness of their 
%% telescopes the AAS Journals has created a group of keywords for telescope 
%% facilities.
%
%% Following the acknowledgments section, use the following syntax and the
%% \facility{} or \facilities{} macros to list the keywords of facilities used 
%% in the research for the paper.  Each keyword is check against the master 
%% list during copy editing.  Individual instruments can be provided in 
%% parentheses, after the keyword, but they are not verified.

\vspace{5mm}
\facilities{TESS, SONG, SOPHIE, NRES ELP, NRES TLV, FIES, EXPRES, LCOGT, MuSCAT2, MASCARA}

%% Similar to \facility{}, there is the optional \software command to allow 
%% authors a place to specify which programs were used during the creation of 
%% the manuscript. Authors should list each code and include either a
%% citation or url to the code inside ()s when available.

%\software{astropy \citep{2013A&A...558A..33A},  
%          Cloudy \citep{2013RMxAA..49..137F}, 
%          SExtractor \citep{1996A&AS..117..393B}
%          }

\software{AstroImageJ \citep{2017AJ....153...77C}, TAPIR \citep{Jensen:2013}, \texttt{Allesfitter} \citep{allesfitter-paper,allesfitter-code}, EXOFASTv2 \citep{2013PASP..125...83E,2017ascl.soft10003E,2019arXiv190709480E}, Astropy \citep{astropy}, Matplotlib \citep{matplotlib}}

%\clearpage

\bibliography{references}{}
\bibliographystyle{aasjournal}

%%%%%%%%%%%%%%%%% APPENDICES %%%%%%%%%%%%%%%%%%%%%

\clearpage
\newpage

\appendix

\begin{deluxetable*}{lccccccc}
\tabletypesize{\scriptsize}
%\tablewidth{\columnwidth}
%\tablecolumns{3}
\tablecaption{A summary of the ground-based transit follow-up observations taken of TOI-1431. \label{tab:SG1_summary}}
\tablehead{
\colhead{Telescope}  & \colhead{Camera}  & \colhead{Filter}  & \colhead{UT Date}  & \colhead{Coverage}  & \colhead{Precision (ppt)}  & \colhead{Joint Fit}  & \colhead{Comments}
}
\startdata
CDK14           & STXL-6303E     & $z'$                     & 2019-12-24    & Full      & 2.62    & Yes   & Minimal systematics  \\
CDK14           & STXL-6303E     & $g'$                     & 2019-12-24    & Full      & 2.32    & Yes   & Minimal systematics  \\
TCS             & MuSCAT2        & $g'$                     & 2020-05-16    & Full      & 0.93    & Yes  & Minimal systematics  \\
TCS             & MuSCAT2        & $r'$                     & 2020-05-16    & Full      & 0.97    & Yes  & Minimal systematics  \\
TCS             & MuSCAT2        & $i'$                     & 2020-05-16    & Full      & 0.94    & Yes  & Minimal systematics  \\
TCS             & MuSCAT2        & $z_s$                    & 2020-05-16    & Full      & 1.01    & Yes  & Minimal systematics  \\
TCS             & MuSCAT2        & $g'$                     & 2020-05-23    & Full      & 1.23    & No   & Limited pre-ingress data and systematics  \\
TCS             & MuSCAT2        & $r'$                     & 2020-05-23    & Full      & 1.07    & No   & Limited pre-ingress data and systematics  \\
TCS             & MuSCAT2        & $i'$                     & 2020-05-23    & Full      & 1.08    & No   & Limited pre-ingress data and systematics  \\
TCS             & MuSCAT2        & $z_s$                    & 2020-05-23    & Full      & 1.09    & No   & Limited pre-ingress data and systematics  \\
AUKR            & ALTA U47       & $z'$                     & 2020-06-16    & Full      & 3.00    & Yes   & Some systematics  \\
LCOGT-McD       & Sinistro       & PANSTARRS Y              & 2020-07-13    & Ingress   & 0.48    & No   & Ingress only  \\
SCT             & ST7XME         & \textit{TESS} band       & 2020-08-08    & Full      & 1.49    & Yes   & Minimal systematics  \\
ULMT            & STX 16803      & $i'$                     & 2020-09-20    & Full      & 1.51    & Yes   & Some systematics  \\
ULMT            & STX 16803      & $i'$                     & 2020-09-28    & Full      & 3.41    & No   & Significant systematics  \\
LCOGT-McD       & Sinistro       & PANSTARRS Y              & 2020-10-14    & Full      & 1.18    & Yes  & Minimal systematics  \\
\enddata
\end{deluxetable*}

\begin{figure*}
  \includegraphics[width=\linewidth]{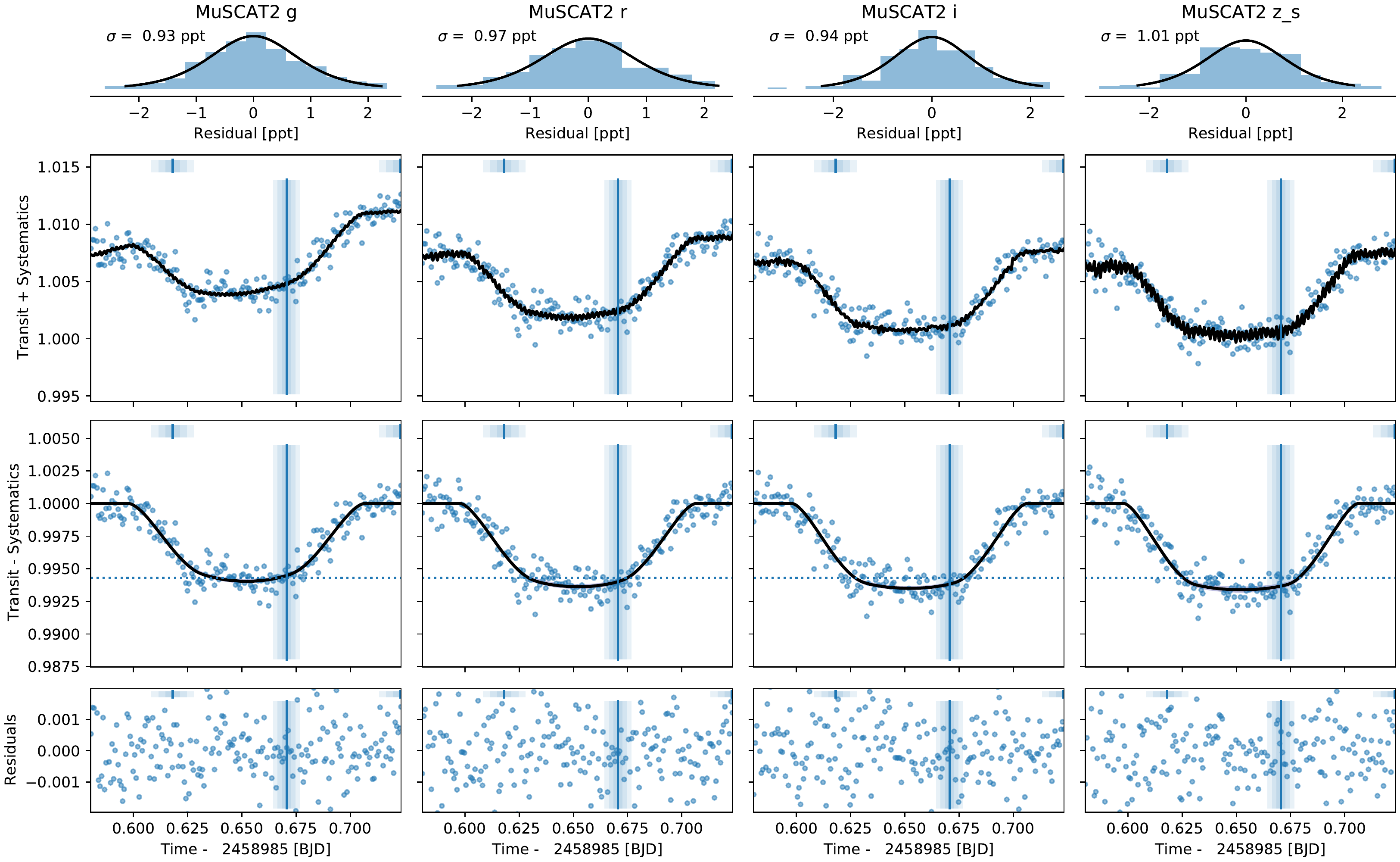}
  \caption{An observed transit on 16 May 2020 by MuSCAT2 using simultaneous multi-color photometry in $g'$, $r'$, $i'$, and $z_s$ bands. The top panel shows the distribution of residuals to the fit of the light curve in each band. The second panel from the top shows the fit to the transit plus systematics. The third panel shows the fit to the transit with the systematics removed. The three blue vertical bars in the second, third, and fourth panels represent the predicted ingress, mid-transit, and egress times along with their $1\sigma$ uncertainties based on the ephemeris used at the time. The horizontal dotted blue lines in the third panel show the predicted transit depth. The bottom panel shows the resulting residuals to the fit of the light curves.}
  \label{fig:Muscat2}
\end{figure*}

\startlongtable
\begin{deluxetable}{cccc}
\tabletypesize{\scriptsize}
\tablecaption{Radial Velocity Measurements\label{tab:vels}}
\tablehead{
\colhead{Date\tablenotemark{a}}  &  \colhead{RV \tablenotemark{b}}   &   \colhead{$\sigma_\mathrm{{RV}}$} & \colhead{Instrument}
\\
\colhead{BJD - 2400000} & \colhead{($\mathrm{m\ s^{-1}}$)} & \colhead{($\mathrm{m\ s^{-1}}$)} & \colhead{}
}
\startdata
58805.564199 & -25696.6 & 14.5  & ELP    \\
58806.565002 & -25761.0 & 15.3  & ELP    \\
58807.566340 & -25346.6 & 13.3  & ELP    \\
58808.570467 & -25830.7 & 29.9  & ELP    \\
58814.663425 & -25557.9 & 26.2  & ELP    \\
58815.307593 & 9.2      & 14.2  & FIES   \\
58818.618086 & -25457.5 & 18.0  & ELP    \\
58836.224800 & -25039.0 & 7.0   & SOPHIE \\
58838.252200 & -25441.0 & 7.0   & SOPHIE \\
58840.399100 & -25532.0 & 45.0  & SOPHIE \\
58841.235500 & -25222.0 & 7.0   & SOPHIE \\
58841.352100 & -25152.0 & 6.0   & SOPHIE \\
58859.232100 & -25528.0 & 7.0   & SOPHIE \\
58860.267700 & -24987.0 & 6.0   & SOPHIE \\
58861.259600 & -25374.0 & 6.0   & SOPHIE \\
58914.768952 & -25465.4 & 53.1  & SONG   \\
58919.763375 & -25434.9 & 81.7  & SONG   \\
58921.764647 & -24969.6 & 74.1  & SONG   \\
58942.562272 & -25283.8 & 19.2  & TLV    \\
58943.567282 & -25788.4 & 15.4  & TLV    \\
58948.564452 & -25514.9 & 16.4  & TLV    \\
58952.584719 & -25528.9 & 16.3  & TLV    \\
58955.532830 & -25299.7 & 28.7  & TLV    \\
58956.668739 & -25303.1 & 139.4 & SONG   \\
58961.645589 & -25067.3 & 56.1  & SONG   \\
58965.721756 & -25219.0 & 44.1  & SONG   \\
58980.495174 & -25900.9 & 19.9  & TLV    \\
58982.464359 & -25256.1 & 27.5  & TLV    \\
58983.637401 & -25477.9 & 64.2  & SONG   \\
58987.442635 & -25265.9 & 22.1  & TLV    \\
58988.470752 & -25659.4 & 27.9  & TLV    \\
58988.619270 & -25374.0 & 26.6  & SONG   \\
58989.466532 & -25588.9 & 42.2  & TLV    \\
58990.449283 & -25278.6 & 20.1  & TLV    \\
58990.567802 & -24.1    & 7.9   & FIES   \\
58991.456242 & -25896.5 & 26.1  & TLV    \\
58991.562807 & -581.4   & 15.6  & FIES   \\
58991.594498 & -565.9   & 12.9  & FIES   \\
58991.642527 & -25435.4 & 18.0  & SONG   \\
58991.698593 & -565.4   & 11.0  & FIES   \\
58992.606914 & -69.5    & 12.5  & FIES   \\
58992.615274 & -75.2    & 18.9  & FIES   \\
58993.658735 & -318.3   & 11.6  & FIES   \\
58993.662679 & -339.5   & 15.6  & FIES   \\
58993.666687 & -314.2   & 14.1  & FIES   \\
58993.670878 & -329.4   & 18.3  & FIES   \\
58993.674862 & -315.8   & 11.2  & FIES   \\
58993.678937 & -333.8   & 13.4  & FIES   \\
58993.686045 & -334.8   & 9.8   & FIES   \\
58993.690092 & -329.0   & 13.9  & FIES   \\
58993.694137 & -337.2   & 13.8  & FIES   \\
58993.698146 & -345.5   & 9.7   & FIES   \\
58994.571980 & -500.4   & 9.3   & FIES   \\
58994.621049 & -471.6   & 8.0   & FIES   \\
58994.704411 & -429.4   & 11.7  & FIES   \\
58995.443468 & -25324.0 & 28.7  & TLV    \\
58995.570851 & 15.5     & 11.5  & FIES   \\
58995.620876 & 13.7     & 7.2   & FIES   \\
58995.703476 & 0.0      & 11.9  & FIES   \\
58996.502720 & -25732.7 & 26.1  & TLV    \\
58997.422153 & -25702.1 & 38.1  & TLV    \\
58998.422132 & -25287.5 & 23.5  & TLV    \\
58999.424288 & -25873.9 & 19.0  & TLV    \\
58999.624293 & -25458.0 & 76.4  & SONG   \\
59001.471298 & -25606.2 & 20.1  & TLV    \\
59001.641856 & -25224.1 & 28.8  & SONG   \\
59002.428683 & -25802.8 & 21.2  & TLV    \\
59002.787631 & -55.3 & 2.5 & EXPRES      \\
59003.954076 & 167.7 & 1.8 & EXPRES      \\
59004.452546 & -25721.2 & 16.9  & TLV    \\
59004.602433 & -25375.3 & 24.6  & SONG   \\
59004.953876 & -279.8 & 2.0 & EXPRES     \\
59005.450710 & -25560.8 & 18.8  & TLV    \\
59006.382260 & -25320.8 & 19.6  & TLV    \\
59007.421885 & -25873.3 & 17.2  & TLV    \\
59007.604415 & -25444.7 & 39.9  & SONG   \\
59008.456417 & -25235.3 & 34.6  & TLV    \\
59009.487869 & -25570.9 & 18.7  & TLV    \\
59009.566599 & -25187.8 & 24.2  & SONG   \\
59010.502769 & -25751.8 & 19.5  & TLV    \\
59010.624528 & -25295.0 & 24.9  & SONG   \\
59012.426456 & -25777.6 & 16.5  & TLV    \\
59012.640398 & -25419.6 & 29.3  & SONG   \\
59012.955625 & -271.6 & 1.9 & EXPRES     \\
59018.520808 & -447.3   & 17.6  & FIES   \\
59018.597262 & -406.4   & 15.7  & FIES   \\
59018.693433 & -330.5   & 8.5   & FIES   \\
59019.513133 & 34.7     & 8.4   & FIES   \\
59019.594250 & 2.8      & 7.5   & FIES   \\
59019.694408 & -25.8    & 9.3   & FIES   \\
59020.514030 & -516.7   & 10.5  & FIES   \\
59020.590422 & -546.9   & 9.0   & FIES   \\
59020.688977 & -552.9   & 12.6  & FIES   \\
59023.610494 & -25418.5 & 21.3  & SONG   \\
59027.569361 & -24874.0 & 24.4  & SONG   \\
59029.510059 & -25065.6 & 20.3  & SONG   \\
59033.957319 & -287.7 & 2.2 & EXPRES     \\
59059.915400 & -36.2 & 2.1 & EXPRES      \\
59059.919605 & -44.0 & 2.0 & EXPRES      \\
59059.923762 & -38.9 & 2.0 & EXPRES      \\
59059.927997 & -45.9 & 2.2 & EXPRES      \\
59059.932407 & -51.7 & 2.2 & EXPRES      \\
59059.936606 & -50.3 & 2.3 & EXPRES      \\
59059.940835 & -50.7 & 2.2 & EXPRES      \\
59059.945055 & -58.2 & 2.1 & EXPRES      \\
59059.949247 & -59.5 & 1.9 & EXPRES      \\
59059.953490 & -62.4 & 2.3 & EXPRES
\enddata
\tablenotetext{a}{The dates for each observation are reported as BJD at the UTC time at the midpoint of the exposure.}
\tablenotetext{b}{FIES and EXPRES radial velocities are given at an arbitrary zero point.}
\end{deluxetable}

\begin{figure}
   \includegraphics[width=\linewidth]{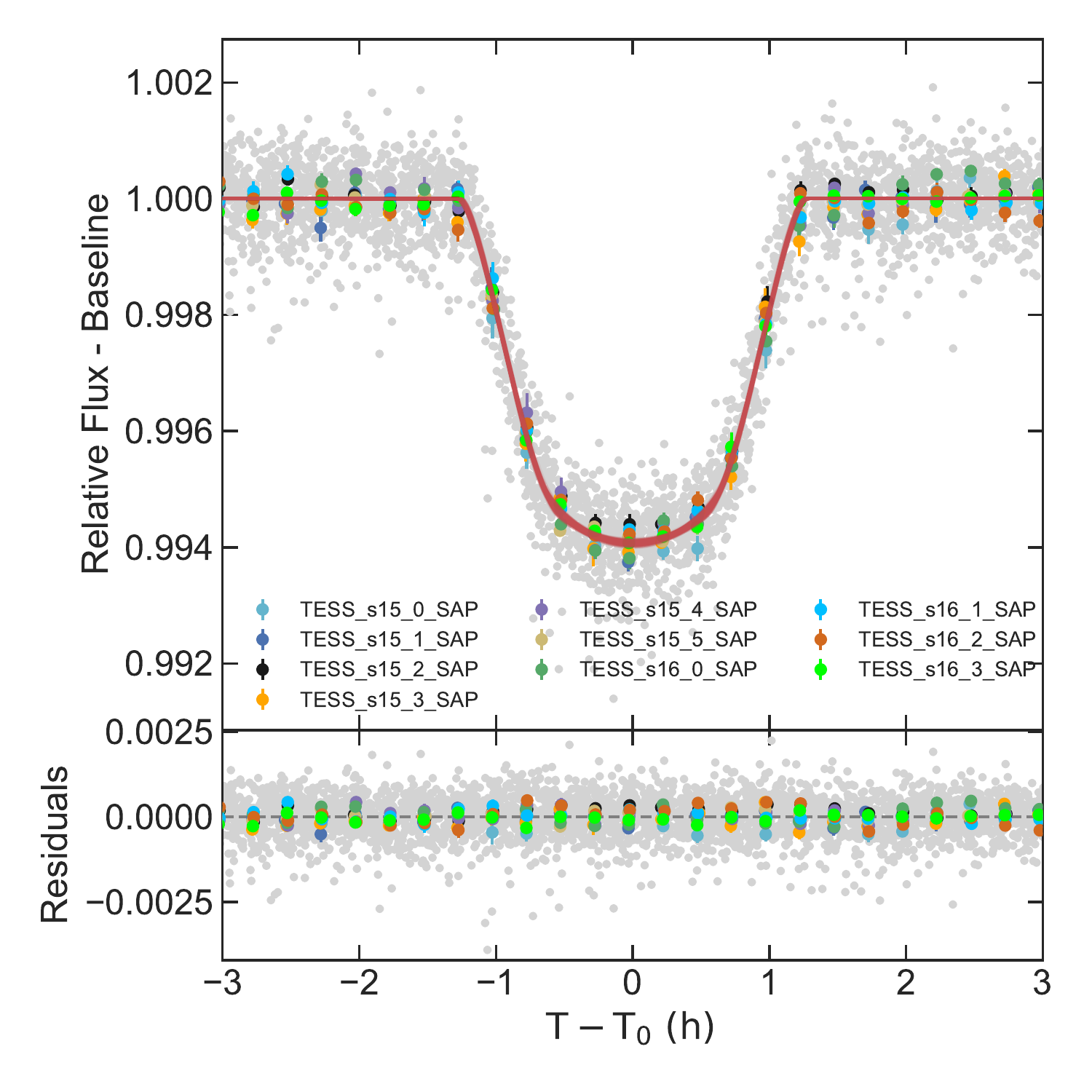}
   \caption{Phase-folded and detrended light curve of TOI-1431 from the simultaneous hybrid cubic spline detrending model using the SAP data. The colored points represent the individual phased light curve segments across Sectors 15 and 16 that have been binned at a cadence of 15\,m. The red solid lines are 20 light curve models drawn from the posteriors of the Nested Sampling analysis in \texttt{Allesfitter}.}
   \label{fig:TESS_LC_SAP}
\end{figure}

\begin{figure*}
   \includegraphics[width=\linewidth]{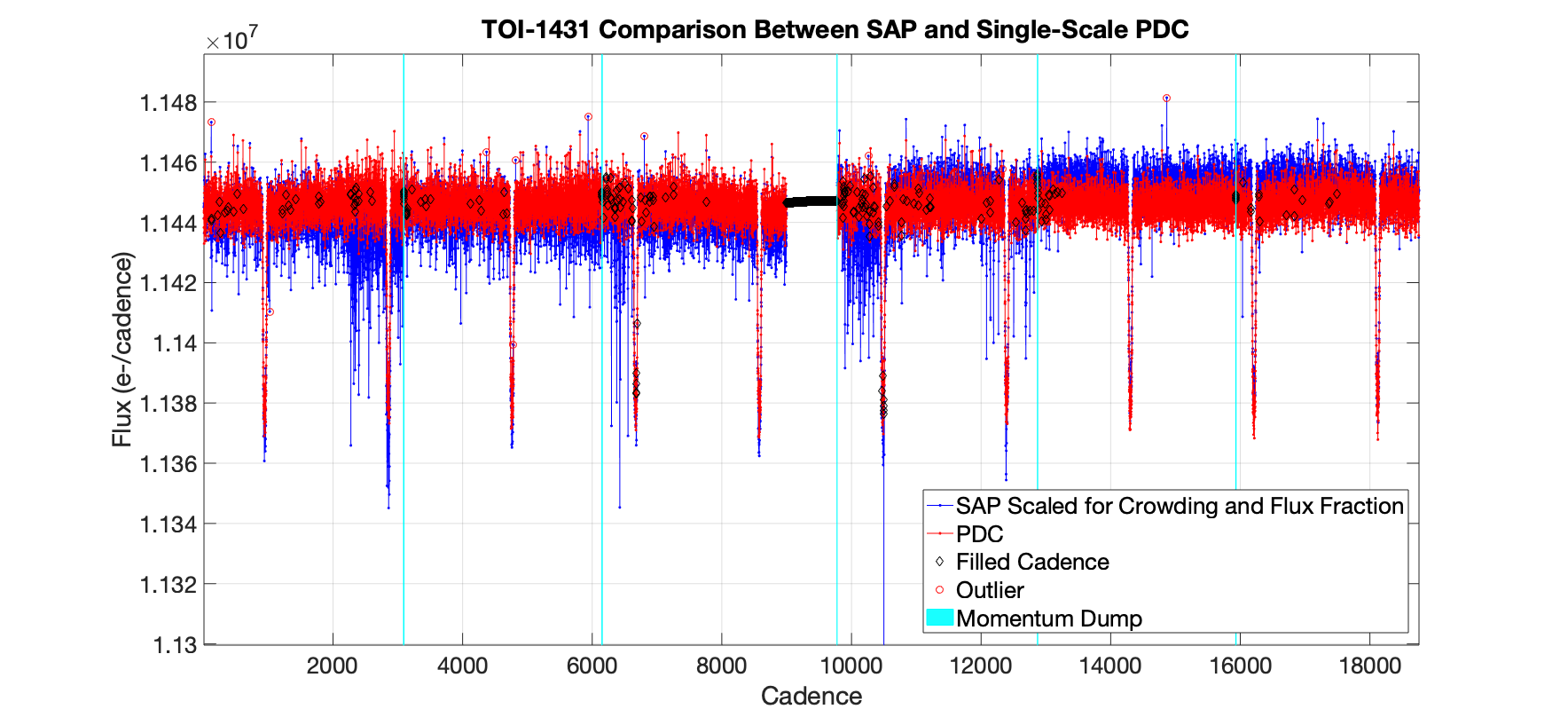}
   \caption{\textit{TESS} Sector 15 photometry of TOI-1431 from the SAP (blue points) and single-scale MAP PDC (red points) light curves. The SAP light curve shows `whisker'-like flux dips that are caused by brief spacecraft pointing excursions, some of which overlap with the transit events. The correction applied to the single-scale MAP PDC light curve from the use of the Co-trending Basis Vectors has removed these features.}
   \label{fig:TESS_SAP_vs_PDC}
\end{figure*}

\begin{figure}
  \includegraphics[width=\linewidth]{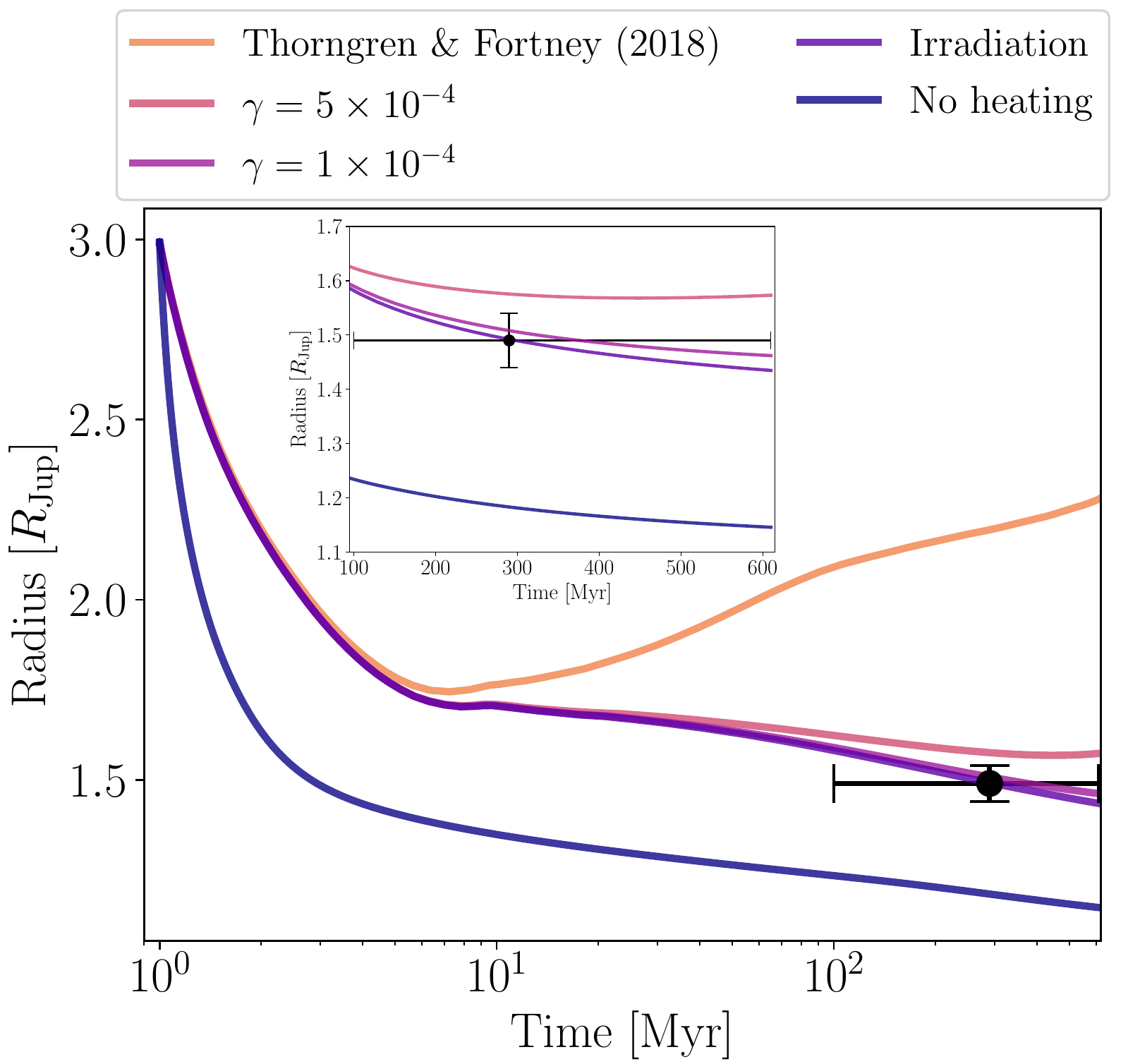}
  \caption{Radius evolution of TOI-1431b for various assumptions about the presence of irradiation and deep deposited heating. The curve labeled  \cite{Thorngren:2017} includes irradiation and applies deposited heating in the deep interior with a strength that varies with equilibrium temperature, as in Equation (34) of \cite{Thorngren:2017}. The models labeled $\gamma = 5 \times 10^{-4}$ and $\gamma = 1 \times 10^{-4}$ respectively assume that $0.05\%$ and $0.01\%$ of the incident stellar power is converted to heat deposited deep in the planetary interior and include irradiation. The model labeled ``irradiation'' considers only irradiation with no deposited heating, and the model labeled ``no heating'' does not include irradiation or deposited heating. Evolution models are conducted with {\tt MESA} \citep{Paxton:2011,Paxton:2013,Paxton:2015,Paxton:2018aa,Paxton:2019aa} using the model setup described in \cite{2020ApJ...893...36K} and stellar evolution tracks from \cite{Choi:2016aa} and \cite{Dotter:2016aa}. The present-day radius of TOI-1431b can be fit by planetary evolution models including only irradiation slowing cooling, but a small amount of deep deposited heating corresponding to $< 0.05\%$ of the incident stellar power converted to heat is allowed.}
  \label{fig:radius_time}
\end{figure}

\begin{figure*}
  \includegraphics[width=\linewidth]{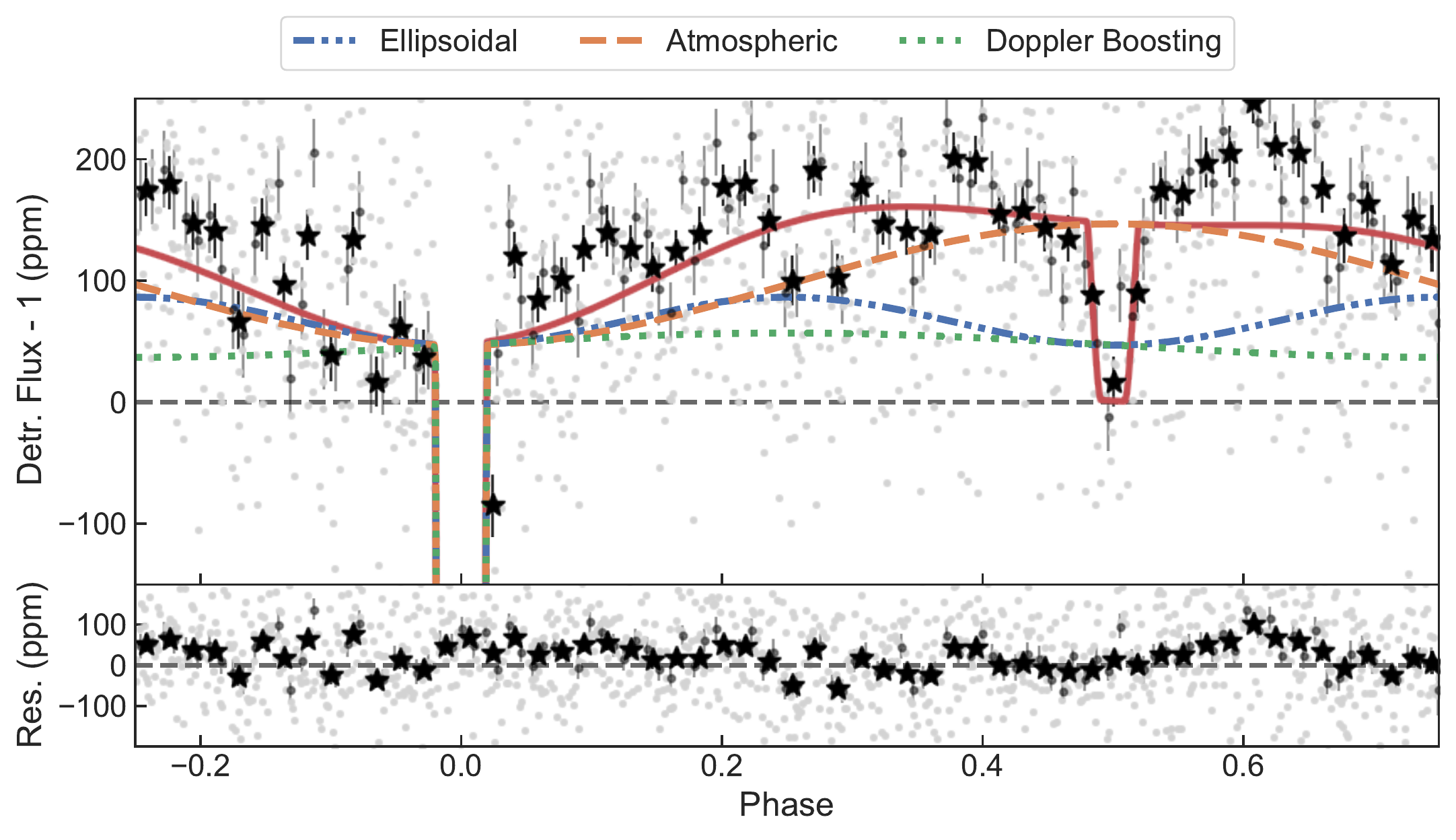}
  \caption{The phase-folded \textit{TESS} SAP light curve for TOI-1431\,b, zoomed in to show the phase curve and secondary eclipse, similar to Figure~\ref{fig:TESS_phase_curve}.}
  \label{fig:TESS_phase_curve_SAP}
\end{figure*}

\end{document}